\documentclass[aps,pra,twocolumn,showpacs,superscriptaddress,amsmath,amssymb,floatfix,10pt,reprint]{revtex4-1}  
\usepackage{graphicx}  
\usepackage{dcolumn}   
\usepackage{bm}        
\usepackage{amssymb}   
\usepackage{framed}
\usepackage{amsmath}
\usepackage{hhline}
\usepackage{adjustbox}
\usepackage{color}
\usepackage{subfigure,amsmath,verbatim,moreverb}
\usepackage{tabularx}
\usepackage{lipsum}
\usepackage{longtable}
\usepackage{booktabs}
\usepackage{rotating}
\usepackage[normalem]{ulem}
\usepackage{amsfonts}
\usepackage{booktabs}
\usepackage{color,soul}
\usepackage{etoolbox}
\AtBeginEnvironment{align}{\setcounter{subeqn}{0}}
\newcounter{subeqn} %

\newcommand{\R}{\mathbf{r}}

\newcommand{\Eq}[1]{Eq.~(\ref{#1})}

\newcommand{\Refs}[1]{Refs. \onlinecite{#1}}
\newcommand{\Fig}[1]{Fig. \ref{#1}}
\newcommand{\Tab}[1]{Tab. \ref{#1}}

\begin{document}

\title{Semilocal Meta-GGA Exchange-Correlation Approximation From Adiabatic Connection Formalism: Extent and Limitations}

\author{Subrata Jana}
\email{subrata.niser@gmail.com}
\affiliation{Department of Chemistry \& Biochemistry, The Ohio State University, Columbus, OH 43210, USA}
\affiliation{Present address: Department of Molecular Chemistry and Materials Science, Weizmann Institute of Science, Rehovoth 76100, Israel}
\author{Szymon \'Smiga}
\email{szsmiga@fizyka.umk.pl}
\affiliation{Institute of Physics, Faculty of Physics, Astronomy and Informatics, Nicolaus Copernicus University in Toru\'n,ul. Grudzik'{a}dzka 5, 87-100 Toru\'n, Poland}
\author{Lucian A. Constantin}
\affiliation{Istituto di Nanoscienze, Consiglio Nazionale delle Ricerche CNR-NANO, 41125 Modena,Italy}
\affiliation{Present address: Institute for Microelectronics and Microsystems (CNR-IMM),
Via Monteroni, Campus Unisalento, 73100 Lecce, Italy}
\author{Prasanjit Samal}
\affiliation{School of Physical Sciences, National Institute of Science Education and Research, HBNI,
Bhubaneswar 752050, India}

\date{\today}

\begin{abstract}

The incorporation of a strong interaction regime within the approximate, semilocal exchange-correlation functionals still remains a very challenging task for density functional theory. One of the promising attempts in this direction is the recently proposed adiabatic connection semilocal correlation (ACSC) approach [\textit{Phys. Rev. B} \textbf{2019}, \textit{99}, 085117] allowing to construct the correlation energy functionals by interpolation of the high and low-density limits for the given semi-local approximation.  The current study extends the ACSC method to the meta-GGA level of theory, providing some new insights. As an example, we construct the correlation energy functional base on the high and low-density limits of the Tao-Perdew-Starverov-Scuseria (TPSS) functional. Arose in this way TPSS-ACSC functional is one electron self-interaction free, accurate for the strictly correlated, and quasi-two-dimensional regimes. Based on simple examples, we show the advantages and disadvantages of ACSC semi-local functionals and provide some new guidelines for future developments in this context.

\end{abstract}

\maketitle

\section{Introduction}

The electronic structure calculations of quantum chemistry, solid-state physics, and material sciences become
enormously
simple since the advent of the Kohn-Sham (KS)~\cite{kohn1965self,hohenberg1964inhomogeneous} density functional
theory (DFT)~\cite{burke2012perspective}. In DFT, the development of efficient yet accurate exchange-correlation
(XC)
functional, which contains all the many-body quantum effects beyond the Hartree method, is one of the main
research topics since the last couple of
decades and continues to be the same in recent times. The accuracy of the ground-state properties
of electronic systems
depends on the XC functional approximation (density functional approximation - DFA). The non-empirical XC
functionals are developed by satisfying many quantum mechanical exact
constraints~\cite{levy2010simple,levy2016mathematical,sun2015strongly,tao2016accurate}
such as:
density scaling rules of XC functionals due to coordinate transformations \cite{levy1985hellmann,
levy2016mathematical,gorling1992requirements,fabiano2013relevance}, second (and fourth) order gradient expansion
of exchange and correlation energies
\cite{svendsenPRB96,antoniewiczPRB85,huPRB86,bruecknerPR68,Argaman2022leading,daas2022gradient,Daas2022arxiv}
, low density, and high
density
limit of the correlation energy functional\cite{gorling1994exact,gorling1993correlation,gorling1995hardness},
asymptotic behavior of the XC energy density or
potential \cite{dellasalaPRL02,engelZPD92,horowitz2009position,constantin2011adiabatic,constantin2016semilocal,
niquet2003asymptotic,almbladh1985exact,umrigar1994accurate}, quasi-2D behavior of the
XC energy \cite{pollack2000evaluating,kaplan2018collapse,constantin2016simple,constantin2008dimensional}, and
exact properties of the XC hole ~\cite{tao2016accurate,tao2008exact,pvrecechtvelovaJCP14,pvrecechtvelovaJCP15}.

Different rungs of Jacob's ladder \cite{perdew2001jacob} classification of non-empirical XC approximations
are developed based on the use of various ingredients, from the simple spin densities and their gradients,
until the occupied and unoccupied KS orbitals and energies~\cite{grimme2006semiempirical,C8CP03852J,bartlett:2005:abinit2,grabowski2014orbital,SmigaJCP2020,SOSa,seidl2000simulation}.
The first rung of the ladder is the local density approximations (LDA)\cite{kohn1965self}. Next rungs are
represented by semilocal functionals, such as generalized gradient
approximations (GGA)~\cite{perdewPRL96,scuseriaREVIEW05} and meta-GGA ~\cite{taoPRL03,
sun2015strongly,tao2016accurate,jana2019improving,patra2019relevance,jana2021szs,patra2020way,jana2021accurate}. Higher rungs are known as 3.5 rung XC functionals
\cite{janesko2012nonspherical,janesko2013rung,janesko2010rung,janesko2012nonempirical,
janesko2018long,constantin2016hartree,constantin2017modified},
hybrids and hyper-GGAs \cite{perdew2008density,perdew2005prescription,odashima2009nonempirical,
arbuznikov2011advances,jaramillo2003local,kummel2008orbital,becke2005real,becke2007unified,becke2003real,
becke2013density,patra2018long,jana2018efficient,jana2019screened,jana2018cpl,jana2020screened,jana2019long,jana2022solid}, double hybrids~\cite{grimme2007doublehybrid,neil2014construction,scan0,sharkas2011double,
souvi2014double,doi:10.1063/1.3640019},
and adiabatic connection (AC) random-phase approximation (RPA) like methods and
DFT version of the coupled-cluster theory~\cite{toulouse2009adiabatic,ruzsinszky2016kernel,ruzsinszky2010rpa,
bates2017convergence,huPRB86,terentjev2018gradient,constantin2016simple,
corradini1998analytical,erhard2016power,patrick2015adiabatic,bartlett2005exchange,bartlett2005ab,
grabowski2013optimized,grabowski2014orbital}.

Specifically, we recall that the AC formalism \cite{langreth1975exchange,gunnarsson1976exchange,savin2003adiabatic,
cohen2007assessment,ernzerhof1996construction,burke1997adiabatic,colonna1999correlation}, used in various
sophisticated XC functionals ~\cite{savin2003adiabatic,ernzerhof1996construction,burke1997adiabatic,
toulouse2009adiabatic,patrick2015adiabatic,cohen2007assessment,colonna1999correlation,
adamo1998exchange,perdew2001exploring,liu2009adiabatic,magyar2003accurate,
sun2009extension,seidl2010adiabatic,vuckovic2016exchange,fabiano2018investigation,
vuckovic2018restoring,kooi2018local,seidl2018communication,
constantin2019correlation}, is based on the coupling constant (or interaction strength) integral formula~\cite{langreth1975exchange, savin2003adiabatic,cohen2007assessment,ernzerhof1996construction,burke1997adiabatic}
\begin{eqnarray}
& E_{xc}[n]=\int_0^1~d\alpha~W_{\alpha}[n]~, \nonumber \\
& W_{\alpha}[n]=\langle\Psi_\alpha[n]|\hat{V}_{ee}|\Psi_\alpha[n]\rangle-U[n],
 \label{ineq1}
\end{eqnarray}
%
where $\hat{V}_{ee}$ is the
Coulomb operator, $U[n]$ is the Hartree energy, $\Psi_\alpha[n]$ is the anti-symmetric
wave function that yields the density $n(\R)$ and minimizes
the expectation value $<\hat{T}+\alpha \hat{V}_{ee}>$,
with $\hat{T}$ being the kinetic energy operator, and $\alpha$ the coupling constant.
Eq.~(\ref{ineq1}) can be seen as the exact definition of the XC functional, and it connects a non-interacting
single particle system ($\alpha=0$) to a fully
interacting
one
($\alpha=1$). Note that the $\alpha\to 0$ limit is known as the weak-interaction limit ( or high-density or $r_s\to 0$ limit, where $r_s$ is the local Seitz radius), where
the perturbative approach is valid.
Thus, the well-known second-order G\"{o}rling-Levy perturbation theory
(GL2)~\cite{gorling1994exact,gorling1993correlation,
gorling1995hardness,gorling1998exact} can be applied in the weak-interaction limit, and $W_{\alpha}[n]$ can be expanded as~\cite{seidl2000simulation}
\begin{equation}
 W{_{\alpha\to 0}}[n] = W_0[n] + W_0'[n]\alpha~ + \dots
 \label{ineq2}
\end{equation}
where $W_0=E_x$ and $W_0'[n]=2E_c^{GL2}[n]$. On the other hand, the
strong-interaction limit ( or low-density or $r_s\to \infty$ limit) of $W_\alpha[n]$ is given as~\cite{seidl2000density,seidl2000simulation,gori2009electronic,liu2009adiabatic}
\begin{equation}
 {\textcolor{black}{W{_{\alpha\to \infty}}[n] = W_\infty[n] + W_\infty'[n]\alpha^{-1/2} + O(\alpha^{-p})\dots,~~~p\geq 3/4}}
 \label{ineq3}
\end{equation}
where $W_\infty[n]$ and $W_\infty'[n]$ have a highly non-local density dependence, captured by the strictly-correlated electrons (SCE) limit \cite{seidl07,gorigiorgi09,MalMirCreReiGor-PRB-13}, and their exact evaluation in general cases is a non-trivial problem.

In particular, one of the successful attempts at practical usability of the AC DFAs came through the interaction strength
interpolation (ISI) method by Seidl and
coworkers~\cite{seidl2000simulation,seidl2000density,perdew2001exploring,seidl2010adiabatic,gori2009electronic,gori2010density,fabiano2016interaction,seidl1999strictly,seidl2018communication, seidl2016challenging} where the DFA formula is built by interpolating between the weak- and strong interaction regimes. The $\alpha \to\infty$ limit is approximated by semilocal gradient expansions (GEA) derived within the point-charge-plus-continuum (PC) model~\cite{seidl2000simulation,seidl2000density,perdew2001exploring,seidl2010adiabatic}.
Based on this form, the ISI has been tested for various
applications~\cite{fabiano2016interaction,giarrusso2018assessment,fabiano2018investigation}. Also, several modifications of the ISI have been suggested~\cite{mirtschink2012energy,gori2009electronic,liu2009adiabatic, SeiPerLev-PRA-99, NCIISI} as well as the PC model itself such as the hPC\cite{SCFISI} or modified PC (mPC)~\cite{constantin2019correlation} which was found to be more robust for the
quasi-two dimensional (quasi-2D) density regime.

Recently, based on the ISI formula, the adiabatic connection semilocal correlation (ACSC) method was introduced\cite{constantin2019correlation}, showing the alternative path of construction of semilocal correlation energy functionals. The ACSC formula interpolates the high and low-density limit for the given semi-local DFA directly, contrary to the standard path where the interpolation is done at the local LDA level and then corrected by gradient or meta-GGA corrections\cite{perdewPRL96,taoPRL03}. We recall that in \Refs{constantin2019correlation}, the ACSC functional was built using Perdew-Burke-Ernzerhof (PBE)\cite{perdewPRL96} high-density formula and mPC model showing similar or improved accuracy over its PBE precursor proving in the same time the evidence for the robustness of ACSC construction.

\begin{figure}[t]
\includegraphics[width=\columnwidth]{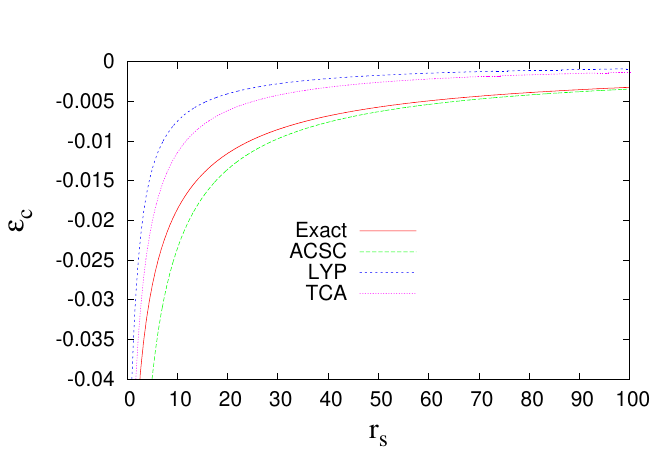}
\caption{Correlation energy per particle $\epsilon_c$
versus the bulk parameter $r_s=(\frac{3}{4\pi n})^{1/3}$, for the uniform electron gas. See text for details of the methods and exact reference curve.}
\label{f1}
\end{figure}

Motivated by the progress in this direction, this paper extends the ACSC method at the meta-GGA level and provides new insights in this context.

In the following, we briefly recall some aspects of ACSC functional construction and investigate a few available approximations for the high- and low-density regimes.
Based on that, we propose an extension of the ACSC method to the meta-GGA level using the high and low-density limits of the Tao-Perdew-Staroverov-Scuseria (TPSS)~\cite{taoPRL03} DFA. Following that, we apply ACSC correlation energy functionals to some model systems (Hooke's atom and H$_2$ molecule) and real calculations (the atomization energies of several small molecules) to show some advantages and current limitations of ACSC functional construction. Lastly, we conclude by discussing the possible advances of the present construction.

\section{Theory}

\subsection{Background of the Adiabatic Connection Semilocal Correlation (ACSC)}
\label{sec-theory}

Following Ref.~\cite{constantin2019correlation}, the ACSC correlation energy per particle is given as (Eq.(15) of ref.~\cite{constantin2019correlation}),
\begin{eqnarray}
&& n\epsilon_c^{ACSC}(\R)=\int_0^1d\alpha\; w_{c,\alpha}(\R)  \label{eq1} \\
&& w_{c,\alpha}(\R) = w_\infty(\R)-w_0(\R) + \frac{ \mathcal{X}(\R)}{\sqrt{1+  \mathcal{Y}(\R)\alpha} +  \mathcal{Z}(\R)} \label{eq1a}
\end{eqnarray}

\begin{eqnarray}
&& n\epsilon_c^{ACSC}(\R)= w_\infty(\R)-w_0(\R) \nonumber\\
&& +\frac{2\mathcal{X}(\R)}{\mathcal{Y}(\R)}[\sqrt{1+\mathcal{Y}(\R)}-1-\mathcal{Z}(\R)
\ln(\frac{\sqrt{1+\mathcal{Y}(\R)}+\mathcal{Z}(\R)}{1+\mathcal{Z}(\R)})]~.\nonumber\\
\label{eq1b}
\end{eqnarray}
The above expression represents a general form for the correlation energy density derived
from the ISI formula~\cite{seidl2000simulation,seidl2000density,perdew2001exploring,seidl2010adiabatic,
 gori2009electronic,gori2010density,fabiano2016interaction,seidl1999strictly,seidl2018communication,
seidl2016challenging} with %
\begin{eqnarray}
& \mathcal{X}(\R)=-2 w'_0(\R) (w'_\infty(\R))^2/ (w_0(\R)-w_\infty(\R))^2,\nonumber\\
& \mathcal{Y}(\R)=4 (w'_0(\R))^2 (w'_\infty(\R))^2/ (w_0(\R)-w_\infty(\R))^4,\nonumber\\
& \mathcal{Z}(\R)=-1-2 w'_0(\R) (w'_\infty(\R))^2/ (w_0(\R)-w_\infty(\R))^3.\nonumber\\
\label{eq2}
\end{eqnarray}
and where $w_0$, $w'_0$, and $w_\infty$, $w'_\infty$ denote the approximation for energy densities for high- ( or weak-interaction) and low-density ( or strong-interaction) limits, respectively.

Considering the accuracy of \Eq{eq1}, it depends on three main aspects:
\begin{itemize}
    \item[i)] the interpolation formula is used to define the $w_{c,\alpha}(\R)$ integrand in \Eq{eq1}. In \Refs{constantin2019correlation} (and here \Eq{eq1a}), the ISI interpolation formula was utilized to define ACSC. We note, however, that for this choice the $W_\alpha[n]$ contains a  spurious term proportional to $\alpha^{-1}$ in its strong-interaction limit ($\alpha\to \infty$)~\cite{seidl2000density} which has been corrected in refs.~\cite{liu2009adiabatic,gorigiorgi09}. To be consistent with our previous work, we stuck with the ISI formula. Nonetheless, other possibilities also exist\cite{mirtschink2012energy,gori2009electronic,liu2009adiabatic, SeiPerLev-PRA-99}.

    \item[ii)] the approximation for $\alpha \to \infty$ limit. Several possibilities exist, e.g.,
    (exact treatment by employing SCE formulas (numerically expensive but feasible) or much less time consuming variants such as
    mPC\cite{constantin2019correlation}, hPC\cite{SCFISI} or the ones derived from semi-local DFA via the procedure described in \Refs{seidl2000density}. Note that by choosing different $\alpha \to \infty$ limits, one can incorporate in ACSC formula different physics, e.g., good performance for the quasi-2D regime.
    \item[iii)] the approximation for $\alpha \to 0$ limit. In principle, this limit can be taken into account exactly by considering the exact exchange (EXX) and  
    GL2 limit\cite{gorling1993correlation,kooi2018local,Gauge3}. However,  evaluation 
    of the GL2 correlation energy density on the numerical grid would likely be computationally expensive. Hence, in \Refs{constantin2019correlation}, the non-local contributions have been substituted by semi-local high-density counterparts obtained from PBE functional\cite{perdewPRL96}.
\end{itemize}

This work extends the ACSC DFA by considering all input quantities at the semi-local (SL) meta-GGA level. For instance the  $w_0(\R)$ and $w_0'(\R)$ approximations are constructed as
\begin{eqnarray}
& w_0(\R)=n(\R)\epsilon_x^{SL}(n(\R),\nabla n(\R),\nabla^2
n(\R),\tau(\R)),\nonumber\\
& w'_0(\R)=2n(\R)\epsilon_c^{SL-GL2}(n(\R),\nabla n(\R),\nabla^2
n(\R),\tau(\R),\zeta(\R)),\nonumber\\
\label{eq6}
\end{eqnarray}
using
SL form of the GL2 correlation energy density
(SL-GL2)~\cite{constantin2019correlation}, where $\tau(\R)=\sum_{j}^{occ}|\nabla\phi_j(\R)|^2/2$ is the KS non-interacting kinetic energy density, with $\phi_j(\R)$ being the one-particle $j$-th occupied KS orbital. We underline that the Laplacian of the
density ($\nabla^2 n$) contains information that is
is already encapsulated in $\tau$ \cite{smiga2017laplacian}, such that many meta-GGA XC functionals do not consider $\nabla^2 n$ as an ingredient.

There are also two prime motivations behind the extension of ACSC functionals to the meta-GGA level:

\begin{itemize}
    \item[i)]  many of SL-GL2 correlation energy functionals, such as
    TPSS-GL2 (and all TPSS-like GL2 functionals) have already been derived~\cite{perdew1996generalized,perdew2008density}; thus, they can be easily applied in the present construction. The quantitative comparison of the accuracy of these SL-GL2 models with reference second-order GL2 correlation energy data was reported in \Refs{LUCISI} in Table S12.
    \item[ii)] the meta-GGA SL-GL2, such as TPSS-GL2 DFA, is one electron self-interaction free, giving precisely zero for the hydrogen atom, which is not the case for PBE-GL2.

\end{itemize}

In the next section, we will address the choice of $w_\infty(\R)$ and $w'_\infty(\R)$.

\subsection{TPSS-ACSC correlation functionals formula}

To construct ACSC meta-GGA DFA, we fix the $w_0(\R)$ and $w'_0(\R)$ (where the energy density $w_{\alpha}(\R)$ is defined by $W_\alpha = \int~d\R~ w_\alpha(\R)$) in the form of TPSS exchange ($w_0(\R)=n(\R)\epsilon_x^{TPSS}$) and TPSS-GL2~\cite{perdew2008density} ($w'_0(\R)=2n(\R)\epsilon_c^{TPSS-GL2}$) , respectively. In the case of $w_\infty(\R)$ and $w'_\infty(\R)$, the choice is not so simple due to various variants available in the literature. As was noted before, the form of $w_\infty(\R)$ and $w'_\infty(\R)$ implies the incorporation of important physics in the ACSC formula, i.e., the quasi-2D regime via mPC\cite{constantin2019correlation} model or very accurate performance for weak and strong-interaction regime via hPC model developed recently\cite{SCFISI}. However, both mPC and hPC are simple GGA level approximations of SCE formulas, which are not one-electron self-interaction free\cite{SCFISI}. Therefore, utilizing these GGA models might impact the performance of ACSC meta-GGA DFA. To overcome this limitation, one can develop the meta-GGA model for TPSS strong-interaction\cite{perdew2004meta} regime as was done in appendix D in \Refs{seidl2000density}. Thus, for clarity of this paper, we recall that for any approximate XC energy DFA ($E^{DFA}_{xc}  = E^{DFA}_{x} + E^{DFA}_{c}$), the corresponding coupling-constant integrand $W^{DFA}_{\alpha}$ can be derived from the following formula.
\begin{eqnarray}
  W_{\alpha}^{DFA}[n_\uparrow,n_\downarrow] =  &&E_{x}^{DFA}[n_\uparrow,n_\downarrow]\nonumber\\
&+& \frac{d}{d\alpha}\Big(\alpha^2 E_{c}^{DFA}[n_{\uparrow,1/\alpha},n_{\downarrow,1/\alpha}] \Big) \; .\nonumber\\
\label{eq4a}
\end{eqnarray}
by considering the  strictly correlated $\alpha \to \infty$ limit.

Thus, for the low-density limit of the TPSS functional, we obtain the $W_{\infty}^{TPSS}$ (Eq.~(\ref{err1})) and $W_{\infty}^{'TPSS}$ (Eq.~(\ref{eer})) expressions with their corresponding energy densities $w_\infty(\R)$ and $w'_\infty(\R)$, respectively. The latter quantities
incorporate all physically meaningful features, i.e., canceling one-electron self-interaction and proper behavior for the quasi-2D regime (shown later), which was also the case for the mPC model\cite{constantin2019correlation}.  Based on the above consideration, we construct the TPSS-ACSC correlation functional using Eq.~(\ref{eq1a}) with TPSS variants of $w_0$, $w'_0$ and $w_\infty$, $w'_\infty$ energy densities.

The final TPSS-ACSC formula diverges to $-\infty$ when $s \rightarrow 0$ ($w'_0\to -\infty$), (e.g., for the case of the uniform electron gas (UEG) model) behaving in this limit as~\cite{constantin2019correlation}
\begin{equation}
    \begin{split}
         \lim_{w'_0\to -\infty} [n\epsilon^{ACSC}_c]= w_\infty - w_0 +  2 w'_\infty \\
 - \frac{2(w'_\infty)^2}{w_0-w_\infty} \{\ln( 1 + \frac{w_0-w_\infty}{w'_\infty})\}~,
\label{szs1a}
    \end{split}
\end{equation}
which reveals the ACSC DFAs accuracy for UEG (see also Fig. 3 in \Refs{constantin2019correlation} for PBE-ACSC functional). On the other hand
for $w'_0\rightarrow 0$ it gives
\begin{equation}
\lim_{w'_0\to 0} [n\epsilon^{ACSC}_c]=
\frac{1}{2}w'_0+\frac {1}{3}\frac{1}{(w_0-w_\infty)}(w'_0)^2+\mathcal{O}((w'_0)^3),
\label{eq5}
\end{equation}
such that $E_c^{TPSS-ACSC}=0$ whenever $E_c^{TPSS-GL2}=0$.
The TPSS-GL2 correlation energy density vanish whenever $\tau=\tau^W$, and $\zeta=1$, where
$\tau^W$ is the von Weizs\"{a}cker kinetic energy density \cite{weizsacker1935theorie,della2015kohn}.
Thus, for one-electron systems, where $E_c^{TPSS-GL2}=0$, the TPSS-ACSC correlation energy is exact, showing that functional is one-electron self-correlation free.

At this point, the analysis of the behavior of TPSS-ACSC correlation energy is required. In Fig. (\ref{f1}), we show the UEG correlation energies per particle of
the exact LDA\cite{perdew1992accurate} (shown by the exact line in Fig. (\ref{f1}). We recall that for UEG, the reduced gradient $s=0$, thus $w_0$ reduces to LDA exchange energy density and $w'_0\to -\infty$ thus for ACSC functional we utilize the ACSC limit for UEG given by Eq. (\ref{szs1a}).
 For comparison we also show Lee-Yang-Parr (LYP) \cite{lee1988development}, and Tognetti-Cortona-Adamo (TCA) correlation
\cite{fabiano2015assessment,tognetti2008new,ragot2004correlation} energy densities. One can note that the ACSC formula is accurate in the low-density limit ($r_s\ge 20$), while in the high-density limit ($r_s\rightarrow 0$) diverges as $\sim r_s^{-1/2}$, thus faster than the exact behavior ($\sim \ln(r_s)$). Nevertheless, in the high-density limit, the exchange energy dominates over the correlation, so the proper choice of the exchange functional part should compensate for this failure of the ACSC correlation. This can be considered a
drawback of ACSC construction because it might lead to some issues with a lack of compatibility between standard semi-local exchange functionals and the  ACSC correlation functionals (mutual error cancellation effect). We will address this issue in the following.

\section{Results \& Discussion}

We first test the accuracy of $W_{\infty}^{TPSS} = \int~d\R~ w^{TPSS}_{\infty}(\R)$  and $W_{\infty}^{'TPSS} = \int~d\R~ w'^{TPSS}_{\infty}(\R)$ expressions, which is reported in Table~\ref{tab1} for real atoms. For comparison, we also present the data obtained for the exact SCE method~\cite{seidl07,gorigiorgi09, Daas2022large}, PC, mPC, hPC as well as PBE ($W_{\infty}^{PBE}$,  $W_{\infty}^{'PBE}$) formulas from \Refs{seidl2000density}. In the case of $W_{\infty}$, the TPSS approximation gives the best performance measured w.r.t. SCE values (even for Ar, Kr and Xr data reported recently\cite{Daas2022large}) being almost three times better than the one obtained for very accurate hPC model. This is partially because the former correctly removes one-electron self-interaction in $W_{\infty}^{TPSS}$, which is taken into account in all GGA $W_{\infty}$ approximations. Nonetheless, even without Hydrogen atom contribution (reported in parenthesis), the MARE of TPSS  $W_{\infty}$ presents the best performance for this model (MARE=0.44\%), closely followed by hPC (MARE=0.48\%) that are twice better than original PC variant.

In the case of $W'_{\infty}$, the overall performance of the TPSS model is worse than the one observed for hPC, in line with the results reported for PC. This can be since $W_{\infty}^{'TPSS}$ in the slowly varying density limit does not recover correctly the gradient expansion of the PC model. The problem lies in the $H_2$ function (Eq.~\ref{eer}), which when $t \to 0$ gives rise to the term proportional to $t^6$, in comparison to the PC model, which yields here term proportional $t^2$. One crucial difference, however, can be noted for the TPSS formula: it correctly recovers the SCE value for the H atom, which is impossible by any GGA variant.

\begin{figure}[hbt]
\begin{center}
\includegraphics[width=1\columnwidth]{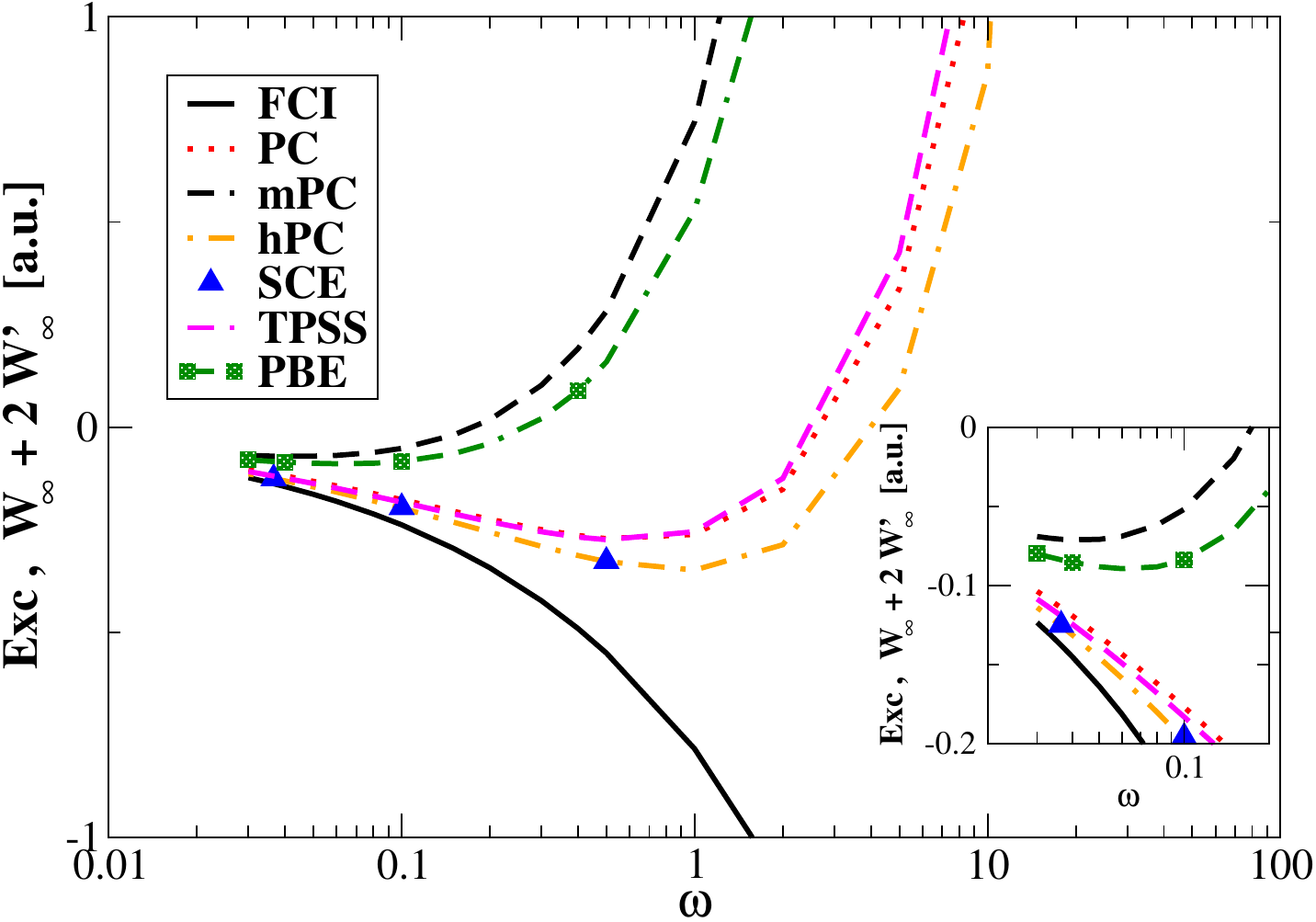}
\end{center}
\caption{\label{fig1h} Comparison of the leading term of the XC energy ($E_{xc} = W_{\infty}  + 2 W'_{\infty}$) in the strong interaction regime of the Hooke's atom calculated using different models with FCI data\cite{LUCISI}.}
\end{figure}

An additional assessment of all models is provided in Table \ref{tab2} and Fig. \ref{fig1h}, where we present results obtained for the Hooke's atom at different confinement strengths $\omega$ (see further text for computational details). Turning our attention to Table \ref{tab2}, we see similar trends to those presented in Table~\ref{tab1} for all values of $\omega$ where exact SCE data are available. Moreover, Fig. \ref{fig1h} shows that in the small $\omega$ range (strong-interaction limit of the Hooke's atom), hPC and TPSS yield the best estimation of the XC energy $E_{xc} = W_{\infty}  + 2 W'_{\infty}$, being slightly better than those obtained from PC model, while the mPC and PBE methods fail. Actually, mPC and PBE $W_{\infty}$ and $W'_{\infty}$ perform very similarly in all investigated cases giving rise to large errors.

\begin{figure}[hbt]
\begin{center}
\includegraphics[width=1\columnwidth]{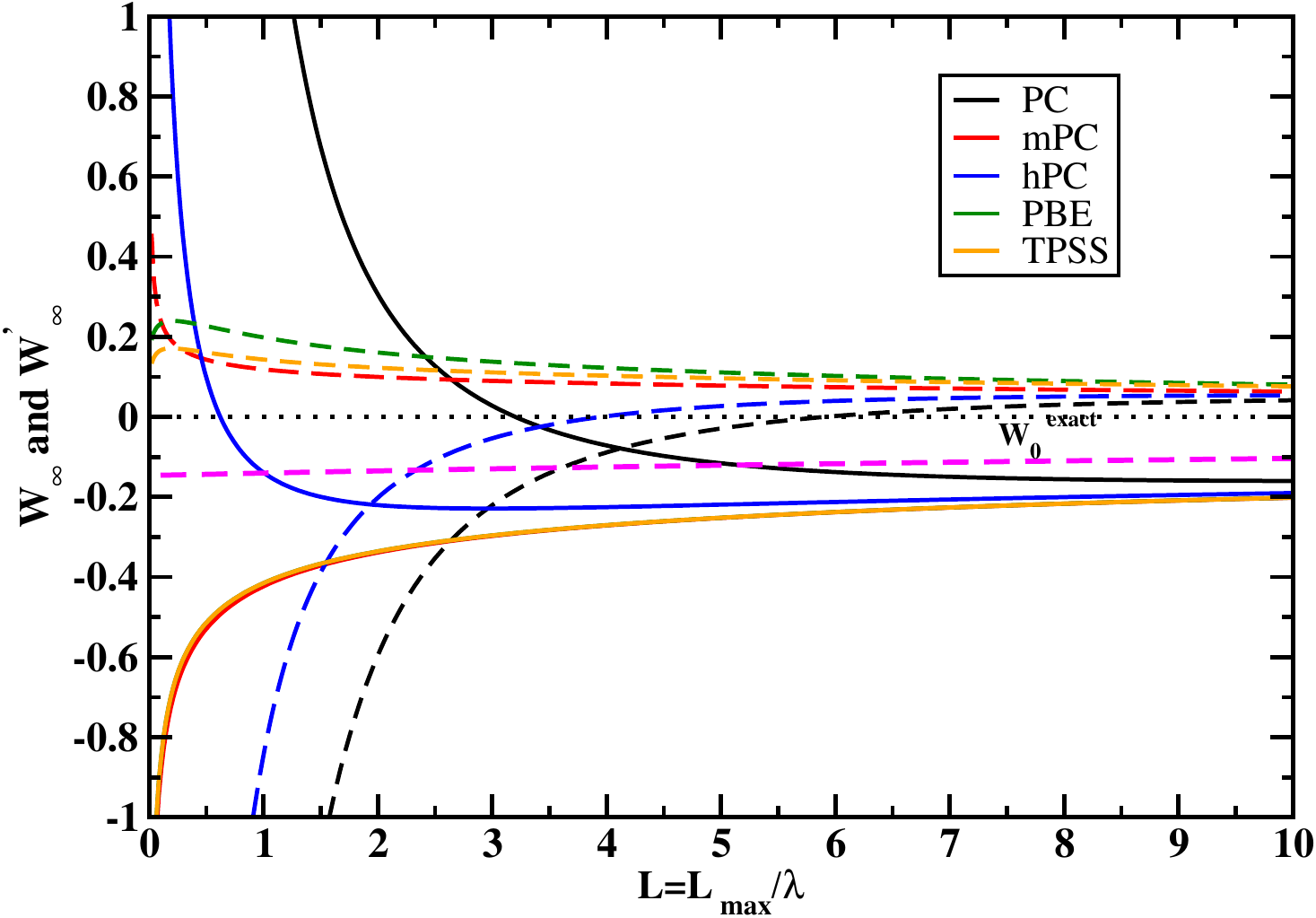}
\end{center}
\caption{\label{figibm}  Comparison of $W_\infty$ (solid line) and $W'_\infty$ (dashed line) behaviors for an IBM quasi-2D electron gas of fixed 2D electron density ($r_s^{2D} = 4$) as a function of
the quantum-well thickness $L$. Also shown is the exact exchange  $W_0$. PC and mPC are obtained from PBE, and results are taken from ref.~\cite{constantin2019correlation}. For PBE the $W_\infty$ and $W'_\infty$ expression from ref.~\cite{seidl2000density}. For TPSS, $W_\infty$ and $W'_\infty$ expression are given in Eq.~\ref{err1} and Eq.~\ref{eer}, respectively. }
\end{figure}

Further, we perform the comparison of $W_\infty$ and $W'_\infty$ behaviors for all studied models for an infinite barrier model (IBM) quasi-2D electron gas of fixed 2D electron density ($r_s^{2D} = 4$) as a function of
the quantum-well thickness $L$ as was also done in \Refs{constantin2019correlation}. The quasi-2D is very useful for the XC functional development, being the exact constraints in several modern density functional approximations~\cite{perdew2014gedanken,sun2015strongly}.Under uniform density limit to the quasi-2D limit, density behaves as $n_\lambda^z(x,y,z)=\lambda n(x,y,\lambda z)$ and the system approaches the 2D limit when $\lambda \to \infty$. In this limit  the XC energy is finite and negative i.e., $\lim_{\lambda\to\infty}E_{xc}[n_\lambda^z(x,y,z)]>-\infty$.
We report this in \Fig{figibm}. One can note that PC and hPC models change signs even for a mild quasi-2D regime. This feature is not allowable because it can lead to non-physical positive correlation energy or total failure of ISI or ACSC correlation energy expressions in quasi-2D regimes. On the other hand, the mPC, PBE, and TPSS  $W_\infty$ and $W'_\infty$ give correct behavior for a whole range of quantum-well thickness $L$.

\begin{table*}[hbt]
\begin{center}
\caption{\label{tab1} The values of $W_{\infty}$ and $W'_{\infty}$ for several atoms obtained from different models and using EXX densities. We use atomic units. The results which agree best with SCE values\cite{seidl07,gorigiorgi09} are highlighted in bold (for Ar, Kr, and Xe the SCE $W_\infty$ values are taken from ref.~\cite{Daas2022large}). The last line of each panel reports the mean absolute relative error (MARE) [(for $W_\infty$ (in parenthesis) and $W'_\infty$ we report the results where H results are excluded]. The $W'^{SCE}_{\infty}$ reference data are reported with the same precision as in the \Refs{gorigiorgi09}.}
\begin{tabular}{lccccccc}
\hline \hline
 & SCE & PC & hPC & mPC & PBE && TPSS \\
 \hline
\multicolumn{8}{c}{$W_\infty$} \\
H & -0.3125 & {-0.3128} & -0.3293 & -0.4000 & -0.4169 && \textbf{-0.3125} \\
He & -1.500 & -1.463 & \textbf{-1.492} & -1.671 & -1.6888 && -1.5122 \\
Be & {-4.021} & -3.943 & {-3.976} & -4.380 & -4.4203 && \textbf{-3.9803} \\
Ne & -20.035 & \textbf{-20.018} & -20.079 & -21.022 & -21.2983 && -19.9792\\
Ar&-51.555 & \textbf{-51.5473} & -51.6158 & -53.2709 & -53.9322&  & -51.3799\\
Kr&-166.850& -167.3561 & -167.4387 & -170.3279 & -172.0157&  & \textbf{-166.7765}\\
Xe&-322.835& -324.5206 & -324.6190& -328.6846 & -331.3261 & & \textbf{-323.3446}\\
MARE[\%] & & 0.78 (0.89) & 1.18 (0.48) & 8.64 (5.41) & 10.36 (6.53)  && \textbf{0.38 (0.44)} \\
& & & & & & &\\
\multicolumn{8}{c}{$W'_\infty$} \\
H & 0 & 0.0426 & {0.0255} & 0.2918  & 0.243 && \textbf{0} \\
He & 0.621 & 0.729 & \textbf{0.646} & 1.728 & 1.517 && 0.728 \\
Be & 2.59 & 2.919 & \textbf{2.600} & 6.167 & 5.442 && 2.713 \\
Ne & 22 & 24.425 & \textbf{23.045} & 38.644 & 35.307 && 23.835 \\
MARE[\%] & & 13.71 & \textbf{3.05}& 130.67 & 104.94 && 10.10\\
\hline\hline
\end{tabular}
\end{center}
\end{table*}

A summary of all essential features of strong-interaction models is given in Table~\ref{tab3}. One can note that TPSS $W_\infty$ and $W'_\infty$ reproduce reference SCE data with quite a good accuracy and some other important features, e.g., good performance in the quasi-2D regime, removes one electron self-interaction. This possibly indicates that the description of all non-local features of the SCE model can be done only by utilizing non-local ingredients such as $\tau$.  This is the first important finding of the present study.

\begin{table}[hbt]
\caption{\label{tab2} The $W_\infty$ and $W'_\infty$ energies (in Ha) for three values of $\omega$  for which the Hooke's atom has analytical solutions\cite{Tau-PRA-93} and exact SCE reference data are available\cite{kooi2018local}. The last line of each panel reports the mean absolute relative error (MARE).}
\begin{center}
\begin{tabular}{lccccccc}
\hline \hline
 & SCE & PC & hPC & mPC & PBE && TPSS \\
 \hline
\multicolumn{8}{c}{$W_\infty$} \\
0.0365373 & -0.170 & -0.156 & -0.167 & -0.191 &  -0.191 && \textbf{-0.170}\\
0.1 & -0.304 & -0.284 & \textbf{-0.303} & -0.344 & -0.344 && -0.308 \\
0.5 & -0.743 & -0.702 & \textbf{-0.743} & -0.841 & -0.843 && -0.754 \\
MARE & & 6.78\% & \textbf{0.70\%} & 12.90\% & 12.98\% && 0.96\% \\
& & & & \\
\multicolumn{8}{c}{$W'_\infty$} \\
0.0365373 & 0.022 & \textbf{0.021} & \textbf{0.021} & 0.060 & 0.053 && 0.026 \\
0.1 & 0.054 & \textbf{0.054} & 0.053 & 0.146 & 0.130 && 0.062 \\
0.5 & 0.208 & 0.215 & \textbf{0.208} & 0.562 & 0.501 && 0.240 \\
MARE & & 2.64\% & \textbf{2.13\%} & 171.10\% & 139.81\% && 14.70\%\\
\hline\hline
\end{tabular}
\end{center}
\end{table}

\begingroup
\begin{table*}[htbp]
\caption{\label{tab3} Brief summarize of properties of $w_\infty(\R)$ and $w'_\infty(\R)$ from various semi-local models. }
{
\begin{tabular}{cccccc}
\hline
&PC~\cite{seidl2000simulation,seidl2000density} & mPC~\cite{constantin2019correlation} &  hPC~\cite{SCFISI} & PBE ~\cite{seidl2000density} & TPSS\cite{perdew2004meta} \\
\hline
level of theory & GEA & GGA & GGA & GGA & meta-GGA\\
accurate for the strictly correlated regime & $\checkmark$ & $\times$ & $\checkmark$ & $\times$ & $\checkmark$ \\
quasi-2D regime & $\times$ & $\checkmark$ & $\times$ & $\checkmark$ & $\checkmark$ \\
self-consistent calculations & $\times$ &  $\checkmark$ &  $\checkmark$ & $\checkmark$&  $\checkmark$ \\
one-electron self interaction free & $\times$ & $\times$ & $\times$ &$\times$ &$\checkmark$\\
\hline
\end{tabular}
}
\end{table*}
\endgroup

\begin{table}
\begin{center}
\caption{\label{tax}
The TPSS and TPSS-ACSC correlation energies (mHa) divided by the number of electrons ($N_e$) for $10$ atoms  (computed using Hartree-Fock analytic orbitals and densities~\cite{davidson1991ground,chakravorty1993ground,clementi1997note}) and eight molecules (computed using Hartree-Fock orbitals and densities obtained with uncontracted cc-pVTZ\cite{dunning:1989:bas} basis sets).}
\begin{tabular}{ccccccccc}
\hline
Atoms&$N_e$	&	TPSS	&	TPSS-ACSC	&	Ref.~\cite{davidson1991ground,chakravorty1993ground,clementi1997note}	\\
	\hline
H	&	1	&	\textbf{0.0}	&	\textbf{0.0}	&	0	\\
He	&	2	&	\textbf{-21.5}	&	-20.2	&	-21	\\
Li	&	3	&	-16.5	&	\textbf{-15.9}	&	-15.1	\\
Be	&	4	&	\textbf{-21.7}	&	-20.8	&	-23.6	\\
N	&	7	&	\textbf{-26.5}	&	-25.9	&	-26.9	\\
Ne	&	10	&	\textbf{-35.4}	&	-35.3	&	-39.1	\\
Ar	&	18	&	-39.5	&	\textbf{-39.8}	&	-40.1	\\
Kr	&	36	&	-49.2	&	\textbf{-49.9}	&	-57.4	\\
Zn	&	30	&	-47.0	&	\textbf{-47.8}	&	-56.2	\\
Xe	&	54	&	-54.1	&	\textbf{-55.3}	&	-57.2	\\
MAE$_{atm}$	&		&	2.6		&     \textbf{2.5}	&		\\ \hline
Molecules\footnote{the geometries has been taken from \Refs{grabowski2014orbital,SmigaJCP2020}}		&$N_e$&	TPSS	&	TPSS-ACSC     	& 	Ref.\footnote{correlation energies obtained at CCSD(T) with uncontracted cc-pVTZ level of theory}	\\ \hline
H$_2$ &	2	&	-21.1		&    \textbf{-19.9} 	&	-19.8	\\
LiH &	4	&	-21.5		&   \textbf{-20.2}  	&	-18.5	\\
Li$_2$ &	6	&	-21.0		& \textbf{-20.0}    	&	-17.9	\\
H$_2$O & 10		&	-33.2		& \textbf{-33.1}    	&	-32.9	\\
NH$_3$ & 10		&	\textbf{-31.9}		& -32.1    	&	-30.6	\\
HF &	10	&	-34.3		&  \textbf{-34.1}   	&	-33.7	\\
CO &	14	&	\textbf{-32.4}		&  \textbf{-32.4}   	&	-34.0	\\
N$_2$ &	14	&	-32.7		&  \textbf{-32.1}   	&	-30.6	\\
MAE$_{mol}$ &		&	1.7		&  \textbf{1.2}   	&		\\
\hline
\end{tabular}
\end{center}
\end{table}

Now, let us focus on the numerical performance of the TPSS-ACSC functional itself. In Table \ref{tax}, we report the correlation energies for small atoms and molecules obtained with TPSS-ACSC and TPSS functionals energy expression.
In the case of atoms, the calculations are performed using the Hartree-Fock (HF) analytic orbitals of Clementi and Roetti~\cite{clementi1974roothaan}. For molecules, we have performed the HF calculations in ACESII\cite{acesII} program using the uncontracted cc-pVTZ\cite{dunning:1989:bas} basis sets and geometries taken from refs.~\cite{grabowski2014orbital,SmigaJCP2020}.

We recall that using self-interaction free HF orbitals allows us to test the error specifically related to the functional construction itself, namely the functional-driven error\cite{BurkeFD_DD}. As was shown in \Refs{hernandez2023new}, the utilization of HF densities can sometimes lead to the worsening of predictions of DFAs or improving them for the wrong reasons. This could happen when a density-driven error gives a significant contribution not canceled totally by applying the HF densities.  In these cases, utilization of more accurate, correlated densities is required\cite{hernandez2023new}. However, in most semilocal DFAs, the total error is predominated by functional-driven error, meaning that HF densities are sufficiently accurate to perform such analysis.

\begin{figure}[t]
\includegraphics[width=\columnwidth]{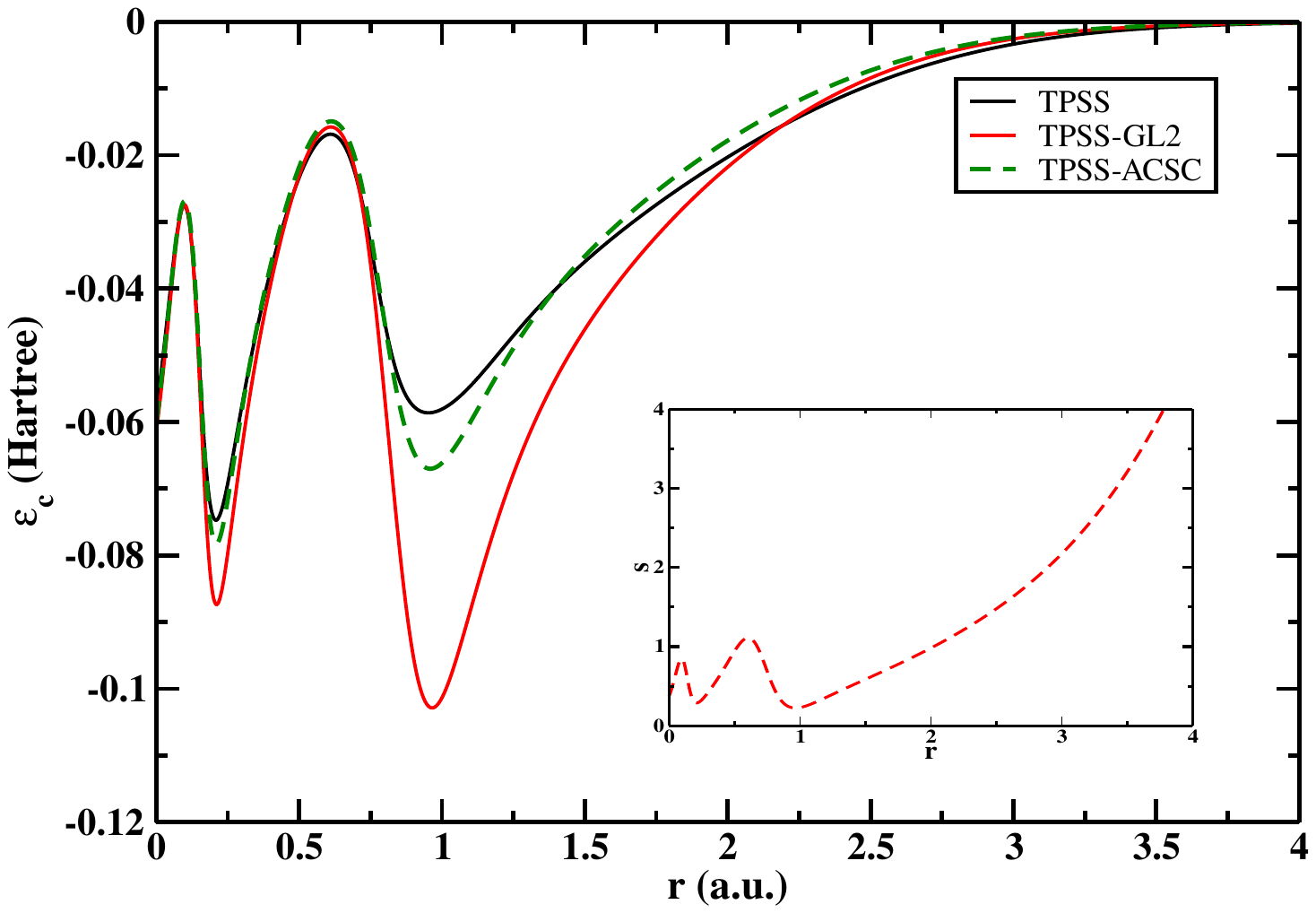}
\caption{Correlation energy per particle $\epsilon_c$
versus the radial distance from the nucleus $r$, for the Ar atom.
(computed using Hartree-Fock analytic orbitals and densities~\cite{davidson1991ground,chakravorty1993ground,clementi1997note}).
In the inset, we show the reduced gradient $s=\frac{|\nabla n|}{2k_F n}$.}
\label{f2}
\end{figure}

As noted before, both considered correlation functionals are one-electron self-interaction-free,
which is visible in the case of the H atom. In most cases, TPSS-ACSC performs in line with its TPSS counterpart, indicating that the correlation effects are well represented in ACSC energy expression. To visualize the correlation densities, in Fig. \ref{f2}, we show a comparison between $\epsilon_c^{TPSS}$, $\epsilon_c^{TPSS-GL2}$, and $\epsilon_c^{TPSS-ACSC}$ for Ar atom. Whenever $s$ is small, $\epsilon_c^{TPSS-GL2}$ starts to depart from $\epsilon_c^{TPSS}$, diverging when $s=0$. However, $\epsilon_c^{TPSS-ACSC}$ is well-behaved everywhere.

As to the molecules, we note that TPSS-ACSC performs very well for most systems, slightly better than TPSS functional. This again confirms
the robustness of correlation functional construction.
%

%

\begin{figure}
\begin{center}
\includegraphics[width=\columnwidth]{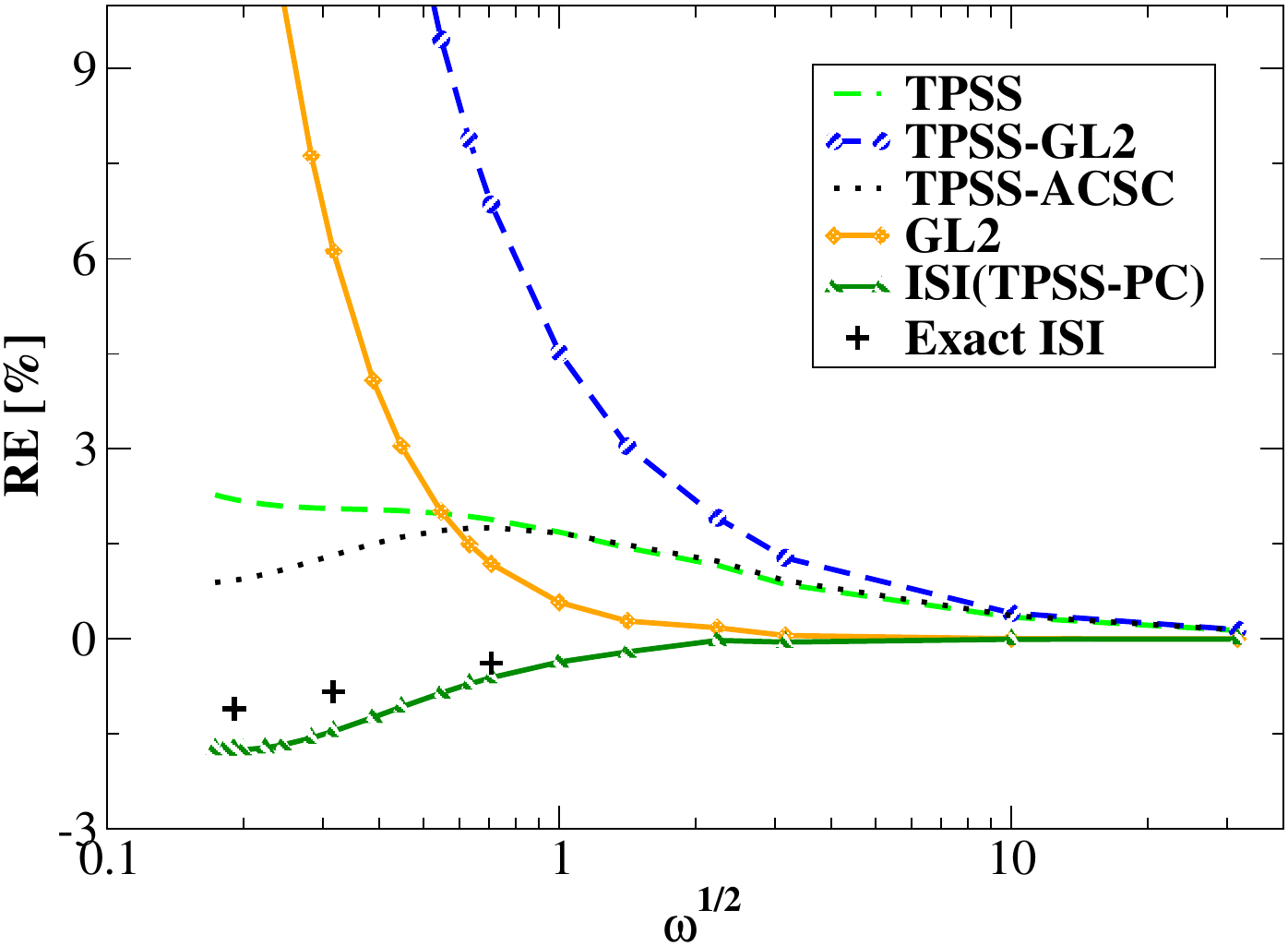}
\end{center}
\caption{\label{fighh} Relative error on XC energies of harmonium atoms for various values of $\omega$ computed at @EXX orbitals for several functional using computational setup from \Refs{LUCISI}. The errors have been computed with respect to FCI data obtained in the same basis set\cite{B926389F,LUCISI}. For all TPSS-like results have been obtained together with the TPSS exchange energy functional. The GL2 and ISI(TPSS) XC correlation results are obtained with the exact GL2\cite{gorling1994exact} formula combined with EXX energy expression. The ISI formula utilizes the $W_\infty$ and $W'_\infty$ given by \Eq{err1} and \Eq{eer}. Exact ISI data are taken from \Refs{kooi2018local}.}
\end{figure}

In \Fig{fighh}, we report the relative error (RE) on XC energy computed for two-electron Hooke's atoms model for various values of confinement strength $\omega$ ($\omega \in [0.03, 1000]$). The errors are computed with respect to Full Configuration Interaction (FCI) results from \Refs{LUCISI}. The calculations have been performed using an identical computational setup as in our previous study\cite{LUCISI, PRBSub, jana2021szs} using EXX reference orbitals. We recall that the system is strongly correlated for small values of confinement strength $\omega$, whereas we enter a weak-interaction regime for large values of $\omega$. Thus, the model provides an excellent tool for testing functional performance in these two regimes. We underline that in all following calculations, all TPSS-like correlation functionals have been combined with the TPSS exchange energy functional to obtain XC energies.

For medium and large values of $\omega$, the TPSS and TPSS-ACSC functionals perform very similarly, giving in the weak interacting region a very small relative error (RE) similar to exact GL2 and ISI XC functionals. In a strong-interaction regime, the TPSS-ACSC improves over its TPSS precursor. We note that in the latter regime, the TPSS-ACSC functional should recover, in principle, the ISI functional data due to the inclusion in both energy expressions the $W_\infty$ and $W'_\infty$ in the form given by  \Eq{err1} and \Eq{eer}.
Although qualitatively, they behave very similarly, there is a sizeable quantitative difference between these two curves. This is most probably related to the significant impact of the GL2 term, which enters both formulas. We recall that the ISI formula utilized the exact GL2 energy expression, whereas the TPSS-ACSC approximated the SL variant. Although they both diverge when $\omega$ tends to zero, the origin of that behavior is different. The exact GL2 energy diverges due to closing the HOMO-LUMO gap in this regime, whereas TPSS-GL2 is due to vanishing reduced gradient, which leads to a much faster divergence. This feature of TPSS-GL2 energy expression governs the behavior of TPSS-ACSC DFA in a small $\omega$ regime. Thus, we might conclude that the quantitative difference between ISI and TPSS-ACSC DFAs comes mainly from the inaccuracy of the SL-GL2 formula used in the later expression.

\begin{figure}
\begin{center}
\includegraphics[width=\columnwidth]{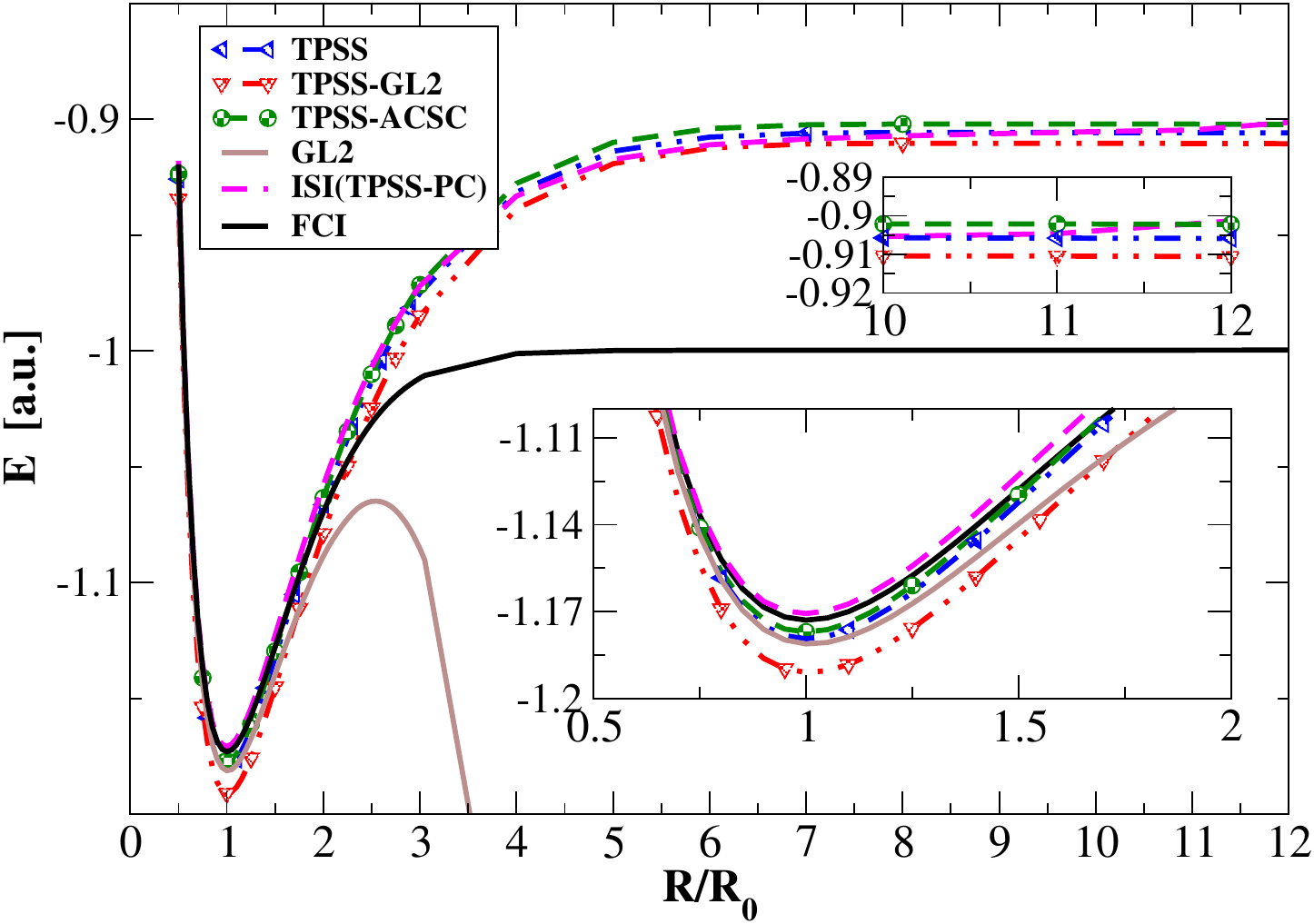}
\end{center}
\caption{\label{figh2}The total energy of the stretched H$_2$ molecule as it is calculated with the various methods. The insets present the same data around the equilibrium distance  ($R/R_0 = 1$) and large $R/R_0 > 10$ values. }
\end{figure}

Now we turn attention to another two electron example where we may encounter a strong-interaction limit, namely, the potential energy surface for the dissociation of the H$_2$ molecule, in a restricted formalism \cite{cohen12}, which is one of the main DFT challenges \cite{cohen12,science21,Peach2007modeling}. This is reported in Fig. \ref{figh2}. All energies have been obtained using EXX orbitals and densities. We want to underline that restricted HF density could give rise to substantial errors in the mid-bond region when the H$_2$ molecule is largely stretched. As pointed out earlier, the functional-driven error dominates most semilocal DFAs. Thus, the utilization of HF densities still gives a valid picture of the performance of semilocal DFAs for the whole range of distances of H$_2$.

One can note that, in general, TPSS-ACSC functional performs very similarly to TPSS, especially near equilibrium distance. More visible differences between these two DFAs can be seen for larger distances $R/R_0 > 3$. Asymptotically, the TPSS-ACSC energy goes almost to the same value as the ISI method with \Eq{err1} and \Eq{eer} employed to describe $W_\infty$ and $W'_\infty$. This is an interesting finding, possibly suggesting the dominant role of strong-interaction limit (see \Eq{szs1a}) for large separation of hydrogen atoms. We note, however, that the TPSS-GL2 total energy gives much more stable results in the asymptotic region in comparison to the exact GL2 curve, which diverges due to the closing HOMO-LUMO gap. This indicates that the proper behavior investigated here ISI and TPSS-ACSC DFAs have a different origin. In the former, the exact GL2 diverges ( $E_{GL2}\rightarrow -\infty$), leading in the asymptotic limit to $E_{xc}^{ISI} \rightarrow W_{\infty} + 2W'_{\infty} (1-\frac{1}{q}\ln(1+q))$ with $q=(E_x-W_{\infty})/W'_{\infty}$ \cite{fabiano2016interaction,SCFISI}. In the latter, the asymptotic limit is governed rather by the mutual error cancellation effect in TPSS-ACSC energy expression. This is because TPSS-GL2 energy expressions do not diverge for large $R/R_0$, meaning that at the asymptotic region Eq.~\ref{szs1a} do not hold.
One possible way to recover Eq.~ref{szs1a} within the TPSS-ACSC formula could be realized via proper incorporation of local gap model\cite{fabiano2014generalized,krieger1999electron,PhysRevB.95.115153} within SL-GL2 formula.

Let us focus on self-consistent results (@SCF) obtained within the generalized KS (gKS) scheme. As an example, we report in \Tab{ta-ae6} AE6~\cite{lynch2003small,haunschild2012theoretical} atomization energies of $6$ small size molecules, obtained using SCF orbitals and densities. One can note that TPSS-ACSC functional, in general, gives results that are twice worse (MAE = 18.4 kcal/mol) than for the TPSS counterpart, which yields MAE of 7.6 kcal/mol. The same trend for the AE6 benchmark occurs when we feed TPSS-ACSC and TPSS total energy expressions with HF orbitals. This indicates the following things:

\begin{itemize}

\item The major part of the error for the TPSS-ACSC functional is related to functional-driven error\cite{BurkeFD_DD}. This is most possibly related to the ACSC model itself, which was not designed to be accurate in the high-density limit where most of the chemical application takes place;

\item because both TPSS and TPSS-ACSC utilize the same semi-local TPSS exchange, the larger error observed in the latter might suggest the lack of compatibility between exchange and correlation functionals (there is no error cancellation effect). The correlation energies are pretty accurate, as shown in \Tab{tax}.
This might indicate that the correct behavior of TPSS-ACSC functional can be restored by proper design of compatible exchange functional;

\end{itemize}

To test this possibility, we have performed \textit{ad hoc} modification of TPSS exchange functional~\cite{taoPRL03} by calibration of second-order gradient expansion parameter ($\mu=0.235$). We note that this parameter might generally vary based on the nature of the localized (such as atoms) or de-localized systems (solids).
Using $\mu = 0.40$, we have observed a significant reduction of MAE for AE6 obtained at @SCF densities to 6.63 kcal/mol.

Finally, the performance of the constructed functionals is also benchmarked for other molecular test cases such as atomization energies, barrier heights and week, and covalent interactions. These results are reported in
Table~\ref{tabenchmark}. A noticeable improvement is observed from TPSS-ACSC ($\mu=0.40$) than TPSS-ACSC, especially for atomization energies.
Interestingly, In other cases, TPSS-ACSC performs slightly better or similarly to TPSS-ACSC ($\mu=0.40$). This indicates that probably some more sophisticated modification of the TPSS exchange functional is required in order to improve the accuracy of the method for all benchmarked cases. One may note that for CT7, W17, and S22, we do not include the dispersion correction as including a functional-specific dispersion interaction is beyond the scope of the present paper.

\begingroup
\begin{table*}
\caption{\label{ta-ae6} AE6 atomization energies (in kcal/mol) computed using
self-consistent (@SCF) and Hartree-Fock (@HF) orbitals and densities, and TPSSx semi-local exchange and TPSS or TPSS-ACSC correlation functionals.  The mean absolute
error (MAE, in kcal/mol) is shown in the last row. The def2-QZVP basis set is used. All calculations are performed using
Q-Chem code~\cite{qchem}.}
{
\begin{tabular}{cccccccccccccccccccccccccccccccccc}
\hline
        &TPSS@SCF &TPSS-ACSC@SCF &TPSS-ACSC@SCF &Ref.~\cite{haunschild2012theoretical}  \\
        & & &($\mu=0.40$) &  \\
\hline
SiH$_4$	&	334.2	&	337.9	&332.0	&323.1	\\
SiO	&	187.1	&	189.4	&179.3	&191.5	\\
S$_2$	&	109.0	&	114.6	&106.2	&101.9	\\
C$_3$H$_4$	&	707.8	&	724.0	&699.8	&701.0	\\
C$_2$H$_2$O$_2$	&	634.1	&	648.8	&619.0	&630.4	\\
C$_4$H$_8$	&	1155.8	&	1182.8	&1141.6	&1143.4	\\
\hline
	MAE&	7.6	&	18.4	&6.6		\\
	&		&		&		\\
       &TPSS@HF &TPSS-ACSC@HF&TPSS-ACSC@HF &Ref.~\cite{haunschild2012theoretical}  \\
       & && ($\mu=0.40$) &  \\
\hline
SiH$_4$	&	\textbf{331.9}	&	337.0	&332.6	&323.1	\\
SiO	&	179.7	&	\textbf{182.2}	&173.1	&191.5	\\
S$_2$	&	\textbf{103.5}	&	109.5	&101.5	&101.9	\\
C$_3$H$_4$	&	\textbf{702.0}	&	719.9&698.4	&701.0	\\
C$_2$H$_2$O$_2$	&	\textbf{621.1}	&	637.2&610.3	&	630.4	\\
C$_4$H$_8$	&	\textbf{1148.3}	&	1179.0	&1142.8	&1143.4	\\
\hline
	MAE &	6.2 &	15.3	&8.5		\\

\hline
\end{tabular}
}
\end{table*}
\endgroup

\begingroup
\begin{table}
\caption{\label{tabenchmark} Mean absolute errors (MAEs in kcal/mol) for the benchmark molecular tests,
obtained using different methods. All calculations are performed self-consistently using a def2-QZVP basis set with Q-Chem code~\cite{qchem}.}
{
\begin{tabular}{cccccccccccccccccccccccccccccccccc}
\hline
        &TPSS&TPSS-ACSC& TPSS-ACSC \\

              & & &($\mu=0.40$) &  \\
\hline
G2/148$^a$	&	5.5	&	15.7	&	7.8	\\
BH6$^b$ &8.2&8.3&8.4&\\
HTBH38$^c$	&	7.7	&	8.3	&	7.1	\\
NHTBH38$^c$	&	9.2	&	9.2	&	9.1	\\
CT7$^d$	&	2.0	&	1.7	&	1.1	\\
WI7$^d$	&	0.24	&	0.26	&	0.12	\\
S22$^e$	&	3.4	&	4.1	&	5.5	\\
\hline
\end{tabular}
\begin{flushleft}
 $^a$atomization energies of 148
molecules~\cite{curtiss1997assessment},$^b$ 6 barrier heights~\cite{haunschild2012theoretical}, $^c$38
 hydrogen (HTBH38) and 38 non-hydrogen bonded reaction barrier heights
(NHTBH38)~\cite{zhao2005benchmark}, $^d$7 charge transfer molecules, and 7 weekly interacting test set~\cite{zhao2005design}, $^e$22 non-covalent interacting systems~\cite{goerigk2017look}.
\end{flushleft}
}
\end{table}
\endgroup

\section{Conclusions}


In this work, we have constructed a semilocal meta-GGA correlation energy functional, based on the ACSC method proposed in \Refs{constantin2019correlation}. The correlation functional, denoted as TPSS-ACSC, interpolates the high and low-density limit of the popular TPSS correlation energy functional showing some direction on how to incorporate a strong interaction regime within the approximate, semilocal exchange-correlation formula.

The new correlation TPSS-ACSC functional is non-empirical, one electron self-interaction free accurate for small atoms and molecules. We provide a careful assessment of TPSS-ACSC functional base on some model systems (the uniform electron gas, Hooke’s atom, stretched H$_2$ molecule) and real-life calculations (atomization energies) showing some advantages and disadvantages of ACSC construction. From this broad perspective, we can conclude that,
although the ACSC method holds a promise for proper description of a strong-interaction regime, it is still in its infancy, which implies that there is still much space for improvement. The most important conclusions of the present study are as follows:

\begin{itemize}
    \item the strong-interaction limit obtained from semilocal TPSS functional formula ($W_{\infty}^{TPSS}$ and $W_{\infty}^{'TPSS}$) reproduce quite well reference SCE data. Moreover, both possess some other important features e.g., good performance in the quasi-2D regime and removing one electron self-interaction. Thus, both formulas could be effectively applied in constructing ACSC and ISI-like formulas.

    \item Although our numerical tests suggest the strong-interaction limit of semilocal TPSS-ACSC correlation is well represented, the semilocal GL2 part may need some amendment (Hooke’s atom, stretched H$_2$ molecule cases) e.g. via proper incorporation of local gap model\cite{fabiano2014generalized,krieger1999electron,PhysRevB.95.115153}.

    \item in order to improve the accuracy of TPSS-ACSC XC functional, it must be combined with the compatible exchange functional, leading to a much better balance in XC term (better mutual error cancelation effect). As was shown, the \textit{ad hoc} modification of TPSS exchange gives some hints in that direction.

\end{itemize}

Some of these new developments in the ACSC context will be addressed in a future study.

\section*{Acknowledgements}
This research was funded in part by National Science Centre, Poland
(grant no. 2021/42/E/ST4/00096). L.A.C. acknowledges the financial support from
ICSC - Centro Nazionale di Ricerca in High Performance
Computing, Big Data, and Quantum Computing, funded by
European Union - NextGenerationEU - PNRR.

\appendix*

\section{Details of TPSS-ACSC correlation functional}
\label{appendix1}

Here we summarized the expressions of energy density $w_{\alpha}$ of Eq.~(\ref{eq1}) $-$ Eq.~(\ref{eq2}) which is defined by $W_\alpha=\int~d^3r~w_\alpha(\bf{r})$. Thus one required following energy densities  $w_0(\bf{r})$, $w_0'(\bf{r})$, $w_{\infty}(\bf{r})$, and $w_{\infty}'(\bf{r})$ to calculate the ACSC correlation.

We take in the following expressions:\\

(i) First, $w_0({\bf{r}})=n({\bf{r}})\epsilon_x^{TPSS}({\bf{r}})$, where $\epsilon_x^{TPSS}({\bf{r}}$ is the TPSS exchange energy per particle~\cite{taoPRL03} given by,
\begin{equation}
 \epsilon_x^{TPSS}({\bf{r}}) =-A_x n({\bf{r}})F_x^{TPSS},~F_x^{TPSS}=1+\kappa-\kappa/(1+x/\kappa)~
\end{equation}
with $\kappa=0.804$. See ref.~\cite{taoPRL03} for the details of the TPSS exchange enhancement factor ($F_x^{TPSS}$).

(ii) Second, $w_0'({\bf{r}})=2n({\bf{r}})\epsilon_c^{TPSS-GL2}$, where $\epsilon_\mathrm{c}^\text{TPSS-GL2}$ is the G\"{o}rling-Levy second-order limit of the TPSS correlation energy~\cite{taoPRL03} per electron which is given by \cite{perdew2008density}
\begin{equation}
 \epsilon_\mathrm{c}^\text{TPSS-GL2}
 = \epsilon_\mathrm{c}^\text{revPKZB-GL2} \left[
 1 + d \epsilon_\mathrm{c}^\text{revPKZB-GL2}
 \left( \frac{\tau_W}{\tau} \right)^3 \right],
\end{equation}
where $d=2.8$ hartree$^{-1}$ is a constant and
\begin{eqnarray}
 & & \hspace*{-0.5cm}
 \epsilon_\mathrm{c}^\text{revPKZB-GL2} \nonumber \\
 & & = \epsilon_\mathrm{c}^\mathrm{PBE-GL2}(n_\uparrow,n_\downarrow,
 \nabla n_\uparrow,\nabla n_\downarrow) \left[
 1 + C(\zeta,\xi) \left( \frac{\tau_W}{\tau} \right)^2 \right]
 \nonumber \\
 & & - \left[ 1 + C(\zeta,\xi) \right]
 \left( \frac{\tau_W}{\tau} \right)^2
 \sum_\sigma \frac{n_\sigma}{n}
 \tilde{\epsilon}_\mathrm{c,\sigma}^\mathrm{PBE-GL2}.  \label{eq:GL2revPKZB}
\end{eqnarray}
In Eq.~(\ref{eq:GL2revPKZB}), $\epsilon_\mathrm{c}^\text{PBE-GL2}$
is the G\"{o}rling--Levy limit of the PBE correlation
energy per electron. It is obtained by replacing $\lambda\bf r$ with
$\bf r$ in the $\lambda\to\infty$ uniform density scaling limit
of the PBE correlation energy per electron, and has the expression
\begin{eqnarray}
 & & \hspace*{-0.5cm}
 \epsilon_\mathrm{c}^\text{PBE-GL2}
 (n_\uparrow,n_\downarrow,\nabla n_\uparrow,\nabla n_\downarrow)
 \nonumber \\
 & & = - \gamma\phi^3 \ln \left[
 1 + \frac{1}{\chi s^2/\phi^2 + (\chi s^2/\phi^2)^2 } \right]~,
\end{eqnarray}
where $\gamma=(1-\ln 2)/\pi^2$, $\phi(\zeta) = \frac{1}{2} \left[
 (1+\zeta)^{2/3} + (1-\zeta)^{2/3} \right]$,
$s=|\nabla n|/2nk_F$ is the reduced density gradient, $k_F=(3\pi^2n)^{1/3}$,
and $\chi=(\beta/\gamma) c^2 e^{-\omega/\gamma} \approx 0.72161$,
where $c = (3\pi^2/16)^{1/3}$, $\beta=0.066725$, and
$\omega=0.046644$.

The spin-dependent function
$\tilde{\epsilon}_\mathrm{c,\sigma}^\mathrm{PBE-GL2}$ is
defined as
\begin{eqnarray}
 \tilde{\epsilon}_\mathrm{c,\sigma}^\mathrm{PBE-GL2}
 & = & \max \big[ \epsilon_\mathrm{c}^\mathrm{PBE-GL2}
 (n_\sigma,0,\nabla n_\sigma,0), \nonumber \\
 & & \epsilon_\mathrm{c}^\mathrm{PBE-GL2}(n_\uparrow,n_\downarrow,
 \nabla n_\uparrow,\nabla n_\downarrow) \big].
\end{eqnarray}
The function $C(\zeta,\xi)$ is the spin-dependent function, where $\zeta$
is the spin-polarization and
$\xi=|\nabla\zeta|/2k_F$.

(iii) Third, in the case of TPSS XC functional the $W_{\infty}$ is derived as,
\begin{eqnarray}
 W_{\infty}^{TPSS}[n_\uparrow,n_\downarrow]&=&E_x^{TPSS}[n_\uparrow,n_\downarrow]+\int~d^3r~n({\bf{r}})~\Big\{\Big(-\frac{d_0(\zeta)}
 {r_s}\nonumber\\
 &+&H_1(r_s,\zeta,t)\Big)\Big[1+C(\zeta,\xi)\Big(\frac{\tau^W}{\tau}\Big)^2\Big]\nonumber\\
 &-&(1+C(\zeta,\xi))
 \Big(\frac{\tau^W}{\tau}\Big)^2\sum_{\sigma}\frac{n_\sigma}{n}
 \Big(-\frac{d_0(1)}{r_{s,\sigma}}\nonumber\\
 &+&H_1(r_{s,\sigma},1,t_\sigma)\Big)\Big\}
 \nonumber\\
 \label{err1}
 \end{eqnarray}

 (iv) Fourth and finally, $W'_{\infty}$ for TPSS functional reads
 \begin{eqnarray}
 W_{\infty}^{'TPSS}[n_\uparrow,n_\downarrow]&=&\frac{1}{2}\int~d^3r~n({\bf{r}})~\Big\{\Big(\frac{d_1(\zeta)}
 {r_s^{3/2}}\nonumber\\
 &+&H_2(r_s,\zeta,t)\Big)\Big[1+C(\zeta,\xi)\Big(\frac{\tau^W}{\tau}\Big)^2\Big]\nonumber\\
 &-&(1+C(\zeta,\xi))
 \Big(\frac{\tau^W}{\tau}\Big)^2\sum_{\sigma}\frac{n_\sigma}{n}
 \Big(\frac{d_1(1)}{r_{s,\sigma}^{3/2}}\nonumber\\
 &+&H_2(r_{s,\sigma},1,t_\sigma)\Big)\Big\}~,
 \nonumber\\
 \label{eer}
 \end{eqnarray}
where $d_1(\zeta) = 1.5$ (spin-independent) was fixed using the same reasoning as in \Refs{seidl2000density} and $E_x^{TPSS}[n_\uparrow,n_\downarrow]$ is the spin-resolved TPSS exchange \cite{taoPRL03}. $H_1$ and $H_2$ are same as given by Eq. (D11) and Eq. (D12) of \Refs{seidl2000density}. $C(\zeta,\xi)$ is given in Eq. (14) of \Refs{taoPRL03}. We recall that $W_{\infty}^{TPSS}[n_\uparrow,n_\downarrow]$ was already reported in
\Refs{perdew2004meta}. However, the expression of $W_{\infty}^{'TPSS}[n_\uparrow,n_\downarrow]$ can be obtained in the similar fashion as Eq.~(D16) of PKZB expression~\cite{seidl2000density}.
For the details of the parameters and terms see \Refs{seidl2000density} (for $d_0(\zeta)$, $d_1(\zeta)$, and $d_1(1)$) and \Refs{taoPRL03} (for $C(\zeta,\xi)$). One may note that the expressions of TPSS (given in this paper) differ from PKZB (given in \Refs{seidl2000density}) from their correlation point of view. Note that Eq.~\ref{err1} is slightly different from Eq.(38) of ref.~\cite{perdew2004meta}. Similarly, Eq.~\ref{eer} is constructed to ensure its' becomes positive.

\twocolumngrid
\bibliography{newcorr}

\begin{thebibliography}{164}%
\makeatletter
\providecommand \@ifxundefined [1]{%
 \@ifx{#1\undefined}
}%
\providecommand \@ifnum [1]{%
 \ifnum #1\expandafter \@firstoftwo
 \else \expandafter \@secondoftwo
 \fi
}%
\providecommand \@ifx [1]{%
 \ifx #1\expandafter \@firstoftwo
 \else \expandafter \@secondoftwo
 \fi
}%
\providecommand \natexlab [1]{#1}%
\providecommand \enquote  [1]{``#1''}%
\providecommand \bibnamefont  [1]{#1}%
\providecommand \bibfnamefont [1]{#1}%
\providecommand \citenamefont [1]{#1}%
\providecommand \href@noop [0]{\@secondoftwo}%
\providecommand \href [0]{\begingroup \@sanitize@url \@href}%
\providecommand \@href[1]{\@@startlink{#1}\@@href}%
\providecommand \@@href[1]{\endgroup#1\@@endlink}%
\providecommand \@sanitize@url [0]{\catcode `\\12\catcode `\$12\catcode
  `\&12\catcode `\#12\catcode `\^12\catcode `\_12\catcode `\%12\relax}%
\providecommand \@@startlink[1]{}%
\providecommand \@@endlink[0]{}%
\providecommand \url  [0]{\begingroup\@sanitize@url \@url }%
\providecommand \@url [1]{\endgroup\@href {#1}{\urlprefix }}%
\providecommand \urlprefix  [0]{URL }%
\providecommand \Eprint [0]{\href }%
\providecommand \doibase [0]{http://dx.doi.org/}%
\providecommand \selectlanguage [0]{\@gobble}%
\providecommand \bibinfo  [0]{\@secondoftwo}%
\providecommand \bibfield  [0]{\@secondoftwo}%
\providecommand \translation [1]{[#1]}%
\providecommand \BibitemOpen [0]{}%
\providecommand \bibitemStop [0]{}%
\providecommand \bibitemNoStop [0]{.\EOS\space}%
\providecommand \EOS [0]{\spacefactor3000\relax}%
\providecommand \BibitemShut  [1]{\csname bibitem#1\endcsname}%
\let\auto@bib@innerbib\@empty
\bibitem [{\citenamefont {Kohn}\ and\ \citenamefont
  {Sham}(1965)}]{kohn1965self}%
  \BibitemOpen
  \bibfield  {author} {\bibinfo {author} {\bibfnamefont {W.}~\bibnamefont
  {Kohn}}\ and\ \bibinfo {author} {\bibfnamefont {L.~J.}\ \bibnamefont
  {Sham}},\ }\href@noop {} {\bibfield  {journal} {\bibinfo  {journal} {Phys.
  Rev.}\ }\textbf {\bibinfo {volume} {140}},\ \bibinfo {pages} {A1133}
  (\bibinfo {year} {1965})}\BibitemShut {NoStop}%
\bibitem [{\citenamefont {Hohenberg}\ and\ \citenamefont
  {Kohn}(1964)}]{hohenberg1964inhomogeneous}%
  \BibitemOpen
  \bibfield  {author} {\bibinfo {author} {\bibfnamefont {P.}~\bibnamefont
  {Hohenberg}}\ and\ \bibinfo {author} {\bibfnamefont {W.}~\bibnamefont
  {Kohn}},\ }\href@noop {} {\bibfield  {journal} {\bibinfo  {journal} {Phys.
  Rev.}\ }\textbf {\bibinfo {volume} {136}},\ \bibinfo {pages} {B864} (\bibinfo
  {year} {1964})}\BibitemShut {NoStop}%
\bibitem [{\citenamefont {Burke}(2012)}]{burke2012perspective}%
  \BibitemOpen
  \bibfield  {author} {\bibinfo {author} {\bibfnamefont {K.}~\bibnamefont
  {Burke}},\ }\href@noop {} {\bibfield  {journal} {\bibinfo  {journal} {J.
  Chem. Phys.}\ }\textbf {\bibinfo {volume} {136}},\ \bibinfo {pages} {150901}
  (\bibinfo {year} {2012})}\BibitemShut {NoStop}%
\bibitem [{\citenamefont {Levy}(2010)}]{levy2010simple}%
  \BibitemOpen
  \bibfield  {author} {\bibinfo {author} {\bibfnamefont {M.}~\bibnamefont
  {Levy}},\ }\href@noop {} {\bibfield  {journal} {\bibinfo  {journal}
  {International Journal of Quantum Chemistry}\ }\textbf {\bibinfo {volume}
  {110}},\ \bibinfo {pages} {3140} (\bibinfo {year} {2010})}\BibitemShut
  {NoStop}%
\bibitem [{\citenamefont {Levy}(2016)}]{levy2016mathematical}%
  \BibitemOpen
  \bibfield  {author} {\bibinfo {author} {\bibfnamefont {M.}~\bibnamefont
  {Levy}},\ }\href@noop {} {\bibfield  {journal} {\bibinfo  {journal} {Int. J.
  Quantum Chem.}\ }\textbf {\bibinfo {volume} {116}},\ \bibinfo {pages} {802}
  (\bibinfo {year} {2016})}\BibitemShut {NoStop}%
\bibitem [{\citenamefont {Sun}\ \emph {et~al.}(2015)\citenamefont {Sun},
  \citenamefont {Ruzsinszky},\ and\ \citenamefont {Perdew}}]{sun2015strongly}%
  \BibitemOpen
  \bibfield  {author} {\bibinfo {author} {\bibfnamefont {J.}~\bibnamefont
  {Sun}}, \bibinfo {author} {\bibfnamefont {A.}~\bibnamefont {Ruzsinszky}}, \
  and\ \bibinfo {author} {\bibfnamefont {J.~P.}\ \bibnamefont {Perdew}},\
  }\href@noop {} {\bibfield  {journal} {\bibinfo  {journal} {Phys. Rev. Lett.}\
  }\textbf {\bibinfo {volume} {115}},\ \bibinfo {pages} {036402} (\bibinfo
  {year} {2015})}\BibitemShut {NoStop}%
\bibitem [{\citenamefont {Tao}\ and\ \citenamefont
  {Mo}(2016)}]{tao2016accurate}%
  \BibitemOpen
  \bibfield  {author} {\bibinfo {author} {\bibfnamefont {J.}~\bibnamefont
  {Tao}}\ and\ \bibinfo {author} {\bibfnamefont {Y.}~\bibnamefont {Mo}},\
  }\href@noop {} {\bibfield  {journal} {\bibinfo  {journal} {Phys. Rev. Lett.}\
  }\textbf {\bibinfo {volume} {117}},\ \bibinfo {pages} {073001} (\bibinfo
  {year} {2016})}\BibitemShut {NoStop}%
\bibitem [{\citenamefont {Levy}\ and\ \citenamefont
  {Perdew}(1985)}]{levy1985hellmann}%
  \BibitemOpen
  \bibfield  {author} {\bibinfo {author} {\bibfnamefont {M.}~\bibnamefont
  {Levy}}\ and\ \bibinfo {author} {\bibfnamefont {J.~P.}\ \bibnamefont
  {Perdew}},\ }\href@noop {} {\bibfield  {journal} {\bibinfo  {journal} {Phys.
  Rev. A}\ }\textbf {\bibinfo {volume} {32}},\ \bibinfo {pages} {2010}
  (\bibinfo {year} {1985})}\BibitemShut {NoStop}%
\bibitem [{\citenamefont {G{\"o}rling}\ and\ \citenamefont
  {Levy}(1992)}]{gorling1992requirements}%
  \BibitemOpen
  \bibfield  {author} {\bibinfo {author} {\bibfnamefont {A.}~\bibnamefont
  {G{\"o}rling}}\ and\ \bibinfo {author} {\bibfnamefont {M.}~\bibnamefont
  {Levy}},\ }\href@noop {} {\bibfield  {journal} {\bibinfo  {journal} {Phys.
  Rev. A}\ }\textbf {\bibinfo {volume} {45}},\ \bibinfo {pages} {1509}
  (\bibinfo {year} {1992})}\BibitemShut {NoStop}%
\bibitem [{\citenamefont {Fabiano}\ and\ \citenamefont
  {Constantin}(2013)}]{fabiano2013relevance}%
  \BibitemOpen
  \bibfield  {author} {\bibinfo {author} {\bibfnamefont {E.}~\bibnamefont
  {Fabiano}}\ and\ \bibinfo {author} {\bibfnamefont {L.~A.}\ \bibnamefont
  {Constantin}},\ }\href@noop {} {\bibfield  {journal} {\bibinfo  {journal}
  {Phys. Rev. A}\ }\textbf {\bibinfo {volume} {87}},\ \bibinfo {pages} {012511}
  (\bibinfo {year} {2013})}\BibitemShut {NoStop}%
\bibitem [{\citenamefont {Svendsen}\ and\ \citenamefont {von
  Barth}(1996)}]{svendsenPRB96}%
  \BibitemOpen
  \bibfield  {author} {\bibinfo {author} {\bibfnamefont {P.-S.}\ \bibnamefont
  {Svendsen}}\ and\ \bibinfo {author} {\bibfnamefont {U.}~\bibnamefont {von
  Barth}},\ }\href@noop {} {\bibfield  {journal} {\bibinfo  {journal} {Phys.
  Rev. B}\ }\textbf {\bibinfo {volume} {54}},\ \bibinfo {pages} {17402}
  (\bibinfo {year} {1996})}\BibitemShut {NoStop}%
\bibitem [{\citenamefont {Antoniewicz}\ and\ \citenamefont
  {Kleinman}(1985)}]{antoniewiczPRB85}%
  \BibitemOpen
  \bibfield  {author} {\bibinfo {author} {\bibfnamefont {P.~R.}\ \bibnamefont
  {Antoniewicz}}\ and\ \bibinfo {author} {\bibfnamefont {L.}~\bibnamefont
  {Kleinman}},\ }\href@noop {} {\bibfield  {journal} {\bibinfo  {journal}
  {Phys. Rev. B}\ }\textbf {\bibinfo {volume} {31}},\ \bibinfo {pages} {6779}
  (\bibinfo {year} {1985})}\BibitemShut {NoStop}%
\bibitem [{\citenamefont {Hu}\ and\ \citenamefont {Langreth}(1986)}]{huPRB86}%
  \BibitemOpen
  \bibfield  {author} {\bibinfo {author} {\bibfnamefont {C.~D.}\ \bibnamefont
  {Hu}}\ and\ \bibinfo {author} {\bibfnamefont {D.~C.}\ \bibnamefont
  {Langreth}},\ }\href@noop {} {\bibfield  {journal} {\bibinfo  {journal}
  {Phys. Rev. B}\ }\textbf {\bibinfo {volume} {33}},\ \bibinfo {pages} {943}
  (\bibinfo {year} {1986})}\BibitemShut {NoStop}%
\bibitem [{\citenamefont {Ma}\ and\ \citenamefont
  {Brueckner}(1968)}]{bruecknerPR68}%
  \BibitemOpen
  \bibfield  {author} {\bibinfo {author} {\bibfnamefont {S.-K.}\ \bibnamefont
  {Ma}}\ and\ \bibinfo {author} {\bibfnamefont {K.~A.}\ \bibnamefont
  {Brueckner}},\ }\href@noop {} {\bibfield  {journal} {\bibinfo  {journal}
  {Phys. Rev.}\ }\textbf {\bibinfo {volume} {165}},\ \bibinfo {pages} {18}
  (\bibinfo {year} {1968})}\BibitemShut {NoStop}%
\bibitem [{\citenamefont {Argaman}\ \emph {et~al.}(2022)\citenamefont
  {Argaman}, \citenamefont {Redd}, \citenamefont {Cancio},\ and\ \citenamefont
  {Burke}}]{Argaman2022leading}%
  \BibitemOpen
  \bibfield  {author} {\bibinfo {author} {\bibfnamefont {N.}~\bibnamefont
  {Argaman}}, \bibinfo {author} {\bibfnamefont {J.}~\bibnamefont {Redd}},
  \bibinfo {author} {\bibfnamefont {A.~C.}\ \bibnamefont {Cancio}}, \ and\
  \bibinfo {author} {\bibfnamefont {K.}~\bibnamefont {Burke}},\ }\href@noop {}
  {\bibfield  {journal} {\bibinfo  {journal} {Phys. Rev. Lett.}\ }\textbf
  {\bibinfo {volume} {129}},\ \bibinfo {pages} {153001} (\bibinfo {year}
  {2022})}\BibitemShut {NoStop}%
\bibitem [{\citenamefont {Daas}\ \emph
  {et~al.}(2022{\natexlab{a}})\citenamefont {Daas}, \citenamefont {Kooi},
  \citenamefont {Grooteman}, \citenamefont {Seidl},\ and\ \citenamefont
  {Gori-Giorgi}}]{daas2022gradient}%
  \BibitemOpen
  \bibfield  {author} {\bibinfo {author} {\bibfnamefont {T.~J.}\ \bibnamefont
  {Daas}}, \bibinfo {author} {\bibfnamefont {D.~P.}\ \bibnamefont {Kooi}},
  \bibinfo {author} {\bibfnamefont {A.~J. A.~F.}\ \bibnamefont {Grooteman}},
  \bibinfo {author} {\bibfnamefont {M.}~\bibnamefont {Seidl}}, \ and\ \bibinfo
  {author} {\bibfnamefont {P.}~\bibnamefont {Gori-Giorgi}},\ }\href@noop {}
  {\bibfield  {journal} {\bibinfo  {journal} {Journal of Chemical Theory and
  Computation}\ }\textbf {\bibinfo {volume} {18}},\ \bibinfo {pages} {1584}
  (\bibinfo {year} {2022}{\natexlab{a}})}\BibitemShut {NoStop}%
\bibitem [{\citenamefont {Daas}\ \emph
  {et~al.}(2022{\natexlab{b}})\citenamefont {Daas}, \citenamefont {Kooi},
  \citenamefont {Benyahia}, \citenamefont {Seidl},\ and\ \citenamefont
  {Gori-Giorgi}}]{Daas2022arxiv}%
  \BibitemOpen
  \bibfield  {author} {\bibinfo {author} {\bibfnamefont {T.~J.}\ \bibnamefont
  {Daas}}, \bibinfo {author} {\bibfnamefont {D.~P.}\ \bibnamefont {Kooi}},
  \bibinfo {author} {\bibfnamefont {T.}~\bibnamefont {Benyahia}}, \bibinfo
  {author} {\bibfnamefont {M.}~\bibnamefont {Seidl}}, \ and\ \bibinfo {author}
  {\bibfnamefont {P.}~\bibnamefont {Gori-Giorgi}},\ }\href@noop {} {\
  (\bibinfo {year} {2022}{\natexlab{b}})},\ \Eprint
  {http://arxiv.org/abs/2211.07512} {arXiv:2211.07512 [physics.chem-ph]}
  \BibitemShut {NoStop}%
\bibitem [{\citenamefont {G{\"o}rling}\ and\ \citenamefont
  {Levy}(1994)}]{gorling1994exact}%
  \BibitemOpen
  \bibfield  {author} {\bibinfo {author} {\bibfnamefont {A.}~\bibnamefont
  {G{\"o}rling}}\ and\ \bibinfo {author} {\bibfnamefont {M.}~\bibnamefont
  {Levy}},\ }\href@noop {} {\bibfield  {journal} {\bibinfo  {journal} {Phys.
  Rev. A}\ }\textbf {\bibinfo {volume} {50}},\ \bibinfo {pages} {196} (\bibinfo
  {year} {1994})}\BibitemShut {NoStop}%
\bibitem [{\citenamefont {G{\"o}rling}\ and\ \citenamefont
  {Levy}(1993)}]{gorling1993correlation}%
  \BibitemOpen
  \bibfield  {author} {\bibinfo {author} {\bibfnamefont {A.}~\bibnamefont
  {G{\"o}rling}}\ and\ \bibinfo {author} {\bibfnamefont {M.}~\bibnamefont
  {Levy}},\ }\href@noop {} {\bibfield  {journal} {\bibinfo  {journal} {Phys.
  Rev. B}\ }\textbf {\bibinfo {volume} {47}},\ \bibinfo {pages} {13105}
  (\bibinfo {year} {1993})}\BibitemShut {NoStop}%
\bibitem [{\citenamefont {G{\"o}rling}\ and\ \citenamefont
  {Levy}(1995)}]{gorling1995hardness}%
  \BibitemOpen
  \bibfield  {author} {\bibinfo {author} {\bibfnamefont {A.}~\bibnamefont
  {G{\"o}rling}}\ and\ \bibinfo {author} {\bibfnamefont {M.}~\bibnamefont
  {Levy}},\ }\href@noop {} {\bibfield  {journal} {\bibinfo  {journal} {Phys.
  Rev. A}\ }\textbf {\bibinfo {volume} {52}},\ \bibinfo {pages} {4493}
  (\bibinfo {year} {1995})}\BibitemShut {NoStop}%
\bibitem [{\citenamefont {{Della Sala}}\ and\ \citenamefont
  {G{\"o}rling}(2002)}]{dellasalaPRL02}%
  \BibitemOpen
  \bibfield  {author} {\bibinfo {author} {\bibfnamefont {F.}~\bibnamefont
  {{Della Sala}}}\ and\ \bibinfo {author} {\bibfnamefont {A.}~\bibnamefont
  {G{\"o}rling}},\ }\href@noop {} {\bibfield  {journal} {\bibinfo  {journal}
  {Phys. Rev. Lett.}\ }\textbf {\bibinfo {volume} {89}},\ \bibinfo {pages}
  {033003} (\bibinfo {year} {2002})}\BibitemShut {NoStop}%
\bibitem [{\citenamefont {Engel}\ \emph {et~al.}(1992)\citenamefont {Engel},
  \citenamefont {Chevary}, \citenamefont {Macdonald},\ and\ \citenamefont
  {Vosko}}]{engelZPD92}%
  \BibitemOpen
  \bibfield  {author} {\bibinfo {author} {\bibfnamefont {E.}~\bibnamefont
  {Engel}}, \bibinfo {author} {\bibfnamefont {J.}~\bibnamefont {Chevary}},
  \bibinfo {author} {\bibfnamefont {L.}~\bibnamefont {Macdonald}}, \ and\
  \bibinfo {author} {\bibfnamefont {S.}~\bibnamefont {Vosko}},\ }\href@noop {}
  {\bibfield  {journal} {\bibinfo  {journal} {Z. Phys. D}\ }\textbf {\bibinfo
  {volume} {23}},\ \bibinfo {pages} {7} (\bibinfo {year} {1992})}\BibitemShut
  {NoStop}%
\bibitem [{\citenamefont {Horowitz}\ \emph {et~al.}(2009)\citenamefont
  {Horowitz}, \citenamefont {Constantin}, \citenamefont {Proetto},\ and\
  \citenamefont {Pitarke}}]{horowitz2009position}%
  \BibitemOpen
  \bibfield  {author} {\bibinfo {author} {\bibfnamefont {C.~M.}\ \bibnamefont
  {Horowitz}}, \bibinfo {author} {\bibfnamefont {L.~A.}\ \bibnamefont
  {Constantin}}, \bibinfo {author} {\bibfnamefont {C.~R.}\ \bibnamefont
  {Proetto}}, \ and\ \bibinfo {author} {\bibfnamefont {J.~M.}\ \bibnamefont
  {Pitarke}},\ }\href@noop {} {\bibfield  {journal} {\bibinfo  {journal} {Phys.
  Rev. B}\ }\textbf {\bibinfo {volume} {80}},\ \bibinfo {pages} {235101}
  (\bibinfo {year} {2009})}\BibitemShut {NoStop}%
\bibitem [{\citenamefont {Constantin}\ and\ \citenamefont
  {Pitarke}(2011)}]{constantin2011adiabatic}%
  \BibitemOpen
  \bibfield  {author} {\bibinfo {author} {\bibfnamefont {L.~A.}\ \bibnamefont
  {Constantin}}\ and\ \bibinfo {author} {\bibfnamefont {J.~M.}\ \bibnamefont
  {Pitarke}},\ }\href@noop {} {\bibfield  {journal} {\bibinfo  {journal}
  {Physical Review B}\ }\textbf {\bibinfo {volume} {83}},\ \bibinfo {pages}
  {075116} (\bibinfo {year} {2011})}\BibitemShut {NoStop}%
\bibitem [{\citenamefont {Constantin}\ \emph
  {et~al.}(2016{\natexlab{a}})\citenamefont {Constantin}, \citenamefont
  {Fabiano}, \citenamefont {Pitarke},\ and\ \citenamefont
  {Della~Sala}}]{constantin2016semilocal}%
  \BibitemOpen
  \bibfield  {author} {\bibinfo {author} {\bibfnamefont {L.~A.}\ \bibnamefont
  {Constantin}}, \bibinfo {author} {\bibfnamefont {E.}~\bibnamefont {Fabiano}},
  \bibinfo {author} {\bibfnamefont {J.~M.}\ \bibnamefont {Pitarke}}, \ and\
  \bibinfo {author} {\bibfnamefont {F.}~\bibnamefont {Della~Sala}},\
  }\href@noop {} {\bibfield  {journal} {\bibinfo  {journal} {Phys. Rev. B}\
  }\textbf {\bibinfo {volume} {93}},\ \bibinfo {pages} {115127} (\bibinfo
  {year} {2016}{\natexlab{a}})}\BibitemShut {NoStop}%
\bibitem [{\citenamefont {Niquet}\ \emph {et~al.}(2003)\citenamefont {Niquet},
  \citenamefont {Fuchs},\ and\ \citenamefont {Gonze}}]{niquet2003asymptotic}%
  \BibitemOpen
  \bibfield  {author} {\bibinfo {author} {\bibfnamefont {Y.~M.}\ \bibnamefont
  {Niquet}}, \bibinfo {author} {\bibfnamefont {M.}~\bibnamefont {Fuchs}}, \
  and\ \bibinfo {author} {\bibfnamefont {X.}~\bibnamefont {Gonze}},\
  }\href@noop {} {\bibfield  {journal} {\bibinfo  {journal} {J. Chem. Phys.}\
  }\textbf {\bibinfo {volume} {118}},\ \bibinfo {pages} {9504} (\bibinfo {year}
  {2003})}\BibitemShut {NoStop}%
\bibitem [{\citenamefont {Almbladh}\ and\ \citenamefont {von
  Barth}(1985)}]{almbladh1985exact}%
  \BibitemOpen
  \bibfield  {author} {\bibinfo {author} {\bibfnamefont {C.-O.}\ \bibnamefont
  {Almbladh}}\ and\ \bibinfo {author} {\bibfnamefont {U.}~\bibnamefont {von
  Barth}},\ }\href@noop {} {\bibfield  {journal} {\bibinfo  {journal} {Phys.
  Rev. B}\ }\textbf {\bibinfo {volume} {31}},\ \bibinfo {pages} {3231}
  (\bibinfo {year} {1985})}\BibitemShut {NoStop}%
\bibitem [{\citenamefont {Umrigar}\ and\ \citenamefont
  {Gonze}(1994)}]{umrigar1994accurate}%
  \BibitemOpen
  \bibfield  {author} {\bibinfo {author} {\bibfnamefont {C.~J.}\ \bibnamefont
  {Umrigar}}\ and\ \bibinfo {author} {\bibfnamefont {X.}~\bibnamefont
  {Gonze}},\ }\href@noop {} {\bibfield  {journal} {\bibinfo  {journal} {Phys.
  Rev. A}\ }\textbf {\bibinfo {volume} {50}},\ \bibinfo {pages} {3827}
  (\bibinfo {year} {1994})}\BibitemShut {NoStop}%
\bibitem [{\citenamefont {Pollack}\ and\ \citenamefont
  {Perdew}(2000)}]{pollack2000evaluating}%
  \BibitemOpen
  \bibfield  {author} {\bibinfo {author} {\bibfnamefont {L.}~\bibnamefont
  {Pollack}}\ and\ \bibinfo {author} {\bibfnamefont {J.}~\bibnamefont
  {Perdew}},\ }\href@noop {} {\bibfield  {journal} {\bibinfo  {journal}
  {Journal of Physics: Condensed Matter}\ }\textbf {\bibinfo {volume} {12}},\
  \bibinfo {pages} {1239} (\bibinfo {year} {2000})}\BibitemShut {NoStop}%
\bibitem [{\citenamefont {Kaplan}\ \emph {et~al.}(2018)\citenamefont {Kaplan},
  \citenamefont {Wagle},\ and\ \citenamefont {Perdew}}]{kaplan2018collapse}%
  \BibitemOpen
  \bibfield  {author} {\bibinfo {author} {\bibfnamefont {A.~D.}\ \bibnamefont
  {Kaplan}}, \bibinfo {author} {\bibfnamefont {K.}~\bibnamefont {Wagle}}, \
  and\ \bibinfo {author} {\bibfnamefont {J.~P.}\ \bibnamefont {Perdew}},\
  }\href@noop {} {\bibfield  {journal} {\bibinfo  {journal} {Phys. Rev. B}\
  }\textbf {\bibinfo {volume} {98}},\ \bibinfo {pages} {085147} (\bibinfo
  {year} {2018})}\BibitemShut {NoStop}%
\bibitem [{\citenamefont {Constantin}(2016)}]{constantin2016simple}%
  \BibitemOpen
  \bibfield  {author} {\bibinfo {author} {\bibfnamefont {L.~A.}\ \bibnamefont
  {Constantin}},\ }\href@noop {} {\bibfield  {journal} {\bibinfo  {journal}
  {Phys. Rev. B}\ }\textbf {\bibinfo {volume} {93}},\ \bibinfo {pages} {121104}
  (\bibinfo {year} {2016})}\BibitemShut {NoStop}%
\bibitem [{\citenamefont {Constantin}(2008)}]{constantin2008dimensional}%
  \BibitemOpen
  \bibfield  {author} {\bibinfo {author} {\bibfnamefont {L.~A.}\ \bibnamefont
  {Constantin}},\ }\href@noop {} {\bibfield  {journal} {\bibinfo  {journal}
  {Phys. Rev. B}\ }\textbf {\bibinfo {volume} {78}},\ \bibinfo {pages} {155106}
  (\bibinfo {year} {2008})}\BibitemShut {NoStop}%
\bibitem [{\citenamefont {Tao}\ \emph {et~al.}(2008)\citenamefont {Tao},
  \citenamefont {Staroverov}, \citenamefont {Scuseria},\ and\ \citenamefont
  {Perdew}}]{tao2008exact}%
  \BibitemOpen
  \bibfield  {author} {\bibinfo {author} {\bibfnamefont {J.}~\bibnamefont
  {Tao}}, \bibinfo {author} {\bibfnamefont {V.~N.}\ \bibnamefont {Staroverov}},
  \bibinfo {author} {\bibfnamefont {G.~E.}\ \bibnamefont {Scuseria}}, \ and\
  \bibinfo {author} {\bibfnamefont {J.~P.}\ \bibnamefont {Perdew}},\
  }\href@noop {} {\bibfield  {journal} {\bibinfo  {journal} {Phys. Rev. A}\
  }\textbf {\bibinfo {volume} {77}},\ \bibinfo {pages} {012509} (\bibinfo
  {year} {2008})}\BibitemShut {NoStop}%
\bibitem [{\citenamefont {P{\v{r}}ececht{\v{e}}lov{\'a}}\ \emph
  {et~al.}(2014)\citenamefont {P{\v{r}}ececht{\v{e}}lov{\'a}}, \citenamefont
  {Bahmann}, \citenamefont {Kaupp},\ and\ \citenamefont
  {Ernzerhof}}]{pvrecechtvelovaJCP14}%
  \BibitemOpen
  \bibfield  {author} {\bibinfo {author} {\bibfnamefont {J.}~\bibnamefont
  {P{\v{r}}ececht{\v{e}}lov{\'a}}}, \bibinfo {author} {\bibfnamefont
  {H.}~\bibnamefont {Bahmann}}, \bibinfo {author} {\bibfnamefont
  {M.}~\bibnamefont {Kaupp}}, \ and\ \bibinfo {author} {\bibfnamefont
  {M.}~\bibnamefont {Ernzerhof}},\ }\href@noop {} {\bibfield  {journal}
  {\bibinfo  {journal} {J. Chem. Phys.}\ }\textbf {\bibinfo {volume} {141}},\
  \bibinfo {pages} {111102} (\bibinfo {year} {2014})}\BibitemShut {NoStop}%
\bibitem [{\citenamefont {P{\v{r}}ececht{\v{e}}lov{\'a}}\ \emph
  {et~al.}(2015)\citenamefont {P{\v{r}}ececht{\v{e}}lov{\'a}}, \citenamefont
  {Bahmann}, \citenamefont {Kaupp},\ and\ \citenamefont
  {Ernzerhof}}]{pvrecechtvelovaJCP15}%
  \BibitemOpen
  \bibfield  {author} {\bibinfo {author} {\bibfnamefont {J.~P.}\ \bibnamefont
  {P{\v{r}}ececht{\v{e}}lov{\'a}}}, \bibinfo {author} {\bibfnamefont
  {H.}~\bibnamefont {Bahmann}}, \bibinfo {author} {\bibfnamefont
  {M.}~\bibnamefont {Kaupp}}, \ and\ \bibinfo {author} {\bibfnamefont
  {M.}~\bibnamefont {Ernzerhof}},\ }\href@noop {} {\bibfield  {journal}
  {\bibinfo  {journal} {J. Chem. Phys.}\ }\textbf {\bibinfo {volume} {143}},\
  \bibinfo {pages} {144102} (\bibinfo {year} {2015})}\BibitemShut {NoStop}%
\bibitem [{\citenamefont {Perdew}\ and\ \citenamefont
  {Schmidt}(2001)}]{perdew2001jacob}%
  \BibitemOpen
  \bibfield  {author} {\bibinfo {author} {\bibfnamefont {J.~P.}\ \bibnamefont
  {Perdew}}\ and\ \bibinfo {author} {\bibfnamefont {K.}~\bibnamefont
  {Schmidt}},\ }in\ \href@noop {} {\emph {\bibinfo {booktitle} {AIP Conference
  Proceedings}}}\ (\bibinfo {organization} {IOP INSTITUTE OF PHYSICS PUBLISHING
  LTD},\ \bibinfo {year} {2001})\ pp.\ \bibinfo {pages} {1--20}\BibitemShut
  {NoStop}%
\bibitem [{\citenamefont {Grimme}(2006)}]{grimme2006semiempirical}%
  \BibitemOpen
  \bibfield  {author} {\bibinfo {author} {\bibfnamefont {S.}~\bibnamefont
  {Grimme}},\ }\href {\doibase 10.1063/1.2148954} {\bibfield  {journal}
  {\bibinfo  {journal} {J. Chem. Phys.}\ }\textbf {\bibinfo {volume} {124}},\
  \bibinfo {pages} {034108} (\bibinfo {year} {2006})},\ \Eprint
  {http://arxiv.org/abs/https://doi.org/10.1063/1.2148954}
  {https://doi.org/10.1063/1.2148954} \BibitemShut {NoStop}%
\bibitem [{\citenamefont {Mehta}\ \emph {et~al.}(2018)\citenamefont {Mehta},
  \citenamefont {Casanova-Páez},\ and\ \citenamefont {Goerigk}}]{C8CP03852J}%
  \BibitemOpen
  \bibfield  {author} {\bibinfo {author} {\bibfnamefont {N.}~\bibnamefont
  {Mehta}}, \bibinfo {author} {\bibfnamefont {M.}~\bibnamefont
  {Casanova-Páez}}, \ and\ \bibinfo {author} {\bibfnamefont {L.}~\bibnamefont
  {Goerigk}},\ }\href {\doibase 10.1039/C8CP03852J} {\bibfield  {journal}
  {\bibinfo  {journal} {Phys. Chem. Chem. Phys.}\ }\textbf {\bibinfo {volume}
  {20}},\ \bibinfo {pages} {23175} (\bibinfo {year} {2018})}\BibitemShut
  {NoStop}%
\bibitem [{\citenamefont {Bartlett}\ \emph
  {et~al.}(2005{\natexlab{a}})\citenamefont {Bartlett}, \citenamefont
  {Grabowski}, \citenamefont {Hirata},\ and\ \citenamefont
  {Ivanov}}]{bartlett:2005:abinit2}%
  \BibitemOpen
  \bibfield  {author} {\bibinfo {author} {\bibfnamefont {R.~J.}\ \bibnamefont
  {Bartlett}}, \bibinfo {author} {\bibfnamefont {I.}~\bibnamefont {Grabowski}},
  \bibinfo {author} {\bibfnamefont {S.}~\bibnamefont {Hirata}}, \ and\ \bibinfo
  {author} {\bibfnamefont {S.}~\bibnamefont {Ivanov}},\ }\href {\doibase
  10.1063/1.1809605} {\bibfield  {journal} {\bibinfo  {journal} {J. Chem.
  Phys.}\ }\textbf {\bibinfo {volume} {122}},\ \bibinfo {eid} {034104}
  (\bibinfo {year} {2005}{\natexlab{a}})}\BibitemShut {NoStop}%
\bibitem [{\citenamefont {Grabowski}\ \emph {et~al.}(2014)\citenamefont
  {Grabowski}, \citenamefont {Fabiano}, \citenamefont {Teale}, \citenamefont
  {{\'S}miga}, \citenamefont {Buksztel},\ and\ \citenamefont
  {Sala}}]{grabowski2014orbital}%
  \BibitemOpen
  \bibfield  {author} {\bibinfo {author} {\bibfnamefont {I.}~\bibnamefont
  {Grabowski}}, \bibinfo {author} {\bibfnamefont {E.}~\bibnamefont {Fabiano}},
  \bibinfo {author} {\bibfnamefont {A.~M.}\ \bibnamefont {Teale}}, \bibinfo
  {author} {\bibfnamefont {S.}~\bibnamefont {{\'S}miga}}, \bibinfo {author}
  {\bibfnamefont {A.}~\bibnamefont {Buksztel}}, \ and\ \bibinfo {author}
  {\bibfnamefont {F.~D.}\ \bibnamefont {Sala}},\ }\href@noop {} {\bibfield
  {journal} {\bibinfo  {journal} {J. Chem. Phys.}\ }\textbf {\bibinfo {volume}
  {141}},\ \bibinfo {pages} {024113} (\bibinfo {year} {2014})}\BibitemShut
  {NoStop}%
\bibitem [{\citenamefont {\'{S}miga}\ \emph {et~al.}(2020)\citenamefont
  {\'{S}miga}, \citenamefont {Marusiak}, \citenamefont {Grabowski},\ and\
  \citenamefont {Fabiano}}]{SmigaJCP2020}%
  \BibitemOpen
  \bibfield  {author} {\bibinfo {author} {\bibfnamefont {S.}~\bibnamefont
  {\'{S}miga}}, \bibinfo {author} {\bibfnamefont {V.}~\bibnamefont {Marusiak}},
  \bibinfo {author} {\bibfnamefont {I.}~\bibnamefont {Grabowski}}, \ and\
  \bibinfo {author} {\bibfnamefont {E.}~\bibnamefont {Fabiano}},\ }\href
  {\doibase 10.1063/1.5128933} {\bibfield  {journal} {\bibinfo  {journal} {J.
  Chem. Phys.}\ }\textbf {\bibinfo {volume} {152}},\ \bibinfo {pages} {054109}
  (\bibinfo {year} {2020})}\BibitemShut {NoStop}%
\bibitem [{\citenamefont {Sieci\'nska}\ \emph {et~al.}(2022)\citenamefont
  {Sieci\'nska}, \citenamefont {\'Smiga}, \citenamefont {Grabowski},
  \citenamefont {Sala},\ and\ \citenamefont {Fabiano}}]{SOSa}%
  \BibitemOpen
  \bibfield  {author} {\bibinfo {author} {\bibfnamefont {S.}~\bibnamefont
  {Sieci\'nska}}, \bibinfo {author} {\bibfnamefont {S.}~\bibnamefont
  {\'Smiga}}, \bibinfo {author} {\bibfnamefont {I.}~\bibnamefont {Grabowski}},
  \bibinfo {author} {\bibfnamefont {F.~D.}\ \bibnamefont {Sala}}, \ and\
  \bibinfo {author} {\bibfnamefont {E.}~\bibnamefont {Fabiano}},\ }\href
  {\doibase 10.1080/00268976.2022.2037771} {\bibfield  {journal} {\bibinfo
  {journal} {Molecular Physics}\ }\textbf {\bibinfo {volume} {0}},\ \bibinfo
  {pages} {e2037771} (\bibinfo {year} {2022})},\ \Eprint
  {http://arxiv.org/abs/https://doi.org/10.1080/00268976.2022.2037771}
  {https://doi.org/10.1080/00268976.2022.2037771} \BibitemShut {NoStop}%
\bibitem [{\citenamefont {Seidl}\ \emph
  {et~al.}(2000{\natexlab{a}})\citenamefont {Seidl}, \citenamefont {Perdew},\
  and\ \citenamefont {Kurth}}]{seidl2000simulation}%
  \BibitemOpen
  \bibfield  {author} {\bibinfo {author} {\bibfnamefont {M.}~\bibnamefont
  {Seidl}}, \bibinfo {author} {\bibfnamefont {J.~P.}\ \bibnamefont {Perdew}}, \
  and\ \bibinfo {author} {\bibfnamefont {S.}~\bibnamefont {Kurth}},\
  }\href@noop {} {\bibfield  {journal} {\bibinfo  {journal} {Phys. Rev. Lett.}\
  }\textbf {\bibinfo {volume} {84}},\ \bibinfo {pages} {5070} (\bibinfo {year}
  {2000}{\natexlab{a}})}\BibitemShut {NoStop}%
\bibitem [{\citenamefont {Perdew}\ \emph
  {et~al.}(1996{\natexlab{a}})\citenamefont {Perdew}, \citenamefont {Burke},\
  and\ \citenamefont {Ernzerhof}}]{perdewPRL96}%
  \BibitemOpen
  \bibfield  {author} {\bibinfo {author} {\bibfnamefont {J.~P.}\ \bibnamefont
  {Perdew}}, \bibinfo {author} {\bibfnamefont {K.}~\bibnamefont {Burke}}, \
  and\ \bibinfo {author} {\bibfnamefont {M.}~\bibnamefont {Ernzerhof}},\
  }\href@noop {} {\bibfield  {journal} {\bibinfo  {journal} {Phys. Rev. Lett.}\
  }\textbf {\bibinfo {volume} {77}},\ \bibinfo {pages} {3865} (\bibinfo {year}
  {1996}{\natexlab{a}})}\BibitemShut {NoStop}%
\bibitem [{\citenamefont {Scuseria}\ and\ \citenamefont
  {Staroverov}(2005)}]{scuseriaREVIEW05}%
  \BibitemOpen
  \bibfield  {author} {\bibinfo {author} {\bibfnamefont {G.~E.}\ \bibnamefont
  {Scuseria}}\ and\ \bibinfo {author} {\bibfnamefont {V.~N.}\ \bibnamefont
  {Staroverov}},\ }in\ \href@noop {} {\emph {\bibinfo {booktitle} {Theory and
  Application of Computational Chemistry: The First 40 Years}}},\ \bibinfo
  {editor} {edited by\ \bibinfo {editor} {\bibfnamefont {C.~E.}\ \bibnamefont
  {Dykstra}}, \bibinfo {editor} {\bibfnamefont {G.}~\bibnamefont {Frenking}},
  \bibinfo {editor} {\bibfnamefont {K.~S.}\ \bibnamefont {Kim}}, \ and\
  \bibinfo {editor} {\bibfnamefont {G.~E.}\ \bibnamefont {Scuseria}}}\
  (\bibinfo  {publisher} {Elsevier: Amsterdam},\ \bibinfo {year} {2005})\ pp.\
  \bibinfo {pages} {669--724}\BibitemShut {NoStop}%
\bibitem [{\citenamefont {Tao}\ \emph {et~al.}(2003)\citenamefont {Tao},
  \citenamefont {Perdew}, \citenamefont {Staroverov},\ and\ \citenamefont
  {Scuseria}}]{taoPRL03}%
  \BibitemOpen
  \bibfield  {author} {\bibinfo {author} {\bibfnamefont {J.}~\bibnamefont
  {Tao}}, \bibinfo {author} {\bibfnamefont {J.~P.}\ \bibnamefont {Perdew}},
  \bibinfo {author} {\bibfnamefont {V.~N.}\ \bibnamefont {Staroverov}}, \ and\
  \bibinfo {author} {\bibfnamefont {G.~E.}\ \bibnamefont {Scuseria}},\
  }\href@noop {} {\bibfield  {journal} {\bibinfo  {journal} {Phys. Rev. Lett.}\
  }\textbf {\bibinfo {volume} {91}},\ \bibinfo {pages} {146401} (\bibinfo
  {year} {2003})}\BibitemShut {NoStop}%
\bibitem [{\citenamefont {Jana}\ \emph
  {et~al.}(2019{\natexlab{a}})\citenamefont {Jana}, \citenamefont {Sharma},\
  and\ \citenamefont {Samal}}]{jana2019improving}%
  \BibitemOpen
  \bibfield  {author} {\bibinfo {author} {\bibfnamefont {S.}~\bibnamefont
  {Jana}}, \bibinfo {author} {\bibfnamefont {K.}~\bibnamefont {Sharma}}, \ and\
  \bibinfo {author} {\bibfnamefont {P.}~\bibnamefont {Samal}},\ }\href@noop {}
  {\bibfield  {journal} {\bibinfo  {journal} {The Journal of Physical Chemistry
  A}\ }\textbf {\bibinfo {volume} {123}},\ \bibinfo {pages} {6356} (\bibinfo
  {year} {2019}{\natexlab{a}})}\BibitemShut {NoStop}%
\bibitem [{\citenamefont {Patra}\ \emph {et~al.}(2019)\citenamefont {Patra},
  \citenamefont {Jana}, \citenamefont {Constantin},\ and\ \citenamefont
  {Samal}}]{patra2019relevance}%
  \BibitemOpen
  \bibfield  {author} {\bibinfo {author} {\bibfnamefont {B.}~\bibnamefont
  {Patra}}, \bibinfo {author} {\bibfnamefont {S.}~\bibnamefont {Jana}},
  \bibinfo {author} {\bibfnamefont {L.~A.}\ \bibnamefont {Constantin}}, \ and\
  \bibinfo {author} {\bibfnamefont {P.}~\bibnamefont {Samal}},\ }\href@noop {}
  {\bibfield  {journal} {\bibinfo  {journal} {Phys. Rev. B}\ }\textbf {\bibinfo
  {volume} {100}},\ \bibinfo {pages} {155140} (\bibinfo {year}
  {2019})}\BibitemShut {NoStop}%
\bibitem [{\citenamefont {Jana}\ \emph
  {et~al.}(2021{\natexlab{a}})\citenamefont {Jana}, \citenamefont {Behera},
  \citenamefont {{\'{S}}miga}, \citenamefont {Constantin},\ and\ \citenamefont
  {Samal}}]{jana2021szs}%
  \BibitemOpen
  \bibfield  {author} {\bibinfo {author} {\bibfnamefont {S.}~\bibnamefont
  {Jana}}, \bibinfo {author} {\bibfnamefont {S.~K.}\ \bibnamefont {Behera}},
  \bibinfo {author} {\bibfnamefont {S.}~\bibnamefont {{\'{S}}miga}}, \bibinfo
  {author} {\bibfnamefont {L.~A.}\ \bibnamefont {Constantin}}, \ and\ \bibinfo
  {author} {\bibfnamefont {P.}~\bibnamefont {Samal}},\ }\href {\doibase
  10.1088/1367-2630/abfd4d} {\bibfield  {journal} {\bibinfo  {journal} {New J.
  Phys.}\ }\textbf {\bibinfo {volume} {23}},\ \bibinfo {pages} {063007}
  (\bibinfo {year} {2021}{\natexlab{a}})}\BibitemShut {NoStop}%
\bibitem [{\citenamefont {Patra}\ \emph {et~al.}(2020)\citenamefont {Patra},
  \citenamefont {Jana},\ and\ \citenamefont {Samal}}]{patra2020way}%
  \BibitemOpen
  \bibfield  {author} {\bibinfo {author} {\bibfnamefont {A.}~\bibnamefont
  {Patra}}, \bibinfo {author} {\bibfnamefont {S.}~\bibnamefont {Jana}}, \ and\
  \bibinfo {author} {\bibfnamefont {P.}~\bibnamefont {Samal}},\ }\href
  {\doibase 10.1063/5.0025173} {\bibfield  {journal} {\bibinfo  {journal} {The
  Journal of Chemical Physics}\ }\textbf {\bibinfo {volume} {153}},\ \bibinfo
  {pages} {184112} (\bibinfo {year} {2020})},\ \Eprint
  {http://arxiv.org/abs/https://doi.org/10.1063/5.0025173}
  {https://doi.org/10.1063/5.0025173} \BibitemShut {NoStop}%
\bibitem [{\citenamefont {Jana}\ \emph
  {et~al.}(2021{\natexlab{b}})\citenamefont {Jana}, \citenamefont {Behera},
  \citenamefont {Śmiga}, \citenamefont {Constantin},\ and\ \citenamefont
  {Samal}}]{jana2021accurate}%
  \BibitemOpen
  \bibfield  {author} {\bibinfo {author} {\bibfnamefont {S.}~\bibnamefont
  {Jana}}, \bibinfo {author} {\bibfnamefont {S.~K.}\ \bibnamefont {Behera}},
  \bibinfo {author} {\bibfnamefont {S.}~\bibnamefont {Śmiga}}, \bibinfo
  {author} {\bibfnamefont {L.~A.}\ \bibnamefont {Constantin}}, \ and\ \bibinfo
  {author} {\bibfnamefont {P.}~\bibnamefont {Samal}},\ }\href {\doibase
  10.1063/5.0051331} {\bibfield  {journal} {\bibinfo  {journal} {The Journal of
  Chemical Physics}\ }\textbf {\bibinfo {volume} {155}},\ \bibinfo {pages}
  {024103} (\bibinfo {year} {2021}{\natexlab{b}})},\ \Eprint
  {http://arxiv.org/abs/https://doi.org/10.1063/5.0051331}
  {https://doi.org/10.1063/5.0051331} \BibitemShut {NoStop}%
\bibitem [{\citenamefont {Janesko}\ and\ \citenamefont
  {Aguero}(2012)}]{janesko2012nonspherical}%
  \BibitemOpen
  \bibfield  {author} {\bibinfo {author} {\bibfnamefont {B.~G.}\ \bibnamefont
  {Janesko}}\ and\ \bibinfo {author} {\bibfnamefont {A.}~\bibnamefont
  {Aguero}},\ }\href@noop {} {\bibfield  {journal} {\bibinfo  {journal} {J.
  Chem. Phys.}\ }\textbf {\bibinfo {volume} {136}},\ \bibinfo {pages} {024111}
  (\bibinfo {year} {2012})}\BibitemShut {NoStop}%
\bibitem [{\citenamefont {Janesko}(2013)}]{janesko2013rung}%
  \BibitemOpen
  \bibfield  {author} {\bibinfo {author} {\bibfnamefont {B.~G.}\ \bibnamefont
  {Janesko}},\ }\href@noop {} {\bibfield  {journal} {\bibinfo  {journal} {Int.
  J. Quantum Chem.}\ }\textbf {\bibinfo {volume} {113}},\ \bibinfo {pages} {83}
  (\bibinfo {year} {2013})}\BibitemShut {NoStop}%
\bibitem [{\citenamefont {Janesko}(2010)}]{janesko2010rung}%
  \BibitemOpen
  \bibfield  {author} {\bibinfo {author} {\bibfnamefont {B.~G.}\ \bibnamefont
  {Janesko}},\ }\href@noop {} {\bibfield  {journal} {\bibinfo  {journal} {J.
  Chem. Phys.}\ }\textbf {\bibinfo {volume} {133}},\ \bibinfo {pages} {104103}
  (\bibinfo {year} {2010})}\BibitemShut {NoStop}%
\bibitem [{\citenamefont {Janesko}(2012)}]{janesko2012nonempirical}%
  \BibitemOpen
  \bibfield  {author} {\bibinfo {author} {\bibfnamefont {B.~G.}\ \bibnamefont
  {Janesko}},\ }\href@noop {} {\bibfield  {journal} {\bibinfo  {journal} {J.
  Chem. Phys.}\ }\textbf {\bibinfo {volume} {137}},\ \bibinfo {pages} {224110}
  (\bibinfo {year} {2012})}\BibitemShut {NoStop}%
\bibitem [{\citenamefont {Janesko}\ \emph {et~al.}(2018)\citenamefont
  {Janesko}, \citenamefont {Proynov}, \citenamefont {Scalmani},\ and\
  \citenamefont {Frisch}}]{janesko2018long}%
  \BibitemOpen
  \bibfield  {author} {\bibinfo {author} {\bibfnamefont {B.~G.}\ \bibnamefont
  {Janesko}}, \bibinfo {author} {\bibfnamefont {E.}~\bibnamefont {Proynov}},
  \bibinfo {author} {\bibfnamefont {G.}~\bibnamefont {Scalmani}}, \ and\
  \bibinfo {author} {\bibfnamefont {M.~J.}\ \bibnamefont {Frisch}},\
  }\href@noop {} {\bibfield  {journal} {\bibinfo  {journal} {J. Chem. Phys.}\
  }\textbf {\bibinfo {volume} {148}},\ \bibinfo {pages} {104112} (\bibinfo
  {year} {2018})}\BibitemShut {NoStop}%
\bibitem [{\citenamefont {Constantin}\ \emph
  {et~al.}(2016{\natexlab{b}})\citenamefont {Constantin}, \citenamefont
  {Fabiano},\ and\ \citenamefont {Della~Sala}}]{constantin2016hartree}%
  \BibitemOpen
  \bibfield  {author} {\bibinfo {author} {\bibfnamefont {L.~A.}\ \bibnamefont
  {Constantin}}, \bibinfo {author} {\bibfnamefont {E.}~\bibnamefont {Fabiano}},
  \ and\ \bibinfo {author} {\bibfnamefont {F.}~\bibnamefont {Della~Sala}},\
  }\href@noop {} {\bibfield  {journal} {\bibinfo  {journal} {J. Chem. Phys.}\
  }\textbf {\bibinfo {volume} {145}},\ \bibinfo {pages} {084110} (\bibinfo
  {year} {2016}{\natexlab{b}})}\BibitemShut {NoStop}%
\bibitem [{\citenamefont {Constantin}\ \emph
  {et~al.}(2017{\natexlab{a}})\citenamefont {Constantin}, \citenamefont
  {Fabiano},\ and\ \citenamefont {Della~Sala}}]{constantin2017modified}%
  \BibitemOpen
  \bibfield  {author} {\bibinfo {author} {\bibfnamefont {L.~A.}\ \bibnamefont
  {Constantin}}, \bibinfo {author} {\bibfnamefont {E.}~\bibnamefont {Fabiano}},
  \ and\ \bibinfo {author} {\bibfnamefont {F.}~\bibnamefont {Della~Sala}},\
  }\href@noop {} {\bibfield  {journal} {\bibinfo  {journal} {J. Chem. Theory
  Comput.}\ }\textbf {\bibinfo {volume} {13}},\ \bibinfo {pages} {4228}
  (\bibinfo {year} {2017}{\natexlab{a}})}\BibitemShut {NoStop}%
\bibitem [{\citenamefont {Perdew}\ \emph {et~al.}(2008)\citenamefont {Perdew},
  \citenamefont {Staroverov}, \citenamefont {Tao},\ and\ \citenamefont
  {Scuseria}}]{perdew2008density}%
  \BibitemOpen
  \bibfield  {author} {\bibinfo {author} {\bibfnamefont {J.~P.}\ \bibnamefont
  {Perdew}}, \bibinfo {author} {\bibfnamefont {V.~N.}\ \bibnamefont
  {Staroverov}}, \bibinfo {author} {\bibfnamefont {J.}~\bibnamefont {Tao}}, \
  and\ \bibinfo {author} {\bibfnamefont {G.~E.}\ \bibnamefont {Scuseria}},\
  }\href@noop {} {\bibfield  {journal} {\bibinfo  {journal} {Phys. Rev. A}\
  }\textbf {\bibinfo {volume} {78}},\ \bibinfo {pages} {052513} (\bibinfo
  {year} {2008})}\BibitemShut {NoStop}%
\bibitem [{\citenamefont {Perdew}\ \emph {et~al.}(2005)\citenamefont {Perdew},
  \citenamefont {Ruzsinszky}, \citenamefont {Tao}, \citenamefont {Staroverov},
  \citenamefont {Scuseria},\ and\ \citenamefont
  {Csonka}}]{perdew2005prescription}%
  \BibitemOpen
  \bibfield  {author} {\bibinfo {author} {\bibfnamefont {J.~P.}\ \bibnamefont
  {Perdew}}, \bibinfo {author} {\bibfnamefont {A.}~\bibnamefont {Ruzsinszky}},
  \bibinfo {author} {\bibfnamefont {J.}~\bibnamefont {Tao}}, \bibinfo {author}
  {\bibfnamefont {V.~N.}\ \bibnamefont {Staroverov}}, \bibinfo {author}
  {\bibfnamefont {G.~E.}\ \bibnamefont {Scuseria}}, \ and\ \bibinfo {author}
  {\bibfnamefont {G.~I.}\ \bibnamefont {Csonka}},\ }\href@noop {} {\bibfield
  {journal} {\bibinfo  {journal} {J. Chem. Phys.}\ }\textbf {\bibinfo {volume}
  {123}},\ \bibinfo {pages} {062201} (\bibinfo {year} {2005})}\BibitemShut
  {NoStop}%
\bibitem [{\citenamefont {Odashima}\ and\ \citenamefont
  {Capelle}(2009)}]{odashima2009nonempirical}%
  \BibitemOpen
  \bibfield  {author} {\bibinfo {author} {\bibfnamefont {M.~M.}\ \bibnamefont
  {Odashima}}\ and\ \bibinfo {author} {\bibfnamefont {K.}~\bibnamefont
  {Capelle}},\ }\href@noop {} {\bibfield  {journal} {\bibinfo  {journal} {Phys.
  Rev. A}\ }\textbf {\bibinfo {volume} {79}},\ \bibinfo {pages} {062515}
  (\bibinfo {year} {2009})}\BibitemShut {NoStop}%
\bibitem [{\citenamefont {Arbuznikov}\ and\ \citenamefont
  {Kaupp}(2011)}]{arbuznikov2011advances}%
  \BibitemOpen
  \bibfield  {author} {\bibinfo {author} {\bibfnamefont {A.~V.}\ \bibnamefont
  {Arbuznikov}}\ and\ \bibinfo {author} {\bibfnamefont {M.}~\bibnamefont
  {Kaupp}},\ }\href@noop {} {\bibfield  {journal} {\bibinfo  {journal} {Int. J.
  Quantum Chem.}\ }\textbf {\bibinfo {volume} {111}},\ \bibinfo {pages} {2625}
  (\bibinfo {year} {2011})}\BibitemShut {NoStop}%
\bibitem [{\citenamefont {Jaramillo}\ \emph {et~al.}(2003)\citenamefont
  {Jaramillo}, \citenamefont {Scuseria},\ and\ \citenamefont
  {Ernzerhof}}]{jaramillo2003local}%
  \BibitemOpen
  \bibfield  {author} {\bibinfo {author} {\bibfnamefont {J.}~\bibnamefont
  {Jaramillo}}, \bibinfo {author} {\bibfnamefont {G.~E.}\ \bibnamefont
  {Scuseria}}, \ and\ \bibinfo {author} {\bibfnamefont {M.}~\bibnamefont
  {Ernzerhof}},\ }\href@noop {} {\bibfield  {journal} {\bibinfo  {journal} {J.
  Chem. Phys.}\ }\textbf {\bibinfo {volume} {118}},\ \bibinfo {pages} {1068}
  (\bibinfo {year} {2003})}\BibitemShut {NoStop}%
\bibitem [{\citenamefont {K{\"u}mmel}\ and\ \citenamefont
  {Kronik}(2008)}]{kummel2008orbital}%
  \BibitemOpen
  \bibfield  {author} {\bibinfo {author} {\bibfnamefont {S.}~\bibnamefont
  {K{\"u}mmel}}\ and\ \bibinfo {author} {\bibfnamefont {L.}~\bibnamefont
  {Kronik}},\ }\href@noop {} {\bibfield  {journal} {\bibinfo  {journal}
  {Reviews of Modern Physics}\ }\textbf {\bibinfo {volume} {80}},\ \bibinfo
  {pages} {3} (\bibinfo {year} {2008})}\BibitemShut {NoStop}%
\bibitem [{\citenamefont {Becke}(2005)}]{becke2005real}%
  \BibitemOpen
  \bibfield  {author} {\bibinfo {author} {\bibfnamefont {A.~D.}\ \bibnamefont
  {Becke}},\ }\href@noop {} {\bibfield  {journal} {\bibinfo  {journal} {J.
  Chem. Phys.}\ }\textbf {\bibinfo {volume} {122}},\ \bibinfo {pages} {064101}
  (\bibinfo {year} {2005})}\BibitemShut {NoStop}%
\bibitem [{\citenamefont {Becke}\ and\ \citenamefont
  {Johnson}(2007)}]{becke2007unified}%
  \BibitemOpen
  \bibfield  {author} {\bibinfo {author} {\bibfnamefont {A.~D.}\ \bibnamefont
  {Becke}}\ and\ \bibinfo {author} {\bibfnamefont {E.~R.}\ \bibnamefont
  {Johnson}},\ }\href@noop {} {\bibfield  {journal} {\bibinfo  {journal} {J.
  Chem. Phys.}\ }\textbf {\bibinfo {volume} {127}},\ \bibinfo {pages} {124108}
  (\bibinfo {year} {2007})}\BibitemShut {NoStop}%
\bibitem [{\citenamefont {Becke}(2003)}]{becke2003real}%
  \BibitemOpen
  \bibfield  {author} {\bibinfo {author} {\bibfnamefont {A.~D.}\ \bibnamefont
  {Becke}},\ }\href@noop {} {\bibfield  {journal} {\bibinfo  {journal} {J.
  Chem. Phys.}\ }\textbf {\bibinfo {volume} {119}},\ \bibinfo {pages} {2972}
  (\bibinfo {year} {2003})}\BibitemShut {NoStop}%
\bibitem [{\citenamefont {Becke}(2013)}]{becke2013density}%
  \BibitemOpen
  \bibfield  {author} {\bibinfo {author} {\bibfnamefont {A.~D.}\ \bibnamefont
  {Becke}},\ }\href@noop {} {\bibfield  {journal} {\bibinfo  {journal} {J.
  Chem. Phys.}\ }\textbf {\bibinfo {volume} {138}},\ \bibinfo {pages} {074109}
  (\bibinfo {year} {2013})}\BibitemShut {NoStop}%
\bibitem [{\citenamefont {Patra}\ \emph {et~al.}(2018)\citenamefont {Patra},
  \citenamefont {Jana},\ and\ \citenamefont {Samal}}]{patra2018long}%
  \BibitemOpen
  \bibfield  {author} {\bibinfo {author} {\bibfnamefont {B.}~\bibnamefont
  {Patra}}, \bibinfo {author} {\bibfnamefont {S.}~\bibnamefont {Jana}}, \ and\
  \bibinfo {author} {\bibfnamefont {P.}~\bibnamefont {Samal}},\ }\href
  {\doibase 10.1039/C8CP00717A} {\bibfield  {journal} {\bibinfo  {journal}
  {Phys. Chem. Chem. Phys.}\ }\textbf {\bibinfo {volume} {20}},\ \bibinfo
  {pages} {8991} (\bibinfo {year} {2018})}\BibitemShut {NoStop}%
\bibitem [{\citenamefont {Jana}\ \emph
  {et~al.}(2018{\natexlab{a}})\citenamefont {Jana}, \citenamefont {Patra},\
  and\ \citenamefont {Samal}}]{jana2018efficient}%
  \BibitemOpen
  \bibfield  {author} {\bibinfo {author} {\bibfnamefont {S.}~\bibnamefont
  {Jana}}, \bibinfo {author} {\bibfnamefont {A.}~\bibnamefont {Patra}}, \ and\
  \bibinfo {author} {\bibfnamefont {P.}~\bibnamefont {Samal}},\ }\href@noop {}
  {\bibfield  {journal} {\bibinfo  {journal} {The Journal of Chemical Physics}\
  }\textbf {\bibinfo {volume} {149}},\ \bibinfo {pages} {094105} (\bibinfo
  {year} {2018}{\natexlab{a}})}\BibitemShut {NoStop}%
\bibitem [{\citenamefont {Jana}\ and\ \citenamefont
  {Samal}(2019)}]{jana2019screened}%
  \BibitemOpen
  \bibfield  {author} {\bibinfo {author} {\bibfnamefont {S.}~\bibnamefont
  {Jana}}\ and\ \bibinfo {author} {\bibfnamefont {P.}~\bibnamefont {Samal}},\
  }\href@noop {} {\bibfield  {journal} {\bibinfo  {journal} {Phys. Chem. Chem.
  Phys.}\ }\textbf {\bibinfo {volume} {21}},\ \bibinfo {pages} {3002} (\bibinfo
  {year} {2019})}\BibitemShut {NoStop}%
\bibitem [{\citenamefont {Jana}\ \emph
  {et~al.}(2018{\natexlab{b}})\citenamefont {Jana}, \citenamefont {Patra},
  \citenamefont {Myneni},\ and\ \citenamefont {Samal}}]{jana2018cpl}%
  \BibitemOpen
  \bibfield  {author} {\bibinfo {author} {\bibfnamefont {S.}~\bibnamefont
  {Jana}}, \bibinfo {author} {\bibfnamefont {B.}~\bibnamefont {Patra}},
  \bibinfo {author} {\bibfnamefont {H.}~\bibnamefont {Myneni}}, \ and\ \bibinfo
  {author} {\bibfnamefont {P.}~\bibnamefont {Samal}},\ }\href@noop {}
  {\bibfield  {journal} {\bibinfo  {journal} {Chemical Physics Letters}\
  }\textbf {\bibinfo {volume} {713}},\ \bibinfo {pages} {1 } (\bibinfo {year}
  {2018}{\natexlab{b}})}\BibitemShut {NoStop}%
\bibitem [{\citenamefont {Jana}\ \emph
  {et~al.}(2020{\natexlab{a}})\citenamefont {Jana}, \citenamefont {Patra},
  \citenamefont {Constantin},\ and\ \citenamefont {Samal}}]{jana2020screened}%
  \BibitemOpen
  \bibfield  {author} {\bibinfo {author} {\bibfnamefont {S.}~\bibnamefont
  {Jana}}, \bibinfo {author} {\bibfnamefont {A.}~\bibnamefont {Patra}},
  \bibinfo {author} {\bibfnamefont {L.~A.}\ \bibnamefont {Constantin}}, \ and\
  \bibinfo {author} {\bibfnamefont {P.}~\bibnamefont {Samal}},\ }\href@noop {}
  {\bibfield  {journal} {\bibinfo  {journal} {The Journal of Chemical Physics}\
  }\textbf {\bibinfo {volume} {152}},\ \bibinfo {pages} {044111} (\bibinfo
  {year} {2020}{\natexlab{a}})}\BibitemShut {NoStop}%
\bibitem [{\citenamefont {Jana}\ \emph
  {et~al.}(2019{\natexlab{b}})\citenamefont {Jana}, \citenamefont {Patra},
  \citenamefont {Constantin}, \citenamefont {Myneni},\ and\ \citenamefont
  {Samal}}]{jana2019long}%
  \BibitemOpen
  \bibfield  {author} {\bibinfo {author} {\bibfnamefont {S.}~\bibnamefont
  {Jana}}, \bibinfo {author} {\bibfnamefont {A.}~\bibnamefont {Patra}},
  \bibinfo {author} {\bibfnamefont {L.~A.}\ \bibnamefont {Constantin}},
  \bibinfo {author} {\bibfnamefont {H.}~\bibnamefont {Myneni}}, \ and\ \bibinfo
  {author} {\bibfnamefont {P.}~\bibnamefont {Samal}},\ }\href@noop {}
  {\bibfield  {journal} {\bibinfo  {journal} {Phys. Rev. A}\ }\textbf {\bibinfo
  {volume} {99}},\ \bibinfo {pages} {042515} (\bibinfo {year}
  {2019}{\natexlab{b}})}\BibitemShut {NoStop}%
\bibitem [{\citenamefont {Jana}\ \emph {et~al.}(2022)\citenamefont {Jana},
  \citenamefont {Constantin}, \citenamefont {Śmiga},\ and\ \citenamefont
  {Samal}}]{jana2022solid}%
  \BibitemOpen
  \bibfield  {author} {\bibinfo {author} {\bibfnamefont {S.}~\bibnamefont
  {Jana}}, \bibinfo {author} {\bibfnamefont {L.~A.}\ \bibnamefont
  {Constantin}}, \bibinfo {author} {\bibfnamefont {S.}~\bibnamefont {Śmiga}},
  \ and\ \bibinfo {author} {\bibfnamefont {P.}~\bibnamefont {Samal}},\
  }\href@noop {} {\bibfield  {journal} {\bibinfo  {journal} {The Journal of
  Chemical Physics}\ }\textbf {\bibinfo {volume} {157}},\ \bibinfo {pages}
  {024102} (\bibinfo {year} {2022})}\BibitemShut {NoStop}%
\bibitem [{\citenamefont {Grimme}\ and\ \citenamefont
  {Neese}(2007)}]{grimme2007doublehybrid}%
  \BibitemOpen
  \bibfield  {author} {\bibinfo {author} {\bibfnamefont {S.}~\bibnamefont
  {Grimme}}\ and\ \bibinfo {author} {\bibfnamefont {F.}~\bibnamefont {Neese}},\
  }\href@noop {} {\bibfield  {journal} {\bibinfo  {journal} {The Journal of
  Chemical Physics}\ }\textbf {\bibinfo {volume} {127}},\ \bibinfo {pages}
  {154116} (\bibinfo {year} {2007})}\BibitemShut {NoStop}%
\bibitem [{\citenamefont {Su}\ and\ \citenamefont
  {Xu}(2014)}]{neil2014construction}%
  \BibitemOpen
  \bibfield  {author} {\bibinfo {author} {\bibfnamefont {N.~Q.}\ \bibnamefont
  {Su}}\ and\ \bibinfo {author} {\bibfnamefont {X.}~\bibnamefont {Xu}},\ }\href
  {\doibase 10.1063/1.4866457} {\bibfield  {journal} {\bibinfo  {journal} {The
  Journal of Chemical Physics}\ }\textbf {\bibinfo {volume} {140}},\ \bibinfo
  {pages} {18A512} (\bibinfo {year} {2014})},\ \Eprint
  {http://arxiv.org/abs/https://doi.org/10.1063/1.4866457}
  {https://doi.org/10.1063/1.4866457} \BibitemShut {NoStop}%
\bibitem [{\citenamefont {Hui}\ and\ \citenamefont {Chai}(2016)}]{scan0}%
  \BibitemOpen
  \bibfield  {author} {\bibinfo {author} {\bibfnamefont {K.}~\bibnamefont
  {Hui}}\ and\ \bibinfo {author} {\bibfnamefont {J.-D.}\ \bibnamefont {Chai}},\
  }\href@noop {} {\bibfield  {journal} {\bibinfo  {journal} {The Journal of
  Chemical Physics}\ }\textbf {\bibinfo {volume} {144}},\ \bibinfo {pages}
  {044114} (\bibinfo {year} {2016})}\BibitemShut {NoStop}%
\bibitem [{\citenamefont {Sharkas}\ \emph {et~al.}(2011)\citenamefont
  {Sharkas}, \citenamefont {Toulouse},\ and\ \citenamefont
  {Savin}}]{sharkas2011double}%
  \BibitemOpen
  \bibfield  {author} {\bibinfo {author} {\bibfnamefont {K.}~\bibnamefont
  {Sharkas}}, \bibinfo {author} {\bibfnamefont {J.}~\bibnamefont {Toulouse}}, \
  and\ \bibinfo {author} {\bibfnamefont {A.}~\bibnamefont {Savin}},\
  }\href@noop {} {\bibfield  {journal} {\bibinfo  {journal} {The Journal of
  chemical physics}\ }\textbf {\bibinfo {volume} {134}},\ \bibinfo {pages}
  {064113} (\bibinfo {year} {2011})}\BibitemShut {NoStop}%
\bibitem [{\citenamefont {Souvi}\ \emph {et~al.}(2014)\citenamefont {Souvi},
  \citenamefont {Sharkas},\ and\ \citenamefont {Toulouse}}]{souvi2014double}%
  \BibitemOpen
  \bibfield  {author} {\bibinfo {author} {\bibfnamefont {S.~M.}\ \bibnamefont
  {Souvi}}, \bibinfo {author} {\bibfnamefont {K.}~\bibnamefont {Sharkas}}, \
  and\ \bibinfo {author} {\bibfnamefont {J.}~\bibnamefont {Toulouse}},\
  }\href@noop {} {\bibfield  {journal} {\bibinfo  {journal} {The Journal of
  chemical physics}\ }\textbf {\bibinfo {volume} {140}},\ \bibinfo {pages}
  {084107} (\bibinfo {year} {2014})}\BibitemShut {NoStop}%
\bibitem [{\citenamefont {Toulouse}\ \emph {et~al.}(2011)\citenamefont
  {Toulouse}, \citenamefont {Sharkas}, \citenamefont {Brémond},\ and\
  \citenamefont {Adamo}}]{doi:10.1063/1.3640019}%
  \BibitemOpen
  \bibfield  {author} {\bibinfo {author} {\bibfnamefont {J.}~\bibnamefont
  {Toulouse}}, \bibinfo {author} {\bibfnamefont {K.}~\bibnamefont {Sharkas}},
  \bibinfo {author} {\bibfnamefont {E.}~\bibnamefont {Brémond}}, \ and\
  \bibinfo {author} {\bibfnamefont {C.}~\bibnamefont {Adamo}},\ }\href
  {\doibase 10.1063/1.3640019} {\bibfield  {journal} {\bibinfo  {journal} {The
  Journal of Chemical Physics}\ }\textbf {\bibinfo {volume} {135}},\ \bibinfo
  {pages} {101102} (\bibinfo {year} {2011})},\ \Eprint
  {http://arxiv.org/abs/https://doi.org/10.1063/1.3640019}
  {https://doi.org/10.1063/1.3640019} \BibitemShut {NoStop}%
\bibitem [{\citenamefont {Toulouse}\ \emph {et~al.}(2009)\citenamefont
  {Toulouse}, \citenamefont {Gerber}, \citenamefont {Jansen}, \citenamefont
  {Savin},\ and\ \citenamefont {Angy{\'a}n}}]{toulouse2009adiabatic}%
  \BibitemOpen
  \bibfield  {author} {\bibinfo {author} {\bibfnamefont {J.}~\bibnamefont
  {Toulouse}}, \bibinfo {author} {\bibfnamefont {I.~C.}\ \bibnamefont
  {Gerber}}, \bibinfo {author} {\bibfnamefont {G.}~\bibnamefont {Jansen}},
  \bibinfo {author} {\bibfnamefont {A.}~\bibnamefont {Savin}}, \ and\ \bibinfo
  {author} {\bibfnamefont {J.~G.}\ \bibnamefont {Angy{\'a}n}},\ }\href@noop {}
  {\bibfield  {journal} {\bibinfo  {journal} {Phys. Rev. Lett.}\ }\textbf
  {\bibinfo {volume} {102}},\ \bibinfo {pages} {096404} (\bibinfo {year}
  {2009})}\BibitemShut {NoStop}%
\bibitem [{\citenamefont {Ruzsinszky}\ \emph {et~al.}(2016)\citenamefont
  {Ruzsinszky}, \citenamefont {Constantin},\ and\ \citenamefont
  {Pitarke}}]{ruzsinszky2016kernel}%
  \BibitemOpen
  \bibfield  {author} {\bibinfo {author} {\bibfnamefont {A.}~\bibnamefont
  {Ruzsinszky}}, \bibinfo {author} {\bibfnamefont {L.~A.}\ \bibnamefont
  {Constantin}}, \ and\ \bibinfo {author} {\bibfnamefont {J.~M.}\ \bibnamefont
  {Pitarke}},\ }\href@noop {} {\bibfield  {journal} {\bibinfo  {journal} {Phys.
  Rev. B}\ }\textbf {\bibinfo {volume} {94}},\ \bibinfo {pages} {165155}
  (\bibinfo {year} {2016})}\BibitemShut {NoStop}%
\bibitem [{\citenamefont {Ruzsinszky}\ \emph {et~al.}(2010)\citenamefont
  {Ruzsinszky}, \citenamefont {Perdew},\ and\ \citenamefont
  {Csonka}}]{ruzsinszky2010rpa}%
  \BibitemOpen
  \bibfield  {author} {\bibinfo {author} {\bibfnamefont {A.}~\bibnamefont
  {Ruzsinszky}}, \bibinfo {author} {\bibfnamefont {J.~P.}\ \bibnamefont
  {Perdew}}, \ and\ \bibinfo {author} {\bibfnamefont {G.~I.}\ \bibnamefont
  {Csonka}},\ }\href@noop {} {\bibfield  {journal} {\bibinfo  {journal}
  {Journal of Chemical Theory and Computation}\ }\textbf {\bibinfo {volume}
  {6}},\ \bibinfo {pages} {127} (\bibinfo {year} {2010})}\BibitemShut {NoStop}%
\bibitem [{\citenamefont {Bates}\ \emph {et~al.}(2017)\citenamefont {Bates},
  \citenamefont {Sensenig},\ and\ \citenamefont
  {Ruzsinszky}}]{bates2017convergence}%
  \BibitemOpen
  \bibfield  {author} {\bibinfo {author} {\bibfnamefont {J.~E.}\ \bibnamefont
  {Bates}}, \bibinfo {author} {\bibfnamefont {J.}~\bibnamefont {Sensenig}}, \
  and\ \bibinfo {author} {\bibfnamefont {A.}~\bibnamefont {Ruzsinszky}},\
  }\href@noop {} {\bibfield  {journal} {\bibinfo  {journal} {Phys. Rev. B}\
  }\textbf {\bibinfo {volume} {95}},\ \bibinfo {pages} {195158} (\bibinfo
  {year} {2017})}\BibitemShut {NoStop}%
\bibitem [{\citenamefont {Terentjev}\ \emph {et~al.}(2018)\citenamefont
  {Terentjev}, \citenamefont {Constantin},\ and\ \citenamefont
  {Pitarke}}]{terentjev2018gradient}%
  \BibitemOpen
  \bibfield  {author} {\bibinfo {author} {\bibfnamefont {A.~V.}\ \bibnamefont
  {Terentjev}}, \bibinfo {author} {\bibfnamefont {L.~A.}\ \bibnamefont
  {Constantin}}, \ and\ \bibinfo {author} {\bibfnamefont {J.~M.}\ \bibnamefont
  {Pitarke}},\ }\href@noop {} {\bibfield  {journal} {\bibinfo  {journal} {Phys.
  Rev. B}\ }\textbf {\bibinfo {volume} {98}},\ \bibinfo {pages} {085123}
  (\bibinfo {year} {2018})}\BibitemShut {NoStop}%
\bibitem [{\citenamefont {Corradini}\ \emph {et~al.}(1998)\citenamefont
  {Corradini}, \citenamefont {Del~Sole}, \citenamefont {Onida},\ and\
  \citenamefont {Palummo}}]{corradini1998analytical}%
  \BibitemOpen
  \bibfield  {author} {\bibinfo {author} {\bibfnamefont {M.}~\bibnamefont
  {Corradini}}, \bibinfo {author} {\bibfnamefont {R.}~\bibnamefont {Del~Sole}},
  \bibinfo {author} {\bibfnamefont {G.}~\bibnamefont {Onida}}, \ and\ \bibinfo
  {author} {\bibfnamefont {M.}~\bibnamefont {Palummo}},\ }\href@noop {}
  {\bibfield  {journal} {\bibinfo  {journal} {Phys. Rev. B}\ }\textbf {\bibinfo
  {volume} {57}},\ \bibinfo {pages} {14569} (\bibinfo {year}
  {1998})}\BibitemShut {NoStop}%
\bibitem [{\citenamefont {Erhard}\ \emph {et~al.}(2016)\citenamefont {Erhard},
  \citenamefont {Bleiziffer},\ and\ \citenamefont
  {G{\"o}rling}}]{erhard2016power}%
  \BibitemOpen
  \bibfield  {author} {\bibinfo {author} {\bibfnamefont {J.}~\bibnamefont
  {Erhard}}, \bibinfo {author} {\bibfnamefont {P.}~\bibnamefont {Bleiziffer}},
  \ and\ \bibinfo {author} {\bibfnamefont {A.}~\bibnamefont {G{\"o}rling}},\
  }\href@noop {} {\bibfield  {journal} {\bibinfo  {journal} {Phys. Rev. Lett.}\
  }\textbf {\bibinfo {volume} {117}},\ \bibinfo {pages} {143002} (\bibinfo
  {year} {2016})}\BibitemShut {NoStop}%
\bibitem [{\citenamefont {Patrick}\ and\ \citenamefont
  {Thygesen}(2015)}]{patrick2015adiabatic}%
  \BibitemOpen
  \bibfield  {author} {\bibinfo {author} {\bibfnamefont {C.~E.}\ \bibnamefont
  {Patrick}}\ and\ \bibinfo {author} {\bibfnamefont {K.~S.}\ \bibnamefont
  {Thygesen}},\ }\href@noop {} {\bibfield  {journal} {\bibinfo  {journal} {J.
  Chem. Phys.}\ }\textbf {\bibinfo {volume} {143}},\ \bibinfo {pages} {102802}
  (\bibinfo {year} {2015})}\BibitemShut {NoStop}%
\bibitem [{\citenamefont {Bartlett}\ \emph
  {et~al.}(2005{\natexlab{b}})\citenamefont {Bartlett}, \citenamefont
  {Grabowski}, \citenamefont {Hirata},\ and\ \citenamefont
  {Ivanov}}]{bartlett2005exchange}%
  \BibitemOpen
  \bibfield  {author} {\bibinfo {author} {\bibfnamefont {R.~J.}\ \bibnamefont
  {Bartlett}}, \bibinfo {author} {\bibfnamefont {I.}~\bibnamefont {Grabowski}},
  \bibinfo {author} {\bibfnamefont {S.}~\bibnamefont {Hirata}}, \ and\ \bibinfo
  {author} {\bibfnamefont {S.}~\bibnamefont {Ivanov}},\ }\href@noop {}
  {\bibfield  {journal} {\bibinfo  {journal} {J. Chem. Phys.}\ }\textbf
  {\bibinfo {volume} {122}},\ \bibinfo {pages} {034104} (\bibinfo {year}
  {2005}{\natexlab{b}})}\BibitemShut {NoStop}%
\bibitem [{\citenamefont {Bartlett}\ \emph
  {et~al.}(2005{\natexlab{c}})\citenamefont {Bartlett}, \citenamefont
  {Lotrich},\ and\ \citenamefont {Schweigert}}]{bartlett2005ab}%
  \BibitemOpen
  \bibfield  {author} {\bibinfo {author} {\bibfnamefont {R.~J.}\ \bibnamefont
  {Bartlett}}, \bibinfo {author} {\bibfnamefont {V.~F.}\ \bibnamefont
  {Lotrich}}, \ and\ \bibinfo {author} {\bibfnamefont {I.~V.}\ \bibnamefont
  {Schweigert}},\ }\href@noop {} {\bibfield  {journal} {\bibinfo  {journal} {J.
  Chem. Phys.}\ }\textbf {\bibinfo {volume} {123}},\ \bibinfo {pages} {062205}
  (\bibinfo {year} {2005}{\natexlab{c}})}\BibitemShut {NoStop}%
\bibitem [{\citenamefont {Grabowski}\ \emph {et~al.}(2013)\citenamefont
  {Grabowski}, \citenamefont {Fabiano},\ and\ \citenamefont
  {Della~Sala}}]{grabowski2013optimized}%
  \BibitemOpen
  \bibfield  {author} {\bibinfo {author} {\bibfnamefont {I.}~\bibnamefont
  {Grabowski}}, \bibinfo {author} {\bibfnamefont {E.}~\bibnamefont {Fabiano}},
  \ and\ \bibinfo {author} {\bibfnamefont {F.}~\bibnamefont {Della~Sala}},\
  }\href@noop {} {\bibfield  {journal} {\bibinfo  {journal} {Phys. Rev. B}\
  }\textbf {\bibinfo {volume} {87}},\ \bibinfo {pages} {075103} (\bibinfo
  {year} {2013})}\BibitemShut {NoStop}%
\bibitem [{\citenamefont {Langreth}\ and\ \citenamefont
  {Perdew}(1975)}]{langreth1975exchange}%
  \BibitemOpen
  \bibfield  {author} {\bibinfo {author} {\bibfnamefont {D.~C.}\ \bibnamefont
  {Langreth}}\ and\ \bibinfo {author} {\bibfnamefont {J.~P.}\ \bibnamefont
  {Perdew}},\ }\href@noop {} {\bibfield  {journal} {\bibinfo  {journal} {Solid
  State Communications}\ }\textbf {\bibinfo {volume} {17}},\ \bibinfo {pages}
  {1425} (\bibinfo {year} {1975})}\BibitemShut {NoStop}%
\bibitem [{\citenamefont {Gunnarsson}\ and\ \citenamefont
  {Lundqvist}(1976)}]{gunnarsson1976exchange}%
  \BibitemOpen
  \bibfield  {author} {\bibinfo {author} {\bibfnamefont {O.}~\bibnamefont
  {Gunnarsson}}\ and\ \bibinfo {author} {\bibfnamefont {B.~I.}\ \bibnamefont
  {Lundqvist}},\ }\href@noop {} {\bibfield  {journal} {\bibinfo  {journal}
  {Phys. Rev. B}\ }\textbf {\bibinfo {volume} {13}},\ \bibinfo {pages} {4274}
  (\bibinfo {year} {1976})}\BibitemShut {NoStop}%
\bibitem [{\citenamefont {Savin}\ \emph {et~al.}(2003)\citenamefont {Savin},
  \citenamefont {Colonna},\ and\ \citenamefont {Pollet}}]{savin2003adiabatic}%
  \BibitemOpen
  \bibfield  {author} {\bibinfo {author} {\bibfnamefont {A.}~\bibnamefont
  {Savin}}, \bibinfo {author} {\bibfnamefont {F.}~\bibnamefont {Colonna}}, \
  and\ \bibinfo {author} {\bibfnamefont {R.}~\bibnamefont {Pollet}},\
  }\href@noop {} {\bibfield  {journal} {\bibinfo  {journal} {Int. J. Quantum
  Chem.}\ }\textbf {\bibinfo {volume} {93}},\ \bibinfo {pages} {166} (\bibinfo
  {year} {2003})}\BibitemShut {NoStop}%
\bibitem [{\citenamefont {Cohen}\ \emph {et~al.}(2007)\citenamefont {Cohen},
  \citenamefont {Mori-S{\'a}nchez},\ and\ \citenamefont
  {Yang}}]{cohen2007assessment}%
  \BibitemOpen
  \bibfield  {author} {\bibinfo {author} {\bibfnamefont {A.~J.}\ \bibnamefont
  {Cohen}}, \bibinfo {author} {\bibfnamefont {P.}~\bibnamefont
  {Mori-S{\'a}nchez}}, \ and\ \bibinfo {author} {\bibfnamefont
  {W.}~\bibnamefont {Yang}},\ }\href@noop {} {\bibfield  {journal} {\bibinfo
  {journal} {J. Chem. Phys.}\ }\textbf {\bibinfo {volume} {127}},\ \bibinfo
  {pages} {034101} (\bibinfo {year} {2007})}\BibitemShut {NoStop}%
\bibitem [{\citenamefont {Ernzerhof}(1996)}]{ernzerhof1996construction}%
  \BibitemOpen
  \bibfield  {author} {\bibinfo {author} {\bibfnamefont {M.}~\bibnamefont
  {Ernzerhof}},\ }\href@noop {} {\bibfield  {journal} {\bibinfo  {journal}
  {Chem. Phys. Lett.}\ }\textbf {\bibinfo {volume} {263}},\ \bibinfo {pages}
  {499} (\bibinfo {year} {1996})}\BibitemShut {NoStop}%
\bibitem [{\citenamefont {Burke}\ \emph {et~al.}(1997)\citenamefont {Burke},
  \citenamefont {Ernzerhof},\ and\ \citenamefont
  {Perdew}}]{burke1997adiabatic}%
  \BibitemOpen
  \bibfield  {author} {\bibinfo {author} {\bibfnamefont {K.}~\bibnamefont
  {Burke}}, \bibinfo {author} {\bibfnamefont {M.}~\bibnamefont {Ernzerhof}}, \
  and\ \bibinfo {author} {\bibfnamefont {J.~P.}\ \bibnamefont {Perdew}},\
  }\href@noop {} {\bibfield  {journal} {\bibinfo  {journal} {Chem. Phys.
  Lett.}\ }\textbf {\bibinfo {volume} {265}},\ \bibinfo {pages} {115} (\bibinfo
  {year} {1997})}\BibitemShut {NoStop}%
\bibitem [{\citenamefont {Colonna}\ and\ \citenamefont
  {Savin}(1999)}]{colonna1999correlation}%
  \BibitemOpen
  \bibfield  {author} {\bibinfo {author} {\bibfnamefont {F.}~\bibnamefont
  {Colonna}}\ and\ \bibinfo {author} {\bibfnamefont {A.}~\bibnamefont
  {Savin}},\ }\href@noop {} {\bibfield  {journal} {\bibinfo  {journal} {J.
  Chem. Phys.}\ }\textbf {\bibinfo {volume} {110}},\ \bibinfo {pages} {2828}
  (\bibinfo {year} {1999})}\BibitemShut {NoStop}%
\bibitem [{\citenamefont {Adamo}\ and\ \citenamefont
  {Barone}(1998)}]{adamo1998exchange}%
  \BibitemOpen
  \bibfield  {author} {\bibinfo {author} {\bibfnamefont {C.}~\bibnamefont
  {Adamo}}\ and\ \bibinfo {author} {\bibfnamefont {V.}~\bibnamefont {Barone}},\
  }\href@noop {} {\bibfield  {journal} {\bibinfo  {journal} {J. Chem. Phys.}\
  }\textbf {\bibinfo {volume} {108}},\ \bibinfo {pages} {664} (\bibinfo {year}
  {1998})}\BibitemShut {NoStop}%
\bibitem [{\citenamefont {Perdew}\ \emph {et~al.}(2001)\citenamefont {Perdew},
  \citenamefont {Kurth},\ and\ \citenamefont {Seidl}}]{perdew2001exploring}%
  \BibitemOpen
  \bibfield  {author} {\bibinfo {author} {\bibfnamefont {J.~P.}\ \bibnamefont
  {Perdew}}, \bibinfo {author} {\bibfnamefont {S.}~\bibnamefont {Kurth}}, \
  and\ \bibinfo {author} {\bibfnamefont {M.}~\bibnamefont {Seidl}},\
  }\href@noop {} {\bibfield  {journal} {\bibinfo  {journal} {Int. J. Mod. Phys.
  B}\ }\textbf {\bibinfo {volume} {15}},\ \bibinfo {pages} {1672} (\bibinfo
  {year} {2001})}\BibitemShut {NoStop}%
\bibitem [{\citenamefont {Liu}\ and\ \citenamefont
  {Burke}(2009)}]{liu2009adiabatic}%
  \BibitemOpen
  \bibfield  {author} {\bibinfo {author} {\bibfnamefont {Z.-F.}\ \bibnamefont
  {Liu}}\ and\ \bibinfo {author} {\bibfnamefont {K.}~\bibnamefont {Burke}},\
  }\href@noop {} {\bibfield  {journal} {\bibinfo  {journal} {Phys. Rev. A}\
  }\textbf {\bibinfo {volume} {79}},\ \bibinfo {pages} {064503} (\bibinfo
  {year} {2009})}\BibitemShut {NoStop}%
\bibitem [{\citenamefont {Magyar}\ \emph {et~al.}(2003)\citenamefont {Magyar},
  \citenamefont {Terilla},\ and\ \citenamefont {Burke}}]{magyar2003accurate}%
  \BibitemOpen
  \bibfield  {author} {\bibinfo {author} {\bibfnamefont {R.}~\bibnamefont
  {Magyar}}, \bibinfo {author} {\bibfnamefont {W.}~\bibnamefont {Terilla}}, \
  and\ \bibinfo {author} {\bibfnamefont {K.}~\bibnamefont {Burke}},\
  }\href@noop {} {\bibfield  {journal} {\bibinfo  {journal} {J. Chem. Phys.}\
  }\textbf {\bibinfo {volume} {119}},\ \bibinfo {pages} {696} (\bibinfo {year}
  {2003})}\BibitemShut {NoStop}%
\bibitem [{\citenamefont {Sun}(2009)}]{sun2009extension}%
  \BibitemOpen
  \bibfield  {author} {\bibinfo {author} {\bibfnamefont {J.}~\bibnamefont
  {Sun}},\ }\href@noop {} {\bibfield  {journal} {\bibinfo  {journal} {J. Chem.
  Theory Comput.}\ }\textbf {\bibinfo {volume} {5}},\ \bibinfo {pages} {708}
  (\bibinfo {year} {2009})}\BibitemShut {NoStop}%
\bibitem [{\citenamefont {Seidl}\ and\ \citenamefont
  {Gori-Giorgi}(2010)}]{seidl2010adiabatic}%
  \BibitemOpen
  \bibfield  {author} {\bibinfo {author} {\bibfnamefont {M.}~\bibnamefont
  {Seidl}}\ and\ \bibinfo {author} {\bibfnamefont {P.}~\bibnamefont
  {Gori-Giorgi}},\ }\href@noop {} {\bibfield  {journal} {\bibinfo  {journal}
  {Phys. Rev. A}\ }\textbf {\bibinfo {volume} {81}},\ \bibinfo {pages} {012508}
  (\bibinfo {year} {2010})}\BibitemShut {NoStop}%
\bibitem [{\citenamefont {Vuckovic}\ \emph
  {et~al.}(2016{\natexlab{a}})\citenamefont {Vuckovic}, \citenamefont {Irons},
  \citenamefont {Savin}, \citenamefont {Teale},\ and\ \citenamefont
  {Gori-Giorgi}}]{vuckovic2016exchange}%
  \BibitemOpen
  \bibfield  {author} {\bibinfo {author} {\bibfnamefont {S.}~\bibnamefont
  {Vuckovic}}, \bibinfo {author} {\bibfnamefont {T.~J.}\ \bibnamefont {Irons}},
  \bibinfo {author} {\bibfnamefont {A.}~\bibnamefont {Savin}}, \bibinfo
  {author} {\bibfnamefont {A.~M.}\ \bibnamefont {Teale}}, \ and\ \bibinfo
  {author} {\bibfnamefont {P.}~\bibnamefont {Gori-Giorgi}},\ }\href@noop {}
  {\bibfield  {journal} {\bibinfo  {journal} {J. Chem. Theory Comput.}\
  }\textbf {\bibinfo {volume} {12}},\ \bibinfo {pages} {2598} (\bibinfo {year}
  {2016}{\natexlab{a}})}\BibitemShut {NoStop}%
\bibitem [{\citenamefont {Fabiano}\ \emph {et~al.}(2019)\citenamefont
  {Fabiano}, \citenamefont {\'Smiga}, \citenamefont {Giarrusso}, \citenamefont
  {Daas}, \citenamefont {Della~Sala}, \citenamefont {Grabowski},\ and\
  \citenamefont {Gori-Giorgi}}]{fabiano2018investigation}%
  \BibitemOpen
  \bibfield  {author} {\bibinfo {author} {\bibfnamefont {E.}~\bibnamefont
  {Fabiano}}, \bibinfo {author} {\bibfnamefont {S.}~\bibnamefont {\'Smiga}},
  \bibinfo {author} {\bibfnamefont {S.}~\bibnamefont {Giarrusso}}, \bibinfo
  {author} {\bibfnamefont {T.~J.}\ \bibnamefont {Daas}}, \bibinfo {author}
  {\bibfnamefont {F.}~\bibnamefont {Della~Sala}}, \bibinfo {author}
  {\bibfnamefont {I.}~\bibnamefont {Grabowski}}, \ and\ \bibinfo {author}
  {\bibfnamefont {P.}~\bibnamefont {Gori-Giorgi}},\ }\href {\doibase
  10.1021/acs.jctc.8b01037} {\bibfield  {journal} {\bibinfo  {journal} {Journal
  of Chemical Theory and Computation}\ }\textbf {\bibinfo {volume} {15}},\
  \bibinfo {pages} {1006} (\bibinfo {year} {2019})},\ \Eprint
  {http://arxiv.org/abs/https://doi.org/10.1021/acs.jctc.8b01037}
  {https://doi.org/10.1021/acs.jctc.8b01037} \BibitemShut {NoStop}%
\bibitem [{\citenamefont {Vuckovic}\ \emph {et~al.}(2018)\citenamefont
  {Vuckovic}, \citenamefont {Gori-Giorgi}, \citenamefont {Della~Sala},\ and\
  \citenamefont {Fabiano}}]{vuckovic2018restoring}%
  \BibitemOpen
  \bibfield  {author} {\bibinfo {author} {\bibfnamefont {S.}~\bibnamefont
  {Vuckovic}}, \bibinfo {author} {\bibfnamefont {P.}~\bibnamefont
  {Gori-Giorgi}}, \bibinfo {author} {\bibfnamefont {F.}~\bibnamefont
  {Della~Sala}}, \ and\ \bibinfo {author} {\bibfnamefont {E.}~\bibnamefont
  {Fabiano}},\ }\href {\doibase 10.1021/acs.jpclett.8b01054} {\bibfield
  {journal} {\bibinfo  {journal} {The Journal of Physical Chemistry Letters}\
  }\textbf {\bibinfo {volume} {9}},\ \bibinfo {pages} {3137} (\bibinfo {year}
  {2018})},\ \bibinfo {note} {pMID: 29787273},\ \Eprint
  {http://arxiv.org/abs/https://doi.org/10.1021/acs.jpclett.8b01054}
  {https://doi.org/10.1021/acs.jpclett.8b01054} \BibitemShut {NoStop}%
\bibitem [{\citenamefont {Kooi}\ and\ \citenamefont
  {Gori-Giorgi}(2018)}]{kooi2018local}%
  \BibitemOpen
  \bibfield  {author} {\bibinfo {author} {\bibfnamefont {D.~P.}\ \bibnamefont
  {Kooi}}\ and\ \bibinfo {author} {\bibfnamefont {P.}~\bibnamefont
  {Gori-Giorgi}},\ }\href@noop {} {\bibfield  {journal} {\bibinfo  {journal}
  {Theor. Chem. Acc.}\ }\textbf {\bibinfo {volume} {137}},\ \bibinfo {pages}
  {166} (\bibinfo {year} {2018})}\BibitemShut {NoStop}%
\bibitem [{\citenamefont {Seidl}\ \emph {et~al.}(2018)\citenamefont {Seidl},
  \citenamefont {Giarrusso}, \citenamefont {Vuckovic}, \citenamefont
  {Fabiano},\ and\ \citenamefont {Gori-Giorgi}}]{seidl2018communication}%
  \BibitemOpen
  \bibfield  {author} {\bibinfo {author} {\bibfnamefont {M.}~\bibnamefont
  {Seidl}}, \bibinfo {author} {\bibfnamefont {S.}~\bibnamefont {Giarrusso}},
  \bibinfo {author} {\bibfnamefont {S.}~\bibnamefont {Vuckovic}}, \bibinfo
  {author} {\bibfnamefont {E.}~\bibnamefont {Fabiano}}, \ and\ \bibinfo
  {author} {\bibfnamefont {P.}~\bibnamefont {Gori-Giorgi}},\ }\href@noop {}
  {\bibfield  {journal} {\bibinfo  {journal} {J. Chem. Phys.}\ }\textbf
  {\bibinfo {volume} {149}},\ \bibinfo {pages} {241101} (\bibinfo {year}
  {2018})}\BibitemShut {NoStop}%
\bibitem [{\citenamefont {Constantin}(2019)}]{constantin2019correlation}%
  \BibitemOpen
  \bibfield  {author} {\bibinfo {author} {\bibfnamefont {L.~A.}\ \bibnamefont
  {Constantin}},\ }\href {\doibase 10.1103/PhysRevB.99.085117} {\bibfield
  {journal} {\bibinfo  {journal} {Phys. Rev. B}\ }\textbf {\bibinfo {volume}
  {99}},\ \bibinfo {pages} {085117} (\bibinfo {year} {2019})}\BibitemShut
  {NoStop}%
\bibitem [{\citenamefont {G{\"o}rling}(1998)}]{gorling1998exact}%
  \BibitemOpen
  \bibfield  {author} {\bibinfo {author} {\bibfnamefont {A.}~\bibnamefont
  {G{\"o}rling}},\ }\href@noop {} {\bibfield  {journal} {\bibinfo  {journal}
  {Int. J. Quantum Chem.}\ }\textbf {\bibinfo {volume} {69}},\ \bibinfo {pages}
  {265} (\bibinfo {year} {1998})}\BibitemShut {NoStop}%
\bibitem [{\citenamefont {Seidl}\ \emph
  {et~al.}(2000{\natexlab{b}})\citenamefont {Seidl}, \citenamefont {Perdew},\
  and\ \citenamefont {Kurth}}]{seidl2000density}%
  \BibitemOpen
  \bibfield  {author} {\bibinfo {author} {\bibfnamefont {M.}~\bibnamefont
  {Seidl}}, \bibinfo {author} {\bibfnamefont {J.~P.}\ \bibnamefont {Perdew}}, \
  and\ \bibinfo {author} {\bibfnamefont {S.}~\bibnamefont {Kurth}},\
  }\href@noop {} {\bibfield  {journal} {\bibinfo  {journal} {Phys. Rev. A}\
  }\textbf {\bibinfo {volume} {62}},\ \bibinfo {pages} {012502} (\bibinfo
  {year} {2000}{\natexlab{b}})}\BibitemShut {NoStop}%
\bibitem [{\citenamefont {Gori-Giorgi}\ \emph
  {et~al.}(2009{\natexlab{a}})\citenamefont {Gori-Giorgi}, \citenamefont
  {Vignale},\ and\ \citenamefont {Seidl}}]{gori2009electronic}%
  \BibitemOpen
  \bibfield  {author} {\bibinfo {author} {\bibfnamefont {P.}~\bibnamefont
  {Gori-Giorgi}}, \bibinfo {author} {\bibfnamefont {G.}~\bibnamefont
  {Vignale}}, \ and\ \bibinfo {author} {\bibfnamefont {M.}~\bibnamefont
  {Seidl}},\ }\href@noop {} {\bibfield  {journal} {\bibinfo  {journal} {J.
  Chem. Theory Comput.}\ }\textbf {\bibinfo {volume} {5}},\ \bibinfo {pages}
  {743} (\bibinfo {year} {2009}{\natexlab{a}})}\BibitemShut {NoStop}%
\bibitem [{\citenamefont {Seidl}\ \emph {et~al.}(2007)\citenamefont {Seidl},
  \citenamefont {Gori-Giorgi},\ and\ \citenamefont {Savin}}]{seidl07}%
  \BibitemOpen
  \bibfield  {author} {\bibinfo {author} {\bibfnamefont {M.}~\bibnamefont
  {Seidl}}, \bibinfo {author} {\bibfnamefont {P.}~\bibnamefont {Gori-Giorgi}},
  \ and\ \bibinfo {author} {\bibfnamefont {A.}~\bibnamefont {Savin}},\ }\href
  {\doibase 10.1103/PhysRevA.75.042511} {\bibfield  {journal} {\bibinfo
  {journal} {Phys. Rev. A}\ }\textbf {\bibinfo {volume} {75}},\ \bibinfo
  {pages} {042511} (\bibinfo {year} {2007})}\BibitemShut {NoStop}%
\bibitem [{\citenamefont {Gori-Giorgi}\ \emph
  {et~al.}(2009{\natexlab{b}})\citenamefont {Gori-Giorgi}, \citenamefont
  {Vignale},\ and\ \citenamefont {Seidl}}]{gorigiorgi09}%
  \BibitemOpen
  \bibfield  {author} {\bibinfo {author} {\bibfnamefont {P.}~\bibnamefont
  {Gori-Giorgi}}, \bibinfo {author} {\bibfnamefont {G.}~\bibnamefont
  {Vignale}}, \ and\ \bibinfo {author} {\bibfnamefont {M.}~\bibnamefont
  {Seidl}},\ }\href {\doibase 10.1021/ct8005248} {\bibfield  {journal}
  {\bibinfo  {journal} {J. Chem. Theory Comput.}\ }\textbf {\bibinfo {volume}
  {5}},\ \bibinfo {pages} {743} (\bibinfo {year}
  {2009}{\natexlab{b}})}\BibitemShut {NoStop}%
\bibitem [{\citenamefont {Malet}\ \emph {et~al.}(2013)\citenamefont {Malet},
  \citenamefont {Mirtschink}, \citenamefont {Cremon}, \citenamefont {Reimann},\
  and\ \citenamefont {Gori-Giorgi}}]{MalMirCreReiGor-PRB-13}%
  \BibitemOpen
  \bibfield  {author} {\bibinfo {author} {\bibfnamefont {F.}~\bibnamefont
  {Malet}}, \bibinfo {author} {\bibfnamefont {A.}~\bibnamefont {Mirtschink}},
  \bibinfo {author} {\bibfnamefont {J.~C.}\ \bibnamefont {Cremon}}, \bibinfo
  {author} {\bibfnamefont {S.~M.}\ \bibnamefont {Reimann}}, \ and\ \bibinfo
  {author} {\bibfnamefont {P.}~\bibnamefont {Gori-Giorgi}},\ }\href {\doibase
  10.1103/PhysRevB.87.115146} {\bibfield  {journal} {\bibinfo  {journal} {Phys.
  Rev. B}\ }\textbf {\bibinfo {volume} {87}},\ \bibinfo {pages} {115146}
  (\bibinfo {year} {2013})}\BibitemShut {NoStop}%
\bibitem [{\citenamefont {Gori-Giorgi}\ and\ \citenamefont
  {Seidl}(2010)}]{gori2010density}%
  \BibitemOpen
  \bibfield  {author} {\bibinfo {author} {\bibfnamefont {P.}~\bibnamefont
  {Gori-Giorgi}}\ and\ \bibinfo {author} {\bibfnamefont {M.}~\bibnamefont
  {Seidl}},\ }\href@noop {} {\bibfield  {journal} {\bibinfo  {journal} {Phys.
  Chem. Chem. Phys.}\ }\textbf {\bibinfo {volume} {12}},\ \bibinfo {pages}
  {14405} (\bibinfo {year} {2010})}\BibitemShut {NoStop}%
\bibitem [{\citenamefont {Fabiano}\ \emph {et~al.}(2016)\citenamefont
  {Fabiano}, \citenamefont {Gori-Giorgi}, \citenamefont {Seidl},\ and\
  \citenamefont {Della~Sala}}]{fabiano2016interaction}%
  \BibitemOpen
  \bibfield  {author} {\bibinfo {author} {\bibfnamefont {E.}~\bibnamefont
  {Fabiano}}, \bibinfo {author} {\bibfnamefont {P.}~\bibnamefont
  {Gori-Giorgi}}, \bibinfo {author} {\bibfnamefont {M.}~\bibnamefont {Seidl}},
  \ and\ \bibinfo {author} {\bibfnamefont {F.}~\bibnamefont {Della~Sala}},\
  }\href@noop {} {\bibfield  {journal} {\bibinfo  {journal} {J. Chem. Theory
  Comput.}\ }\textbf {\bibinfo {volume} {12}},\ \bibinfo {pages} {4885}
  (\bibinfo {year} {2016})}\BibitemShut {NoStop}%
\bibitem [{\citenamefont {Seidl}\ \emph
  {et~al.}(1999{\natexlab{a}})\citenamefont {Seidl}, \citenamefont {Perdew},\
  and\ \citenamefont {Levy}}]{seidl1999strictly}%
  \BibitemOpen
  \bibfield  {author} {\bibinfo {author} {\bibfnamefont {M.}~\bibnamefont
  {Seidl}}, \bibinfo {author} {\bibfnamefont {J.~P.}\ \bibnamefont {Perdew}}, \
  and\ \bibinfo {author} {\bibfnamefont {M.}~\bibnamefont {Levy}},\ }\href@noop
  {} {\bibfield  {journal} {\bibinfo  {journal} {Phys. Rev. A}\ }\textbf
  {\bibinfo {volume} {59}},\ \bibinfo {pages} {51} (\bibinfo {year}
  {1999}{\natexlab{a}})}\BibitemShut {NoStop}%
\bibitem [{\citenamefont {Seidl}\ \emph {et~al.}(2016)\citenamefont {Seidl},
  \citenamefont {Vuckovic},\ and\ \citenamefont
  {Gori-Giorgi}}]{seidl2016challenging}%
  \BibitemOpen
  \bibfield  {author} {\bibinfo {author} {\bibfnamefont {M.}~\bibnamefont
  {Seidl}}, \bibinfo {author} {\bibfnamefont {S.}~\bibnamefont {Vuckovic}}, \
  and\ \bibinfo {author} {\bibfnamefont {P.}~\bibnamefont {Gori-Giorgi}},\
  }\href@noop {} {\bibfield  {journal} {\bibinfo  {journal} {Mol. Phys.}\
  }\textbf {\bibinfo {volume} {114}},\ \bibinfo {pages} {1076} (\bibinfo {year}
  {2016})}\BibitemShut {NoStop}%
\bibitem [{\citenamefont {Giarrusso}\ \emph {et~al.}(2018)\citenamefont
  {Giarrusso}, \citenamefont {Gori-Giorgi}, \citenamefont {Della~Sala},\ and\
  \citenamefont {Fabiano}}]{giarrusso2018assessment}%
  \BibitemOpen
  \bibfield  {author} {\bibinfo {author} {\bibfnamefont {S.}~\bibnamefont
  {Giarrusso}}, \bibinfo {author} {\bibfnamefont {P.}~\bibnamefont
  {Gori-Giorgi}}, \bibinfo {author} {\bibfnamefont {F.}~\bibnamefont
  {Della~Sala}}, \ and\ \bibinfo {author} {\bibfnamefont {E.}~\bibnamefont
  {Fabiano}},\ }\href@noop {} {\bibfield  {journal} {\bibinfo  {journal} {J.
  Chem. Phys.}\ }\textbf {\bibinfo {volume} {148}},\ \bibinfo {pages} {134106}
  (\bibinfo {year} {2018})}\BibitemShut {NoStop}%
\bibitem [{\citenamefont {Mirtschink}\ \emph {et~al.}(2012)\citenamefont
  {Mirtschink}, \citenamefont {Seidl},\ and\ \citenamefont
  {Gori-Giorgi}}]{mirtschink2012energy}%
  \BibitemOpen
  \bibfield  {author} {\bibinfo {author} {\bibfnamefont {A.}~\bibnamefont
  {Mirtschink}}, \bibinfo {author} {\bibfnamefont {M.}~\bibnamefont {Seidl}}, \
  and\ \bibinfo {author} {\bibfnamefont {P.}~\bibnamefont {Gori-Giorgi}},\
  }\href@noop {} {\bibfield  {journal} {\bibinfo  {journal} {J. Chem. Theory
  Comput.}\ }\textbf {\bibinfo {volume} {8}},\ \bibinfo {pages} {3097}
  (\bibinfo {year} {2012})}\BibitemShut {NoStop}%
\bibitem [{\citenamefont {Seidl}\ \emph
  {et~al.}(1999{\natexlab{b}})\citenamefont {Seidl}, \citenamefont {Perdew},\
  and\ \citenamefont {Levy}}]{SeiPerLev-PRA-99}%
  \BibitemOpen
  \bibfield  {author} {\bibinfo {author} {\bibfnamefont {M.}~\bibnamefont
  {Seidl}}, \bibinfo {author} {\bibfnamefont {J.~P.}\ \bibnamefont {Perdew}}, \
  and\ \bibinfo {author} {\bibfnamefont {M.}~\bibnamefont {Levy}},\ }\href@noop
  {} {\bibfield  {journal} {\bibinfo  {journal} {Phys. Rev. A}\ }\textbf
  {\bibinfo {volume} {{59}}},\ \bibinfo {pages} {51} (\bibinfo {year}
  {1999}{\natexlab{b}})}\BibitemShut {NoStop}%
\bibitem [{\citenamefont {Daas}\ \emph {et~al.}(2021)\citenamefont {Daas},
  \citenamefont {Fabiano}, \citenamefont {Della~Sala}, \citenamefont
  {Gori-Giorgi},\ and\ \citenamefont {Vuckovic}}]{NCIISI}%
  \BibitemOpen
  \bibfield  {author} {\bibinfo {author} {\bibfnamefont {T.~J.}\ \bibnamefont
  {Daas}}, \bibinfo {author} {\bibfnamefont {E.}~\bibnamefont {Fabiano}},
  \bibinfo {author} {\bibfnamefont {F.}~\bibnamefont {Della~Sala}}, \bibinfo
  {author} {\bibfnamefont {P.}~\bibnamefont {Gori-Giorgi}}, \ and\ \bibinfo
  {author} {\bibfnamefont {S.}~\bibnamefont {Vuckovic}},\ }\href {\doibase
  10.1021/acs.jpclett.1c01157} {\bibfield  {journal} {\bibinfo  {journal} {The
  Journal of Physical Chemistry Letters}\ }\textbf {\bibinfo {volume} {12}},\
  \bibinfo {pages} {4867} (\bibinfo {year} {2021})},\ \bibinfo {note} {pMID:
  34003655},\ \Eprint
  {http://arxiv.org/abs/https://doi.org/10.1021/acs.jpclett.1c01157}
  {https://doi.org/10.1021/acs.jpclett.1c01157} \BibitemShut {NoStop}%
\bibitem [{\citenamefont {\'Smiga}\ \emph {et~al.}(2022)\citenamefont
  {\'Smiga}, \citenamefont {Della~Sala}, \citenamefont {Gori-Giorgi},\ and\
  \citenamefont {Fabiano}}]{SCFISI}%
  \BibitemOpen
  \bibfield  {author} {\bibinfo {author} {\bibfnamefont {S.}~\bibnamefont
  {\'Smiga}}, \bibinfo {author} {\bibfnamefont {F.}~\bibnamefont {Della~Sala}},
  \bibinfo {author} {\bibfnamefont {P.}~\bibnamefont {Gori-Giorgi}}, \ and\
  \bibinfo {author} {\bibfnamefont {E.}~\bibnamefont {Fabiano}},\ }\href
  {\doibase 10.1021/acs.jctc.2c00352} {\bibfield  {journal} {\bibinfo
  {journal} {Journal of Chemical Theory and Computation}\ }\textbf {\bibinfo
  {volume} {18}},\ \bibinfo {pages} {5936} (\bibinfo {year} {2022})},\ \bibinfo
  {note} {pMID: 36094908},\ \Eprint
  {http://arxiv.org/abs/https://doi.org/10.1021/acs.jctc.2c00352}
  {https://doi.org/10.1021/acs.jctc.2c00352} \BibitemShut {NoStop}%
\bibitem [{\citenamefont {Vuckovic}\ \emph
  {et~al.}(2016{\natexlab{b}})\citenamefont {Vuckovic}, \citenamefont {Irons},
  \citenamefont {Savin}, \citenamefont {Teale},\ and\ \citenamefont
  {Gori-Giorgi}}]{Gauge3}%
  \BibitemOpen
  \bibfield  {author} {\bibinfo {author} {\bibfnamefont {S.}~\bibnamefont
  {Vuckovic}}, \bibinfo {author} {\bibfnamefont {T.~J.~P.}\ \bibnamefont
  {Irons}}, \bibinfo {author} {\bibfnamefont {A.}~\bibnamefont {Savin}},
  \bibinfo {author} {\bibfnamefont {A.~M.}\ \bibnamefont {Teale}}, \ and\
  \bibinfo {author} {\bibfnamefont {P.}~\bibnamefont {Gori-Giorgi}},\ }\href
  {\doibase 10.1021/acs.jctc.6b00177} {\bibfield  {journal} {\bibinfo
  {journal} {Journal of Chemical Theory and Computation}\ }\textbf {\bibinfo
  {volume} {12}},\ \bibinfo {pages} {2598} (\bibinfo {year}
  {2016}{\natexlab{b}})},\ \bibinfo {note} {pMID: 27116427},\ \Eprint
  {http://arxiv.org/abs/https://doi.org/10.1021/acs.jctc.6b00177}
  {https://doi.org/10.1021/acs.jctc.6b00177} \BibitemShut {NoStop}%
\bibitem [{\citenamefont {{\'S}miga}\ \emph {et~al.}(2017)\citenamefont
  {{\'S}miga}, \citenamefont {Fabiano}, \citenamefont {Constantin},\ and\
  \citenamefont {Della~Sala}}]{smiga2017laplacian}%
  \BibitemOpen
  \bibfield  {author} {\bibinfo {author} {\bibfnamefont {S.}~\bibnamefont
  {{\'S}miga}}, \bibinfo {author} {\bibfnamefont {E.}~\bibnamefont {Fabiano}},
  \bibinfo {author} {\bibfnamefont {L.~A.}\ \bibnamefont {Constantin}}, \ and\
  \bibinfo {author} {\bibfnamefont {F.}~\bibnamefont {Della~Sala}},\
  }\href@noop {} {\bibfield  {journal} {\bibinfo  {journal} {The Journal of
  chemical physics}\ }\textbf {\bibinfo {volume} {146}},\ \bibinfo {pages}
  {064105} (\bibinfo {year} {2017})}\BibitemShut {NoStop}%
\bibitem [{\citenamefont {Perdew}\ \emph
  {et~al.}(1996{\natexlab{b}})\citenamefont {Perdew}, \citenamefont {Burke},\
  and\ \citenamefont {Wang}}]{perdew1996generalized}%
  \BibitemOpen
  \bibfield  {author} {\bibinfo {author} {\bibfnamefont {J.~P.}\ \bibnamefont
  {Perdew}}, \bibinfo {author} {\bibfnamefont {K.}~\bibnamefont {Burke}}, \
  and\ \bibinfo {author} {\bibfnamefont {Y.}~\bibnamefont {Wang}},\ }\href@noop
  {} {\bibfield  {journal} {\bibinfo  {journal} {Phys. Rev. B}\ }\textbf
  {\bibinfo {volume} {54}},\ \bibinfo {pages} {16533} (\bibinfo {year}
  {1996}{\natexlab{b}})}\BibitemShut {NoStop}%
\bibitem [{\citenamefont {\'Smiga}\ and\ \citenamefont
  {Constantin}(2020)}]{LUCISI}%
  \BibitemOpen
  \bibfield  {author} {\bibinfo {author} {\bibfnamefont {S.}~\bibnamefont
  {\'Smiga}}\ and\ \bibinfo {author} {\bibfnamefont {L.~A.}\ \bibnamefont
  {Constantin}},\ }\href {\doibase 10.1021/acs.jctc.0c00328} {\bibfield
  {journal} {\bibinfo  {journal} {J. Chem. Theory Comput.}\ }\textbf {\bibinfo
  {volume} {16}},\ \bibinfo {pages} {4983} (\bibinfo {year}
  {2020})}\BibitemShut {NoStop}%
\bibitem [{\citenamefont {Perdew}\ \emph {et~al.}(2004)\citenamefont {Perdew},
  \citenamefont {Tao}, \citenamefont {Staroverov},\ and\ \citenamefont
  {Scuseria}}]{perdew2004meta}%
  \BibitemOpen
  \bibfield  {author} {\bibinfo {author} {\bibfnamefont {J.~P.}\ \bibnamefont
  {Perdew}}, \bibinfo {author} {\bibfnamefont {J.}~\bibnamefont {Tao}},
  \bibinfo {author} {\bibfnamefont {V.~N.}\ \bibnamefont {Staroverov}}, \ and\
  \bibinfo {author} {\bibfnamefont {G.~E.}\ \bibnamefont {Scuseria}},\
  }\href@noop {} {\bibfield  {journal} {\bibinfo  {journal} {J. Chem. Phys.}\
  }\textbf {\bibinfo {volume} {120}},\ \bibinfo {pages} {6898} (\bibinfo {year}
  {2004})}\BibitemShut {NoStop}%
\bibitem [{\citenamefont {von Weizs{\"a}cker}(1935)}]{weizsacker1935theorie}%
  \BibitemOpen
  \bibfield  {author} {\bibinfo {author} {\bibfnamefont {C.~F.}\ \bibnamefont
  {von Weizs{\"a}cker}},\ }\href@noop {} {\bibfield  {journal} {\bibinfo
  {journal} {Zeitschrift f{\"u}r Physik A Hadrons and Nuclei}\ }\textbf
  {\bibinfo {volume} {96}},\ \bibinfo {pages} {431} (\bibinfo {year}
  {1935})}\BibitemShut {NoStop}%
\bibitem [{\citenamefont {Della~Sala}\ \emph {et~al.}(2015)\citenamefont
  {Della~Sala}, \citenamefont {Fabiano},\ and\ \citenamefont
  {Constantin}}]{della2015kohn}%
  \BibitemOpen
  \bibfield  {author} {\bibinfo {author} {\bibfnamefont {F.}~\bibnamefont
  {Della~Sala}}, \bibinfo {author} {\bibfnamefont {E.}~\bibnamefont {Fabiano}},
  \ and\ \bibinfo {author} {\bibfnamefont {L.~A.}\ \bibnamefont {Constantin}},\
  }\href@noop {} {\bibfield  {journal} {\bibinfo  {journal} {Phys. Rev. B}\
  }\textbf {\bibinfo {volume} {91}},\ \bibinfo {pages} {035126} (\bibinfo
  {year} {2015})}\BibitemShut {NoStop}%
\bibitem [{\citenamefont {Perdew}\ and\ \citenamefont
  {Wang}(1992)}]{perdew1992accurate}%
  \BibitemOpen
  \bibfield  {author} {\bibinfo {author} {\bibfnamefont {J.~P.}\ \bibnamefont
  {Perdew}}\ and\ \bibinfo {author} {\bibfnamefont {Y.}~\bibnamefont {Wang}},\
  }\href@noop {} {\bibfield  {journal} {\bibinfo  {journal} {Phys. Rev. B}\
  }\textbf {\bibinfo {volume} {45}},\ \bibinfo {pages} {13244} (\bibinfo {year}
  {1992})}\BibitemShut {NoStop}%
\bibitem [{\citenamefont {Lee}\ \emph {et~al.}(1988)\citenamefont {Lee},
  \citenamefont {Yang},\ and\ \citenamefont {Parr}}]{lee1988development}%
  \BibitemOpen
  \bibfield  {author} {\bibinfo {author} {\bibfnamefont {C.}~\bibnamefont
  {Lee}}, \bibinfo {author} {\bibfnamefont {W.}~\bibnamefont {Yang}}, \ and\
  \bibinfo {author} {\bibfnamefont {R.~G.}\ \bibnamefont {Parr}},\ }\href@noop
  {} {\bibfield  {journal} {\bibinfo  {journal} {Phys. Rev. B}\ }\textbf
  {\bibinfo {volume} {37}},\ \bibinfo {pages} {785} (\bibinfo {year}
  {1988})}\BibitemShut {NoStop}%
\bibitem [{\citenamefont {Fabiano}\ \emph {et~al.}(2015)\citenamefont
  {Fabiano}, \citenamefont {Constantin}, \citenamefont {Terentjevs},
  \citenamefont {Della~Sala},\ and\ \citenamefont
  {Cortona}}]{fabiano2015assessment}%
  \BibitemOpen
  \bibfield  {author} {\bibinfo {author} {\bibfnamefont {E.}~\bibnamefont
  {Fabiano}}, \bibinfo {author} {\bibfnamefont {L.}~\bibnamefont {Constantin}},
  \bibinfo {author} {\bibfnamefont {A.}~\bibnamefont {Terentjevs}}, \bibinfo
  {author} {\bibfnamefont {F.}~\bibnamefont {Della~Sala}}, \ and\ \bibinfo
  {author} {\bibfnamefont {P.}~\bibnamefont {Cortona}},\ }\href@noop {}
  {\bibfield  {journal} {\bibinfo  {journal} {Theor. Chem. Acc.}\ }\textbf
  {\bibinfo {volume} {134}},\ \bibinfo {pages} {139} (\bibinfo {year}
  {2015})}\BibitemShut {NoStop}%
\bibitem [{\citenamefont {Tognetti}\ \emph {et~al.}(2008)\citenamefont
  {Tognetti}, \citenamefont {Cortona},\ and\ \citenamefont
  {Adamo}}]{tognetti2008new}%
  \BibitemOpen
  \bibfield  {author} {\bibinfo {author} {\bibfnamefont {V.}~\bibnamefont
  {Tognetti}}, \bibinfo {author} {\bibfnamefont {P.}~\bibnamefont {Cortona}}, \
  and\ \bibinfo {author} {\bibfnamefont {C.}~\bibnamefont {Adamo}},\
  }\href@noop {} {\bibfield  {journal} {\bibinfo  {journal} {The Journal of
  chemical physics}\ }\textbf {\bibinfo {volume} {128}},\ \bibinfo {pages}
  {034101} (\bibinfo {year} {2008})}\BibitemShut {NoStop}%
\bibitem [{\citenamefont {Ragot}\ and\ \citenamefont
  {Cortona}(2004)}]{ragot2004correlation}%
  \BibitemOpen
  \bibfield  {author} {\bibinfo {author} {\bibfnamefont {S.}~\bibnamefont
  {Ragot}}\ and\ \bibinfo {author} {\bibfnamefont {P.}~\bibnamefont
  {Cortona}},\ }\href@noop {} {\bibfield  {journal} {\bibinfo  {journal} {The
  Journal of chemical physics}\ }\textbf {\bibinfo {volume} {121}},\ \bibinfo
  {pages} {7671} (\bibinfo {year} {2004})}\BibitemShut {NoStop}%
\bibitem [{\citenamefont {Daas}\ \emph
  {et~al.}(2022{\natexlab{c}})\citenamefont {Daas}, \citenamefont {Kooi},
  \citenamefont {Benyahia}, \citenamefont {Seidl},\ and\ \citenamefont
  {Gori-Giorgi}}]{Daas2022large}%
  \BibitemOpen
  \bibfield  {author} {\bibinfo {author} {\bibfnamefont {T.~J.}\ \bibnamefont
  {Daas}}, \bibinfo {author} {\bibfnamefont {D.~P.}\ \bibnamefont {Kooi}},
  \bibinfo {author} {\bibfnamefont {T.}~\bibnamefont {Benyahia}}, \bibinfo
  {author} {\bibfnamefont {M.}~\bibnamefont {Seidl}}, \ and\ \bibinfo {author}
  {\bibfnamefont {P.}~\bibnamefont {Gori-Giorgi}},\ }\href@noop {} {\enquote
  {\bibinfo {title} {Large-z atoms in the strong-interaction limit of dft:
  Implications for gradient expansions and for the lieb-oxford bound},}\ }
  (\bibinfo {year} {2022}{\natexlab{c}}),\ \Eprint
  {http://arxiv.org/abs/arXiv:2211.07512} {arXiv:2211.07512} \BibitemShut
  {NoStop}%
\bibitem [{\citenamefont {Perdew}\ \emph {et~al.}(2014)\citenamefont {Perdew},
  \citenamefont {Ruzsinszky}, \citenamefont {Sun},\ and\ \citenamefont
  {Burke}}]{perdew2014gedanken}%
  \BibitemOpen
  \bibfield  {author} {\bibinfo {author} {\bibfnamefont {J.~P.}\ \bibnamefont
  {Perdew}}, \bibinfo {author} {\bibfnamefont {A.}~\bibnamefont {Ruzsinszky}},
  \bibinfo {author} {\bibfnamefont {J.}~\bibnamefont {Sun}}, \ and\ \bibinfo
  {author} {\bibfnamefont {K.}~\bibnamefont {Burke}},\ }\href@noop {}
  {\bibfield  {journal} {\bibinfo  {journal} {J. Chem. Phys.}\ }\textbf
  {\bibinfo {volume} {140}},\ \bibinfo {pages} {18A533} (\bibinfo {year}
  {2014})}\BibitemShut {NoStop}%
\bibitem [{\citenamefont {Taut}(1993)}]{Tau-PRA-93}%
  \BibitemOpen
  \bibfield  {author} {\bibinfo {author} {\bibfnamefont {M.}~\bibnamefont
  {Taut}},\ }\href@noop {} {\bibfield  {journal} {\bibinfo  {journal} {Phys.
  Rev. A}\ }\textbf {\bibinfo {volume} {{48}}},\ \bibinfo {pages} {3561}
  (\bibinfo {year} {1993})}\BibitemShut {NoStop}%
\bibitem [{\citenamefont {Davidson}\ \emph {et~al.}(1991)\citenamefont
  {Davidson}, \citenamefont {Hagstrom}, \citenamefont {Chakravorty},
  \citenamefont {Umar},\ and\ \citenamefont {Fischer}}]{davidson1991ground}%
  \BibitemOpen
  \bibfield  {author} {\bibinfo {author} {\bibfnamefont {E.~R.}\ \bibnamefont
  {Davidson}}, \bibinfo {author} {\bibfnamefont {S.~A.}\ \bibnamefont
  {Hagstrom}}, \bibinfo {author} {\bibfnamefont {S.~J.}\ \bibnamefont
  {Chakravorty}}, \bibinfo {author} {\bibfnamefont {V.~M.}\ \bibnamefont
  {Umar}}, \ and\ \bibinfo {author} {\bibfnamefont {C.~F.}\ \bibnamefont
  {Fischer}},\ }\href@noop {} {\bibfield  {journal} {\bibinfo  {journal} {Phys.
  Rev. A}\ }\textbf {\bibinfo {volume} {44}},\ \bibinfo {pages} {7071}
  (\bibinfo {year} {1991})}\BibitemShut {NoStop}%
\bibitem [{\citenamefont {Chakravorty}\ \emph {et~al.}(1993)\citenamefont
  {Chakravorty}, \citenamefont {Gwaltney}, \citenamefont {Davidson},
  \citenamefont {Parpia},\ and\ \citenamefont
  {Fischer}}]{chakravorty1993ground}%
  \BibitemOpen
  \bibfield  {author} {\bibinfo {author} {\bibfnamefont {S.~J.}\ \bibnamefont
  {Chakravorty}}, \bibinfo {author} {\bibfnamefont {S.~R.}\ \bibnamefont
  {Gwaltney}}, \bibinfo {author} {\bibfnamefont {E.~R.}\ \bibnamefont
  {Davidson}}, \bibinfo {author} {\bibfnamefont {F.~A.}\ \bibnamefont
  {Parpia}}, \ and\ \bibinfo {author} {\bibfnamefont {C.~F.}\ \bibnamefont
  {Fischer}},\ }\href@noop {} {\bibfield  {journal} {\bibinfo  {journal} {Phys.
  Rev. A}\ }\textbf {\bibinfo {volume} {47}},\ \bibinfo {pages} {3649}
  (\bibinfo {year} {1993})}\BibitemShut {NoStop}%
\bibitem [{\citenamefont {Clementi}\ and\ \citenamefont
  {Corongiu}(1997)}]{clementi1997note}%
  \BibitemOpen
  \bibfield  {author} {\bibinfo {author} {\bibfnamefont {E.}~\bibnamefont
  {Clementi}}\ and\ \bibinfo {author} {\bibfnamefont {G.}~\bibnamefont
  {Corongiu}},\ }\href@noop {} {\bibfield  {journal} {\bibinfo  {journal} {Int.
  J. Quantum Chem.}\ }\textbf {\bibinfo {volume} {62}},\ \bibinfo {pages} {571}
  (\bibinfo {year} {1997})}\BibitemShut {NoStop}%
\bibitem [{\citenamefont {Dunning}(1989)}]{dunning:1989:bas}%
  \BibitemOpen
  \bibfield  {author} {\bibinfo {author} {\bibfnamefont {T.~H.}\ \bibnamefont
  {Dunning}},\ }\href {\doibase 10.1063/1.456153} {\bibfield  {journal}
  {\bibinfo  {journal} {J. Chem. Phys.}\ }\textbf {\bibinfo {volume} {90}},\
  \bibinfo {pages} {1007} (\bibinfo {year} {1989})}\BibitemShut {NoStop}%
\bibitem [{\citenamefont {Clementi}\ and\ \citenamefont
  {Roetti}(1974)}]{clementi1974roothaan}%
  \BibitemOpen
  \bibfield  {author} {\bibinfo {author} {\bibfnamefont {E.}~\bibnamefont
  {Clementi}}\ and\ \bibinfo {author} {\bibfnamefont {C.}~\bibnamefont
  {Roetti}},\ }\href@noop {} {\bibfield  {journal} {\bibinfo  {journal} {Atomic
  data and nuclear data tables}\ }\textbf {\bibinfo {volume} {14}},\ \bibinfo
  {pages} {177} (\bibinfo {year} {1974})}\BibitemShut {NoStop}%
\bibitem [{\citenamefont {Stanton}\ \emph {et~al.}(2007)\citenamefont
  {Stanton}, \citenamefont {Gauss}, \citenamefont {Watts}, \citenamefont
  {Nooijen}, \citenamefont {Oliphant}, \citenamefont {Perera}, \citenamefont
  {Szalay}, \citenamefont {Lauderdale}, \citenamefont {Kucharski},
  \citenamefont {Gwaltney}, \citenamefont {Beck}, \citenamefont
  {Balkov{\'{a}}}, \citenamefont {Bernholdt}, \citenamefont {Baeck},
  \citenamefont {Rozyczko}, \citenamefont {Sekino}, \citenamefont {Hober},\
  and\ \citenamefont {{R. J. Bartlett Integral packages included are VMOL (J.
  Almlöf and P.R. Taylor); VPROPS (P. Taylor) ABACUS; (T. Helgaker, H.J. Aa.
  Jensen, P. J{\"{o}}rgensen, J. Olsen, and P.R. Taylor)}}}]{acesII}%
  \BibitemOpen
  \bibfield  {author} {\bibinfo {author} {\bibfnamefont {J.~F.}\ \bibnamefont
  {Stanton}}, \bibinfo {author} {\bibfnamefont {J.}~\bibnamefont {Gauss}},
  \bibinfo {author} {\bibfnamefont {J.~D.}\ \bibnamefont {Watts}}, \bibinfo
  {author} {\bibfnamefont {M.}~\bibnamefont {Nooijen}}, \bibinfo {author}
  {\bibfnamefont {N.}~\bibnamefont {Oliphant}}, \bibinfo {author}
  {\bibfnamefont {S.~A.}\ \bibnamefont {Perera}}, \bibinfo {author}
  {\bibfnamefont {P.}~\bibnamefont {Szalay}}, \bibinfo {author} {\bibfnamefont
  {W.~J.}\ \bibnamefont {Lauderdale}}, \bibinfo {author} {\bibfnamefont
  {S.}~\bibnamefont {Kucharski}}, \bibinfo {author} {\bibfnamefont
  {S.}~\bibnamefont {Gwaltney}}, \bibinfo {author} {\bibfnamefont
  {S.}~\bibnamefont {Beck}}, \bibinfo {author} {\bibfnamefont {A.}~\bibnamefont
  {Balkov{\'{a}}}}, \bibinfo {author} {\bibfnamefont {D.~E.}\ \bibnamefont
  {Bernholdt}}, \bibinfo {author} {\bibfnamefont {K.~K.}\ \bibnamefont
  {Baeck}}, \bibinfo {author} {\bibfnamefont {P.}~\bibnamefont {Rozyczko}},
  \bibinfo {author} {\bibfnamefont {H.}~\bibnamefont {Sekino}}, \bibinfo
  {author} {\bibfnamefont {C.}~\bibnamefont {Hober}}, \ and\ \bibinfo {author}
  {\bibnamefont {{R. J. Bartlett Integral packages included are VMOL (J.
  Almlöf and P.R. Taylor); VPROPS (P. Taylor) ABACUS; (T. Helgaker, H.J. Aa.
  Jensen, P. J{\"{o}}rgensen, J. Olsen, and P.R. Taylor)}}},\ }\href@noop {}
  {\emph {\bibinfo {title} {ACES II}}}\ (\bibinfo  {publisher} {Quantum Theory
  Project},\ \bibinfo {address} {Gainesville, Florida},\ \bibinfo {year}
  {2007})\BibitemShut {NoStop}%
\bibitem [{\citenamefont {Kim}\ \emph {et~al.}(2013)\citenamefont {Kim},
  \citenamefont {Sim},\ and\ \citenamefont {Burke}}]{BurkeFD_DD}%
  \BibitemOpen
  \bibfield  {author} {\bibinfo {author} {\bibfnamefont {M.-C.}\ \bibnamefont
  {Kim}}, \bibinfo {author} {\bibfnamefont {E.}~\bibnamefont {Sim}}, \ and\
  \bibinfo {author} {\bibfnamefont {K.}~\bibnamefont {Burke}},\ }\href
  {\doibase 10.1103/PhysRevLett.111.073003} {\bibfield  {journal} {\bibinfo
  {journal} {Phys. Rev. Lett.}\ }\textbf {\bibinfo {volume} {111}},\ \bibinfo
  {pages} {073003} (\bibinfo {year} {2013})}\BibitemShut {NoStop}%
\bibitem [{\citenamefont {Hernandez}\ \emph {et~al.}(2023)\citenamefont
  {Hernandez}, \citenamefont {Rettig},\ and\ \citenamefont
  {Head-Gordon}}]{hernandez2023new}%
  \BibitemOpen
  \bibfield  {author} {\bibinfo {author} {\bibfnamefont {D.~J.}\ \bibnamefont
  {Hernandez}}, \bibinfo {author} {\bibfnamefont {A.}~\bibnamefont {Rettig}}, \
  and\ \bibinfo {author} {\bibfnamefont {M.}~\bibnamefont {Head-Gordon}},\
  }\href@noop {} {\bibfield  {journal} {\bibinfo  {journal} {arXiv preprint
  arXiv:2306.15016}\ } (\bibinfo {year} {2023})}\BibitemShut {NoStop}%
\bibitem [{\citenamefont {Jana}\ \emph
  {et~al.}(2020{\natexlab{b}})\citenamefont {Jana}, \citenamefont {Patra},
  \citenamefont {\ifmmode~\acute{S}\else \'{S}\fi{}miga}, \citenamefont
  {Constantin},\ and\ \citenamefont {Samal}}]{PRBSub}%
  \BibitemOpen
  \bibfield  {author} {\bibinfo {author} {\bibfnamefont {S.}~\bibnamefont
  {Jana}}, \bibinfo {author} {\bibfnamefont {B.}~\bibnamefont {Patra}},
  \bibinfo {author} {\bibfnamefont {S.}~\bibnamefont {\ifmmode~\acute{S}\else
  \'{S}\fi{}miga}}, \bibinfo {author} {\bibfnamefont {L.~A.}\ \bibnamefont
  {Constantin}}, \ and\ \bibinfo {author} {\bibfnamefont {P.}~\bibnamefont
  {Samal}},\ }\href {\doibase 10.1103/PhysRevB.102.155107} {\bibfield
  {journal} {\bibinfo  {journal} {Phys. Rev. B}\ }\textbf {\bibinfo {volume}
  {102}},\ \bibinfo {pages} {155107} (\bibinfo {year}
  {2020}{\natexlab{b}})}\BibitemShut {NoStop}%
\bibitem [{\citenamefont {Cohen}\ \emph {et~al.}(2012)\citenamefont {Cohen},
  \citenamefont {Mori-Sánchez},\ and\ \citenamefont {Yang}}]{cohen12}%
  \BibitemOpen
  \bibfield  {author} {\bibinfo {author} {\bibfnamefont {A.~J.}\ \bibnamefont
  {Cohen}}, \bibinfo {author} {\bibfnamefont {P.}~\bibnamefont
  {Mori-Sánchez}}, \ and\ \bibinfo {author} {\bibfnamefont {W.}~\bibnamefont
  {Yang}},\ }\href {\doibase 10.1021/cr200107z} {\bibfield  {journal} {\bibinfo
   {journal} {Chem. Rev.}\ }\textbf {\bibinfo {volume} {112}},\ \bibinfo
  {pages} {289–320} (\bibinfo {year} {2012})},\ \Eprint
  {http://arxiv.org/abs/https://doi.org/10.1021/cr200107z}
  {https://doi.org/10.1021/cr200107z} \BibitemShut {NoStop}%
\bibitem [{\citenamefont {Kirkpatrick}\ \emph {et~al.}(2021)\citenamefont
  {Kirkpatrick}, \citenamefont {McMorrow}, \citenamefont {Turban},
  \citenamefont {Gaunt}, \citenamefont {Spencer}, \citenamefont {Matthews},
  \citenamefont {Obika}, \citenamefont {Thiry}, \citenamefont {Fortunato},
  \citenamefont {Pfau}, \citenamefont {Castellanos}, \citenamefont {Petersen},
  \citenamefont {Nelson}, \citenamefont {Kohli}, \citenamefont {Mori-Sánchez},
  \citenamefont {Hassabis},\ and\ \citenamefont {Cohen}}]{science21}%
  \BibitemOpen
  \bibfield  {author} {\bibinfo {author} {\bibfnamefont {J.}~\bibnamefont
  {Kirkpatrick}}, \bibinfo {author} {\bibfnamefont {B.}~\bibnamefont
  {McMorrow}}, \bibinfo {author} {\bibfnamefont {D.~H.~P.}\ \bibnamefont
  {Turban}}, \bibinfo {author} {\bibfnamefont {A.~L.}\ \bibnamefont {Gaunt}},
  \bibinfo {author} {\bibfnamefont {J.~S.}\ \bibnamefont {Spencer}}, \bibinfo
  {author} {\bibfnamefont {A.~G. D.~G.}\ \bibnamefont {Matthews}}, \bibinfo
  {author} {\bibfnamefont {A.}~\bibnamefont {Obika}}, \bibinfo {author}
  {\bibfnamefont {L.}~\bibnamefont {Thiry}}, \bibinfo {author} {\bibfnamefont
  {M.}~\bibnamefont {Fortunato}}, \bibinfo {author} {\bibfnamefont
  {D.}~\bibnamefont {Pfau}}, \bibinfo {author} {\bibfnamefont {L.~R.}\
  \bibnamefont {Castellanos}}, \bibinfo {author} {\bibfnamefont
  {S.}~\bibnamefont {Petersen}}, \bibinfo {author} {\bibfnamefont {A.~W.~R.}\
  \bibnamefont {Nelson}}, \bibinfo {author} {\bibfnamefont {P.}~\bibnamefont
  {Kohli}}, \bibinfo {author} {\bibfnamefont {P.}~\bibnamefont
  {Mori-Sánchez}}, \bibinfo {author} {\bibfnamefont {D.}~\bibnamefont
  {Hassabis}}, \ and\ \bibinfo {author} {\bibfnamefont {A.~J.}\ \bibnamefont
  {Cohen}},\ }\href {\doibase 10.1126/science.abj6511} {\bibfield  {journal}
  {\bibinfo  {journal} {Science}\ }\textbf {\bibinfo {volume} {374}},\ \bibinfo
  {pages} {1385–1389} (\bibinfo {year} {2021})},\ \Eprint
  {http://arxiv.org/abs/https://www.science.org/doi/pdf/10.1126/science.abj6511}
  {https://www.science.org/doi/pdf/10.1126/science.abj6511} \BibitemShut
  {NoStop}%
\bibitem [{\citenamefont {Peach}\ \emph {et~al.}(2007)\citenamefont {Peach},
  \citenamefont {Teale},\ and\ \citenamefont {Tozer}}]{Peach2007modeling}%
  \BibitemOpen
  \bibfield  {author} {\bibinfo {author} {\bibfnamefont {M.~J.~G.}\
  \bibnamefont {Peach}}, \bibinfo {author} {\bibfnamefont {A.~M.}\ \bibnamefont
  {Teale}}, \ and\ \bibinfo {author} {\bibfnamefont {D.~J.}\ \bibnamefont
  {Tozer}},\ }\href@noop {} {\bibfield  {journal} {\bibinfo  {journal} {The
  Journal of Chemical Physics}\ }\textbf {\bibinfo {volume} {126}},\ \bibinfo
  {pages} {244104} (\bibinfo {year} {2007})}\BibitemShut {NoStop}%
\bibitem [{\citenamefont {Fabiano}\ \emph {et~al.}(2014)\citenamefont
  {Fabiano}, \citenamefont {Trevisanutto}, \citenamefont {Terentjevs},\ and\
  \citenamefont {Constantin}}]{fabiano2014generalized}%
  \BibitemOpen
  \bibfield  {author} {\bibinfo {author} {\bibfnamefont {E.}~\bibnamefont
  {Fabiano}}, \bibinfo {author} {\bibfnamefont {P.~E.}\ \bibnamefont
  {Trevisanutto}}, \bibinfo {author} {\bibfnamefont {A.}~\bibnamefont
  {Terentjevs}}, \ and\ \bibinfo {author} {\bibfnamefont {L.~A.}\ \bibnamefont
  {Constantin}},\ }\href@noop {} {\bibfield  {journal} {\bibinfo  {journal} {J.
  Chem. Theory Comput.}\ }\textbf {\bibinfo {volume} {10}},\ \bibinfo {pages}
  {2016} (\bibinfo {year} {2014})}\BibitemShut {NoStop}%
\bibitem [{\citenamefont {Krieger}\ \emph {et~al.}(1999)\citenamefont
  {Krieger}, \citenamefont {Chen}, \citenamefont {Iafrate}, \citenamefont
  {Savin}, \citenamefont {Gonis},\ and\ \citenamefont
  {Kioussis}}]{krieger1999electron}%
  \BibitemOpen
  \bibfield  {author} {\bibinfo {author} {\bibfnamefont {J.}~\bibnamefont
  {Krieger}}, \bibinfo {author} {\bibfnamefont {J.}~\bibnamefont {Chen}},
  \bibinfo {author} {\bibfnamefont {G.}~\bibnamefont {Iafrate}}, \bibinfo
  {author} {\bibfnamefont {A.}~\bibnamefont {Savin}}, \bibinfo {author}
  {\bibfnamefont {A.}~\bibnamefont {Gonis}}, \ and\ \bibinfo {author}
  {\bibfnamefont {N.}~\bibnamefont {Kioussis}},\ }\href@noop {} {\bibfield
  {journal} {\bibinfo  {journal} {Kluwer Academic, New York}\ ,\ \bibinfo
  {pages} {463}} (\bibinfo {year} {1999})}\BibitemShut {NoStop}%
\bibitem [{\citenamefont {Constantin}\ \emph
  {et~al.}(2017{\natexlab{b}})\citenamefont {Constantin}, \citenamefont
  {Fabiano}, \citenamefont {\ifmmode~\acute{S}\else \'{S}\fi{}miga},\ and\
  \citenamefont {Della~Sala}}]{PhysRevB.95.115153}%
  \BibitemOpen
  \bibfield  {author} {\bibinfo {author} {\bibfnamefont {L.~A.}\ \bibnamefont
  {Constantin}}, \bibinfo {author} {\bibfnamefont {E.}~\bibnamefont {Fabiano}},
  \bibinfo {author} {\bibfnamefont {S.}~\bibnamefont {\ifmmode~\acute{S}\else
  \'{S}\fi{}miga}}, \ and\ \bibinfo {author} {\bibfnamefont {F.}~\bibnamefont
  {Della~Sala}},\ }\href {\doibase 10.1103/PhysRevB.95.115153} {\bibfield
  {journal} {\bibinfo  {journal} {Phys. Rev. B}\ }\textbf {\bibinfo {volume}
  {95}},\ \bibinfo {pages} {115153} (\bibinfo {year}
  {2017}{\natexlab{b}})}\BibitemShut {NoStop}%
\bibitem [{\citenamefont {Matito}\ \emph {et~al.}(2010)\citenamefont {Matito},
  \citenamefont {Cioslowski},\ and\ \citenamefont {Vyboishchikov}}]{B926389F}%
  \BibitemOpen
  \bibfield  {author} {\bibinfo {author} {\bibfnamefont {E.}~\bibnamefont
  {Matito}}, \bibinfo {author} {\bibfnamefont {J.}~\bibnamefont {Cioslowski}},
  \ and\ \bibinfo {author} {\bibfnamefont {S.~F.}\ \bibnamefont
  {Vyboishchikov}},\ }\href {\doibase 10.1039/B926389F} {\bibfield  {journal}
  {\bibinfo  {journal} {Phys. Chem. Chem. Phys.}\ }\textbf {\bibinfo {volume}
  {12}},\ \bibinfo {pages} {6712} (\bibinfo {year} {2010})}\BibitemShut
  {NoStop}%
\bibitem [{\citenamefont {Lynch}\ and\ \citenamefont
  {Truhlar}(2003)}]{lynch2003small}%
  \BibitemOpen
  \bibfield  {author} {\bibinfo {author} {\bibfnamefont {B.~J.}\ \bibnamefont
  {Lynch}}\ and\ \bibinfo {author} {\bibfnamefont {D.~G.}\ \bibnamefont
  {Truhlar}},\ }\href@noop {} {\bibfield  {journal} {\bibinfo  {journal} {J.
  Phys. Chem. A}\ }\textbf {\bibinfo {volume} {107}},\ \bibinfo {pages} {8996}
  (\bibinfo {year} {2003})}\BibitemShut {NoStop}%
\bibitem [{\citenamefont {Haunschild}\ and\ \citenamefont
  {Klopper}(2012)}]{haunschild2012theoretical}%
  \BibitemOpen
  \bibfield  {author} {\bibinfo {author} {\bibfnamefont {R.}~\bibnamefont
  {Haunschild}}\ and\ \bibinfo {author} {\bibfnamefont {W.}~\bibnamefont
  {Klopper}},\ }\href@noop {} {\bibfield  {journal} {\bibinfo  {journal}
  {Theor. Chem. Acc.}\ }\textbf {\bibinfo {volume} {131}},\ \bibinfo {pages}
  {1112} (\bibinfo {year} {2012})}\BibitemShut {NoStop}%
\bibitem [{\citenamefont {Shao}\ \emph {et~al.}(2015)\citenamefont {Shao},
  \citenamefont {Gan}, \citenamefont {Epifanovsky}, \citenamefont {Gilbert},
  \citenamefont {Wormit}, \citenamefont {Kussmann}, \citenamefont {Lange},
  \citenamefont {Behn}, \citenamefont {Deng}, \citenamefont {Feng},
  \citenamefont {Ghosh}, \citenamefont {Goldey}, \citenamefont {Horn},
  \citenamefont {Jacobson}, \citenamefont {Kaliman}, \citenamefont
  {Khaliullin}, \citenamefont {Kuś}, \citenamefont {Landau}, \citenamefont
  {Liu}, \citenamefont {Proynov}, \citenamefont {Rhee}, \citenamefont
  {Richard}, \citenamefont {Rohrdanz}, \citenamefont {Steele}, \citenamefont
  {Sundstrom}, \citenamefont {III}, \citenamefont {Zimmerman}, \citenamefont
  {Zuev}, \citenamefont {Albrecht}, \citenamefont {Alguire}, \citenamefont
  {Austin}, \citenamefont {Beran}, \citenamefont {Bernard}, \citenamefont
  {Berquist}, \citenamefont {Brandhorst}, \citenamefont {Bravaya},
  \citenamefont {Brown}, \citenamefont {Casanova}, \citenamefont {Chang},
  \citenamefont {Chen}, \citenamefont {Chien}, \citenamefont {Closser},
  \citenamefont {Crittenden}, \citenamefont {Diedenhofen}, \citenamefont {Jr.},
  \citenamefont {Do}, \citenamefont {Dutoi}, \citenamefont {Edgar},
  \citenamefont {Fatehi}, \citenamefont {Fusti-Molnar}, \citenamefont
  {Ghysels}, \citenamefont {Golubeva-Zadorozhnaya}, \citenamefont {Gomes},
  \citenamefont {Hanson-Heine}, \citenamefont {Harbach}, \citenamefont
  {Hauser}, \citenamefont {Hohenstein}, \citenamefont {Holden}, \citenamefont
  {Jagau}, \citenamefont {Ji}, \citenamefont {Kaduk}, \citenamefont
  {Khistyaev}, \citenamefont {Kim}, \citenamefont {Kim}, \citenamefont {King},
  \citenamefont {Klunzinger}, \citenamefont {Kosenkov}, \citenamefont
  {Kowalczyk}, \citenamefont {Krauter}, \citenamefont {Lao}, \citenamefont
  {Laurent}, \citenamefont {Lawler}, \citenamefont {Levchenko}, \citenamefont
  {Lin}, \citenamefont {Liu}, \citenamefont {Livshits}, \citenamefont {Lochan},
  \citenamefont {Luenser}, \citenamefont {Manohar}, \citenamefont {Manzer},
  \citenamefont {Mao}, \citenamefont {Mardirossian}, \citenamefont {Marenich},
  \citenamefont {Maurer}, \citenamefont {Mayhall}, \citenamefont {Neuscamman},
  \citenamefont {Oana}, \citenamefont {Olivares-Amaya}, \citenamefont
  {O’Neill}, \citenamefont {Parkhill}, \citenamefont {Perrine}, \citenamefont
  {Peverati}, \citenamefont {Prociuk}, \citenamefont {Rehn}, \citenamefont
  {Rosta}, \citenamefont {Russ}, \citenamefont {Sharada}, \citenamefont
  {Sharma}, \citenamefont {Small}, \citenamefont {Sodt}, \citenamefont {Stein},
  \citenamefont {Stück}, \citenamefont {Su}, \citenamefont {Thom},
  \citenamefont {Tsuchimochi}, \citenamefont {Vanovschi}, \citenamefont {Vogt},
  \citenamefont {Vydrov}, \citenamefont {Wang}, \citenamefont {Watson},
  \citenamefont {Wenzel}, \citenamefont {White}, \citenamefont {Williams},
  \citenamefont {Yang}, \citenamefont {Yeganeh}, \citenamefont {Yost},
  \citenamefont {You}, \citenamefont {Zhang}, \citenamefont {Zhang},
  \citenamefont {Zhao}, \citenamefont {Brooks}, \citenamefont {Chan},
  \citenamefont {Chipman}, \citenamefont {Cramer}, \citenamefont {III},
  \citenamefont {Gordon}, \citenamefont {Hehre}, \citenamefont {Klamt},
  \citenamefont {III}, \citenamefont {Schmidt}, \citenamefont {Sherrill},
  \citenamefont {Truhlar}, \citenamefont {Warshel}, \citenamefont {Xu},
  \citenamefont {Aspuru-Guzik}, \citenamefont {Baer}, \citenamefont {Bell},
  \citenamefont {Besley}, \citenamefont {Chai}, \citenamefont {Dreuw},
  \citenamefont {Dunietz}, \citenamefont {Furlani}, \citenamefont {Gwaltney},
  \citenamefont {Hsu}, \citenamefont {Jung}, \citenamefont {Kong},
  \citenamefont {Lambrecht}, \citenamefont {Liang}, \citenamefont {Ochsenfeld},
  \citenamefont {Rassolov}, \citenamefont {Slipchenko}, \citenamefont
  {Subotnik}, \citenamefont {Voorhis}, \citenamefont {Herbert}, \citenamefont
  {Krylov}, \citenamefont {Gill},\ and\ \citenamefont {Head-Gordon}}]{qchem}%
  \BibitemOpen
  \bibfield  {author} {\bibinfo {author} {\bibfnamefont {Y.}~\bibnamefont
  {Shao}}, \bibinfo {author} {\bibfnamefont {Z.}~\bibnamefont {Gan}}, \bibinfo
  {author} {\bibfnamefont {E.}~\bibnamefont {Epifanovsky}}, \bibinfo {author}
  {\bibfnamefont {A.~T.}\ \bibnamefont {Gilbert}}, \bibinfo {author}
  {\bibfnamefont {M.}~\bibnamefont {Wormit}}, \bibinfo {author} {\bibfnamefont
  {J.}~\bibnamefont {Kussmann}}, \bibinfo {author} {\bibfnamefont {A.~W.}\
  \bibnamefont {Lange}}, \bibinfo {author} {\bibfnamefont {A.}~\bibnamefont
  {Behn}}, \bibinfo {author} {\bibfnamefont {J.}~\bibnamefont {Deng}}, \bibinfo
  {author} {\bibfnamefont {X.}~\bibnamefont {Feng}}, \bibinfo {author}
  {\bibfnamefont {D.}~\bibnamefont {Ghosh}}, \bibinfo {author} {\bibfnamefont
  {M.}~\bibnamefont {Goldey}}, \bibinfo {author} {\bibfnamefont {P.~R.}\
  \bibnamefont {Horn}}, \bibinfo {author} {\bibfnamefont {L.~D.}\ \bibnamefont
  {Jacobson}}, \bibinfo {author} {\bibfnamefont {I.}~\bibnamefont {Kaliman}},
  \bibinfo {author} {\bibfnamefont {R.~Z.}\ \bibnamefont {Khaliullin}},
  \bibinfo {author} {\bibfnamefont {T.}~\bibnamefont {Kuś}}, \bibinfo {author}
  {\bibfnamefont {A.}~\bibnamefont {Landau}}, \bibinfo {author} {\bibfnamefont
  {J.}~\bibnamefont {Liu}}, \bibinfo {author} {\bibfnamefont {E.~I.}\
  \bibnamefont {Proynov}}, \bibinfo {author} {\bibfnamefont {Y.~M.}\
  \bibnamefont {Rhee}}, \bibinfo {author} {\bibfnamefont {R.~M.}\ \bibnamefont
  {Richard}}, \bibinfo {author} {\bibfnamefont {M.~A.}\ \bibnamefont
  {Rohrdanz}}, \bibinfo {author} {\bibfnamefont {R.~P.}\ \bibnamefont
  {Steele}}, \bibinfo {author} {\bibfnamefont {E.~J.}\ \bibnamefont
  {Sundstrom}}, \bibinfo {author} {\bibfnamefont {H.~L.~W.}\ \bibnamefont
  {III}}, \bibinfo {author} {\bibfnamefont {P.~M.}\ \bibnamefont {Zimmerman}},
  \bibinfo {author} {\bibfnamefont {D.}~\bibnamefont {Zuev}}, \bibinfo {author}
  {\bibfnamefont {B.}~\bibnamefont {Albrecht}}, \bibinfo {author}
  {\bibfnamefont {E.}~\bibnamefont {Alguire}}, \bibinfo {author} {\bibfnamefont
  {B.}~\bibnamefont {Austin}}, \bibinfo {author} {\bibfnamefont {G.~J.~O.}\
  \bibnamefont {Beran}}, \bibinfo {author} {\bibfnamefont {Y.~A.}\ \bibnamefont
  {Bernard}}, \bibinfo {author} {\bibfnamefont {E.}~\bibnamefont {Berquist}},
  \bibinfo {author} {\bibfnamefont {K.}~\bibnamefont {Brandhorst}}, \bibinfo
  {author} {\bibfnamefont {K.~B.}\ \bibnamefont {Bravaya}}, \bibinfo {author}
  {\bibfnamefont {S.~T.}\ \bibnamefont {Brown}}, \bibinfo {author}
  {\bibfnamefont {D.}~\bibnamefont {Casanova}}, \bibinfo {author}
  {\bibfnamefont {C.-M.}\ \bibnamefont {Chang}}, \bibinfo {author}
  {\bibfnamefont {Y.}~\bibnamefont {Chen}}, \bibinfo {author} {\bibfnamefont
  {S.~H.}\ \bibnamefont {Chien}}, \bibinfo {author} {\bibfnamefont {K.~D.}\
  \bibnamefont {Closser}}, \bibinfo {author} {\bibfnamefont {D.~L.}\
  \bibnamefont {Crittenden}}, \bibinfo {author} {\bibfnamefont
  {M.}~\bibnamefont {Diedenhofen}}, \bibinfo {author} {\bibfnamefont
  {R.~A.~D.}\ \bibnamefont {Jr.}}, \bibinfo {author} {\bibfnamefont
  {H.}~\bibnamefont {Do}}, \bibinfo {author} {\bibfnamefont {A.~D.}\
  \bibnamefont {Dutoi}}, \bibinfo {author} {\bibfnamefont {R.~G.}\ \bibnamefont
  {Edgar}}, \bibinfo {author} {\bibfnamefont {S.}~\bibnamefont {Fatehi}},
  \bibinfo {author} {\bibfnamefont {L.}~\bibnamefont {Fusti-Molnar}}, \bibinfo
  {author} {\bibfnamefont {A.}~\bibnamefont {Ghysels}}, \bibinfo {author}
  {\bibfnamefont {A.}~\bibnamefont {Golubeva-Zadorozhnaya}}, \bibinfo {author}
  {\bibfnamefont {J.}~\bibnamefont {Gomes}}, \bibinfo {author} {\bibfnamefont
  {M.~W.}\ \bibnamefont {Hanson-Heine}}, \bibinfo {author} {\bibfnamefont
  {P.~H.}\ \bibnamefont {Harbach}}, \bibinfo {author} {\bibfnamefont {A.~W.}\
  \bibnamefont {Hauser}}, \bibinfo {author} {\bibfnamefont {E.~G.}\
  \bibnamefont {Hohenstein}}, \bibinfo {author} {\bibfnamefont {Z.~C.}\
  \bibnamefont {Holden}}, \bibinfo {author} {\bibfnamefont {T.-C.}\
  \bibnamefont {Jagau}}, \bibinfo {author} {\bibfnamefont {H.}~\bibnamefont
  {Ji}}, \bibinfo {author} {\bibfnamefont {B.}~\bibnamefont {Kaduk}}, \bibinfo
  {author} {\bibfnamefont {K.}~\bibnamefont {Khistyaev}}, \bibinfo {author}
  {\bibfnamefont {J.}~\bibnamefont {Kim}}, \bibinfo {author} {\bibfnamefont
  {J.}~\bibnamefont {Kim}}, \bibinfo {author} {\bibfnamefont {R.~A.}\
  \bibnamefont {King}}, \bibinfo {author} {\bibfnamefont {P.}~\bibnamefont
  {Klunzinger}}, \bibinfo {author} {\bibfnamefont {D.}~\bibnamefont
  {Kosenkov}}, \bibinfo {author} {\bibfnamefont {T.}~\bibnamefont {Kowalczyk}},
  \bibinfo {author} {\bibfnamefont {C.~M.}\ \bibnamefont {Krauter}}, \bibinfo
  {author} {\bibfnamefont {K.~U.}\ \bibnamefont {Lao}}, \bibinfo {author}
  {\bibfnamefont {A.~D.}\ \bibnamefont {Laurent}}, \bibinfo {author}
  {\bibfnamefont {K.~V.}\ \bibnamefont {Lawler}}, \bibinfo {author}
  {\bibfnamefont {S.~V.}\ \bibnamefont {Levchenko}}, \bibinfo {author}
  {\bibfnamefont {C.~Y.}\ \bibnamefont {Lin}}, \bibinfo {author} {\bibfnamefont
  {F.}~\bibnamefont {Liu}}, \bibinfo {author} {\bibfnamefont {E.}~\bibnamefont
  {Livshits}}, \bibinfo {author} {\bibfnamefont {R.~C.}\ \bibnamefont
  {Lochan}}, \bibinfo {author} {\bibfnamefont {A.}~\bibnamefont {Luenser}},
  \bibinfo {author} {\bibfnamefont {P.}~\bibnamefont {Manohar}}, \bibinfo
  {author} {\bibfnamefont {S.~F.}\ \bibnamefont {Manzer}}, \bibinfo {author}
  {\bibfnamefont {S.-P.}\ \bibnamefont {Mao}}, \bibinfo {author} {\bibfnamefont
  {N.}~\bibnamefont {Mardirossian}}, \bibinfo {author} {\bibfnamefont {A.~V.}\
  \bibnamefont {Marenich}}, \bibinfo {author} {\bibfnamefont {S.~A.}\
  \bibnamefont {Maurer}}, \bibinfo {author} {\bibfnamefont {N.~J.}\
  \bibnamefont {Mayhall}}, \bibinfo {author} {\bibfnamefont {E.}~\bibnamefont
  {Neuscamman}}, \bibinfo {author} {\bibfnamefont {C.~M.}\ \bibnamefont
  {Oana}}, \bibinfo {author} {\bibfnamefont {R.}~\bibnamefont
  {Olivares-Amaya}}, \bibinfo {author} {\bibfnamefont {D.~P.}\ \bibnamefont
  {O’Neill}}, \bibinfo {author} {\bibfnamefont {J.~A.}\ \bibnamefont
  {Parkhill}}, \bibinfo {author} {\bibfnamefont {T.~M.}\ \bibnamefont
  {Perrine}}, \bibinfo {author} {\bibfnamefont {R.}~\bibnamefont {Peverati}},
  \bibinfo {author} {\bibfnamefont {A.}~\bibnamefont {Prociuk}}, \bibinfo
  {author} {\bibfnamefont {D.~R.}\ \bibnamefont {Rehn}}, \bibinfo {author}
  {\bibfnamefont {E.}~\bibnamefont {Rosta}}, \bibinfo {author} {\bibfnamefont
  {N.~J.}\ \bibnamefont {Russ}}, \bibinfo {author} {\bibfnamefont {S.~M.}\
  \bibnamefont {Sharada}}, \bibinfo {author} {\bibfnamefont {S.}~\bibnamefont
  {Sharma}}, \bibinfo {author} {\bibfnamefont {D.~W.}\ \bibnamefont {Small}},
  \bibinfo {author} {\bibfnamefont {A.}~\bibnamefont {Sodt}}, \bibinfo {author}
  {\bibfnamefont {T.}~\bibnamefont {Stein}}, \bibinfo {author} {\bibfnamefont
  {D.}~\bibnamefont {Stück}}, \bibinfo {author} {\bibfnamefont {Y.-C.}\
  \bibnamefont {Su}}, \bibinfo {author} {\bibfnamefont {A.~J.}\ \bibnamefont
  {Thom}}, \bibinfo {author} {\bibfnamefont {T.}~\bibnamefont {Tsuchimochi}},
  \bibinfo {author} {\bibfnamefont {V.}~\bibnamefont {Vanovschi}}, \bibinfo
  {author} {\bibfnamefont {L.}~\bibnamefont {Vogt}}, \bibinfo {author}
  {\bibfnamefont {O.}~\bibnamefont {Vydrov}}, \bibinfo {author} {\bibfnamefont
  {T.}~\bibnamefont {Wang}}, \bibinfo {author} {\bibfnamefont {M.~A.}\
  \bibnamefont {Watson}}, \bibinfo {author} {\bibfnamefont {J.}~\bibnamefont
  {Wenzel}}, \bibinfo {author} {\bibfnamefont {A.}~\bibnamefont {White}},
  \bibinfo {author} {\bibfnamefont {C.~F.}\ \bibnamefont {Williams}}, \bibinfo
  {author} {\bibfnamefont {J.}~\bibnamefont {Yang}}, \bibinfo {author}
  {\bibfnamefont {S.}~\bibnamefont {Yeganeh}}, \bibinfo {author} {\bibfnamefont
  {S.~R.}\ \bibnamefont {Yost}}, \bibinfo {author} {\bibfnamefont {Z.-Q.}\
  \bibnamefont {You}}, \bibinfo {author} {\bibfnamefont {I.~Y.}\ \bibnamefont
  {Zhang}}, \bibinfo {author} {\bibfnamefont {X.}~\bibnamefont {Zhang}},
  \bibinfo {author} {\bibfnamefont {Y.}~\bibnamefont {Zhao}}, \bibinfo {author}
  {\bibfnamefont {B.~R.}\ \bibnamefont {Brooks}}, \bibinfo {author}
  {\bibfnamefont {G.~K.}\ \bibnamefont {Chan}}, \bibinfo {author}
  {\bibfnamefont {D.~M.}\ \bibnamefont {Chipman}}, \bibinfo {author}
  {\bibfnamefont {C.~J.}\ \bibnamefont {Cramer}}, \bibinfo {author}
  {\bibfnamefont {W.~A.~G.}\ \bibnamefont {III}}, \bibinfo {author}
  {\bibfnamefont {M.~S.}\ \bibnamefont {Gordon}}, \bibinfo {author}
  {\bibfnamefont {W.~J.}\ \bibnamefont {Hehre}}, \bibinfo {author}
  {\bibfnamefont {A.}~\bibnamefont {Klamt}}, \bibinfo {author} {\bibfnamefont
  {H.~F.~S.}\ \bibnamefont {III}}, \bibinfo {author} {\bibfnamefont {M.~W.}\
  \bibnamefont {Schmidt}}, \bibinfo {author} {\bibfnamefont {C.~D.}\
  \bibnamefont {Sherrill}}, \bibinfo {author} {\bibfnamefont {D.~G.}\
  \bibnamefont {Truhlar}}, \bibinfo {author} {\bibfnamefont {A.}~\bibnamefont
  {Warshel}}, \bibinfo {author} {\bibfnamefont {X.}~\bibnamefont {Xu}},
  \bibinfo {author} {\bibfnamefont {A.}~\bibnamefont {Aspuru-Guzik}}, \bibinfo
  {author} {\bibfnamefont {R.}~\bibnamefont {Baer}}, \bibinfo {author}
  {\bibfnamefont {A.~T.}\ \bibnamefont {Bell}}, \bibinfo {author}
  {\bibfnamefont {N.~A.}\ \bibnamefont {Besley}}, \bibinfo {author}
  {\bibfnamefont {J.-D.}\ \bibnamefont {Chai}}, \bibinfo {author}
  {\bibfnamefont {A.}~\bibnamefont {Dreuw}}, \bibinfo {author} {\bibfnamefont
  {B.~D.}\ \bibnamefont {Dunietz}}, \bibinfo {author} {\bibfnamefont {T.~R.}\
  \bibnamefont {Furlani}}, \bibinfo {author} {\bibfnamefont {S.~R.}\
  \bibnamefont {Gwaltney}}, \bibinfo {author} {\bibfnamefont {C.-P.}\
  \bibnamefont {Hsu}}, \bibinfo {author} {\bibfnamefont {Y.}~\bibnamefont
  {Jung}}, \bibinfo {author} {\bibfnamefont {J.}~\bibnamefont {Kong}}, \bibinfo
  {author} {\bibfnamefont {D.~S.}\ \bibnamefont {Lambrecht}}, \bibinfo {author}
  {\bibfnamefont {W.}~\bibnamefont {Liang}}, \bibinfo {author} {\bibfnamefont
  {C.}~\bibnamefont {Ochsenfeld}}, \bibinfo {author} {\bibfnamefont {V.~A.}\
  \bibnamefont {Rassolov}}, \bibinfo {author} {\bibfnamefont {L.~V.}\
  \bibnamefont {Slipchenko}}, \bibinfo {author} {\bibfnamefont {J.~E.}\
  \bibnamefont {Subotnik}}, \bibinfo {author} {\bibfnamefont {T.~V.}\
  \bibnamefont {Voorhis}}, \bibinfo {author} {\bibfnamefont {J.~M.}\
  \bibnamefont {Herbert}}, \bibinfo {author} {\bibfnamefont {A.~I.}\
  \bibnamefont {Krylov}}, \bibinfo {author} {\bibfnamefont {P.~M.}\
  \bibnamefont {Gill}}, \ and\ \bibinfo {author} {\bibfnamefont
  {M.}~\bibnamefont {Head-Gordon}},\ }\href@noop {} {\bibfield  {journal}
  {\bibinfo  {journal} {Molecular Physics}\ }\textbf {\bibinfo {volume}
  {113}},\ \bibinfo {pages} {184} (\bibinfo {year} {2015})}\BibitemShut
  {NoStop}%
\bibitem [{\citenamefont {Curtiss}\ \emph {et~al.}(1997)\citenamefont
  {Curtiss}, \citenamefont {Raghavachari}, \citenamefont {Redfern},\ and\
  \citenamefont {Pople}}]{curtiss1997assessment}%
  \BibitemOpen
  \bibfield  {author} {\bibinfo {author} {\bibfnamefont {L.~A.}\ \bibnamefont
  {Curtiss}}, \bibinfo {author} {\bibfnamefont {K.}~\bibnamefont
  {Raghavachari}}, \bibinfo {author} {\bibfnamefont {P.~C.}\ \bibnamefont
  {Redfern}}, \ and\ \bibinfo {author} {\bibfnamefont {J.~A.}\ \bibnamefont
  {Pople}},\ }\href@noop {} {\bibfield  {journal} {\bibinfo  {journal} {The
  Journal of Chemical Physics}\ }\textbf {\bibinfo {volume} {106}},\ \bibinfo
  {pages} {1063} (\bibinfo {year} {1997})}\BibitemShut {NoStop}%
\bibitem [{\citenamefont {Zhao}\ and\ \citenamefont
  {Truhlar}(2005{\natexlab{a}})}]{zhao2005benchmark}%
  \BibitemOpen
  \bibfield  {author} {\bibinfo {author} {\bibfnamefont {Y.}~\bibnamefont
  {Zhao}}\ and\ \bibinfo {author} {\bibfnamefont {D.~G.}\ \bibnamefont
  {Truhlar}},\ }\href@noop {} {\bibfield  {journal} {\bibinfo  {journal}
  {Journal of Chemical Theory and Computation}\ }\textbf {\bibinfo {volume}
  {1}},\ \bibinfo {pages} {415} (\bibinfo {year}
  {2005}{\natexlab{a}})}\BibitemShut {NoStop}%
\bibitem [{\citenamefont {Zhao}\ and\ \citenamefont
  {Truhlar}(2005{\natexlab{b}})}]{zhao2005design}%
  \BibitemOpen
  \bibfield  {author} {\bibinfo {author} {\bibfnamefont {Y.}~\bibnamefont
  {Zhao}}\ and\ \bibinfo {author} {\bibfnamefont {D.~G.}\ \bibnamefont
  {Truhlar}},\ }\href@noop {} {\bibfield  {journal} {\bibinfo  {journal} {The
  Journal of Physical Chemistry A}\ }\textbf {\bibinfo {volume} {109}},\
  \bibinfo {pages} {5656} (\bibinfo {year} {2005}{\natexlab{b}})}\BibitemShut
  {NoStop}%
\bibitem [{\citenamefont {Goerigk}\ \emph {et~al.}(2017)\citenamefont
  {Goerigk}, \citenamefont {Hansen}, \citenamefont {Bauer}, \citenamefont
  {Ehrlich}, \citenamefont {Najibi},\ and\ \citenamefont
  {Grimme}}]{goerigk2017look}%
  \BibitemOpen
  \bibfield  {author} {\bibinfo {author} {\bibfnamefont {L.}~\bibnamefont
  {Goerigk}}, \bibinfo {author} {\bibfnamefont {A.}~\bibnamefont {Hansen}},
  \bibinfo {author} {\bibfnamefont {C.}~\bibnamefont {Bauer}}, \bibinfo
  {author} {\bibfnamefont {S.}~\bibnamefont {Ehrlich}}, \bibinfo {author}
  {\bibfnamefont {A.}~\bibnamefont {Najibi}}, \ and\ \bibinfo {author}
  {\bibfnamefont {S.}~\bibnamefont {Grimme}},\ }\href@noop {} {\bibfield
  {journal} {\bibinfo  {journal} {Phys. Chem. Chem. Phys.}\ }\textbf {\bibinfo
  {volume} {19}},\ \bibinfo {pages} {32184} (\bibinfo {year}
  {2017})}\BibitemShut {NoStop}%
\end{thebibliography}%

\end{document}